\documentclass[%
 reprint,
 amsmath,amssymb,
 aps,
floatfix,
]{revtex4-2}
\usepackage{afterpage}
\usepackage{graphicx}
\usepackage{dcolumn}
\usepackage{bm}
\usepackage[tracking=true]{microtype}
\usepackage{braket}
\usepackage{siunitx}
\usepackage[dvipsnames]{xcolor}
\usepackage{hyperref}

\hypersetup{
	colorlinks=true,       
	linkcolor=blue,  
    citecolor=JungleGreen,        
    filecolor=Orchid,      
    urlcolor=black           
}

\newcommand{\RN}[1]{%
  \textup{\uppercase\expandafter{\romannumeral#1}}%
  }
\usepackage{tikz}
\newcommand{\m}[1]{\boldsymbol{#1}}
\newcommand{\T}[1]{\text{#1}}
\newcommand{\dg}{^\dagger}

\usepackage{verbatim}
\newcommand{\vac}{\ket{\text{vac}}}

\numberwithin{equation}{section}
\begin{document}
\preprint{APS/123-QED}

\title{Taking apart squeezed light}
\author{C. Drago}
 \email{christian.drago@mail.utoronto.ca}

\author{J. E. Sipe}%
\email{sipe@physics.utoronto.ca}
\affiliation{%
 Department of Physics, University of Toronto, 60 St. George Street, Toronto, Ontario, Canada, M5S 1A7
}%
\date{\today}
\begin{abstract}
We develop a formalism to describe squeezed light with large spectral-temporal correlations. This description is valid in all regimes, but is especially applicable in the long pulse to continuous-wave limit where the photon density at any particular time is small, although the total number of photons can be quite large. Our method relies on the Whittaker-Shannon interpolation formula applied to the joint temporal amplitude of squeezed light, which allows us to ``take apart'' the squeezed state. This provides a local description of the state and its photon statistics, making the underlying physics more transparent than does the use of the Schmidt decomposition. The formalism can easily be extended to more exotic nonclassical states where a Schmidt decomposition is not possible.
\end{abstract}
\maketitle
\section{Introduction}
Squeezed light is of interest for applications in quantum sensing and imaging \cite{pirandola2018advances,kolobov2007quantum}, and as a resource for quantum computing \cite{bourassa2021blueprint}.
For light propagating in one direction 
in a quasi-1D structure, such as an optical fiber or a channel waveguide in an integrated photonic structure \cite{quesada2022beyond}, or even light propagating in free space under the approximation that diffraction is negligible, a squeezed state can be written as 
\begin{align}
 & \left|\Psi\right\rangle =e^{\frac{\beta}{2}\int d\omega_{1}d\omega_{2}\gamma(\omega_{1},\omega_{2})a^{\dagger}(\omega_{1})a^{\dagger}(\omega_{2})-\text{h.c.}}\left|\text{vac}\right\rangle ,\label{eq:squeezed state}
\end{align}
where for simplicity only one polarization and one transverse mode is considered. We label the lowering operator at a frequency shifted by $\omega$ from a center  reference frequency $\omega_{o}$ 
by $a(\omega)$ (see Appendix \ref{sec:app:field operator}),
\begin{align}
\label{eq:comm}
 & [a(\omega_{1}),a^{\dagger}(\omega_{2})]=\delta(\omega_{1}-\omega_{2}),
\end{align}
and $\left|\text{vac}\right\rangle $ is the vacuum state. Here $\gamma(\omega_{1},\omega_{2})$
is the joint spectral amplitude, 
\begin{align}
\label{eq:jtanorm}
 & \int\left|\gamma(\omega_{1},\omega_{2})\right|^{2}d\omega_{1}d\omega_{2}=1,
\end{align}
and $\beta$ is the squeezing amplitude; unless otherwise indicated,
we take integrals to range from $-\infty$ to $\infty$. 

Often the properties of interest can be captured by simple functions of frequencies, such as correlation functions of the form 
\begin{subequations}
\label{eq:gs}
\begin{gather}
    G^{(1)}(\omega)=\left\langle \Psi|a^{\dagger}(\omega)a(\omega)|\Psi\right\rangle\\
     G^{(2)}(\omega_{1},\omega_{2})=\left\langle \Psi|a^{\dagger}(\omega_{1})a^{\dagger}(\omega_{2})a(\omega_{2})a(\omega_{1})|\Psi\right\rangle,
\end{gather}
\end{subequations}
etc. For a pulse of light where $\left|\beta\right|\ll1$,
the state is only slightly different from the vacuum state, 
\begin{equation}
\label{eq:PsiOmega}
    \begin{split}
        \left|\Psi\right\rangle \approx |\text{vac}\rangle&+\frac{\beta}{2}\int d\omega_{1}d\omega_{2}\gamma(\omega_{1},\omega_{2})a^{\dagger}(\omega_{1})a^{\dagger}(\omega_{2})\left|\text{vac}\right\rangle
    \end{split}
\end{equation}
where higher order terms in $\beta$ have been neglected, and there is only a small probability amplitude for a two-photon state. Then
\begin{subequations}
\label{eq:smallbeta_omega}
    \begin{gather}
    G^{(1)}(\omega)\rightarrow\left|\beta\right|^{2}\int\left|\gamma(\omega,\omega')\right|^{2}d\omega',\\
    G^{(2)}(\omega_{1},\omega_{2})\rightarrow\left|\beta\right|^{2}\left|\gamma(\omega_{1},\omega_{2})\right|^{2}.
    \end{gather}
\end{subequations}

In this paper we consider the evaluation of quantities such as these, even when $\left|\beta\right|$ is not much less than one. A standard strategy in such a situation is to decompose the joint spectral amplitude in terms of Schmidt modes. If there is only one or a few Schmidt modes, as might occur for squeezed light generated by a short 
pump pulse, the expressions for correlation functions of the squeezed light in terms of Schmidt modes can be easily evaluated even if $\left|\beta\right|$ is large, and they immediately identify much of the physics. But large values of $\left|\beta\right|$ can also arise for squeezed light generated by pump pulses that are very long,  and even if their intensities are very weak. Here, although the rate at which pairs of photons are generated 
may be quite small, the total number of pairs of photons generated diverges as the pump pulse approaches CW excitation, and thus both $\left|\beta\right|$ and the Schmidt number diverge.
Our goal is  
to identify strategies that allow for the 
calculation of quantities such as correlation functions to be done quickly for such states, and in a way that makes the physics of the squeezed light clear.

We begin by introducing the temporal representation of the joint spectral amplitude $\overline\gamma(t_{1},t_{2})$ and some of its general features in Section \ref{sec:Joint amplitude}.  Then in Section \ref{sec:Schmidt modes} we consider a natural first approach, which is to use the Schmidt decomposition of $\overline\gamma(t_{1},t_{2})$ even if the Schmidt number is very large. We find that this approach does not directly make the physics of the state $\left|\Psi\right\rangle$ apparent, and this motivates our search for other ways to ``take apart" the joint amplitude of the squeezed light that better elucidate the physics. In section \ref{sec:An approximate Schmidt decomposition} we introduce a new approach suggested by the time correlation functions of the light; it works well if there is significant degeneracy in the amplitudes of the Schmidt decomposition. In section \ref{sec:The Whittaker-Shannon decomposition} we generalize this, based on the Whittaker-Shannon interpolation formula, in a scheme that is applicable even if there is no such degeneracy. We argue that this new way of ``taking apart" the joint amplitude does make the physics of the state more apparent, and in section \ref{sec:Employing the Whittaker-Shannon decomposition} we compare it to the Schmidt decomposition. Then using our formalism we provide a ``local decomposition" of the squeezed state, and demonstrate its use in calculations in section \ref{sec:Local states and calculation of correlation functions}. We give a discussion of the ``strongly squeezed limit'' within our framework in section \ref{sec:The strongly squeezed limit}, and end in section \ref{sec:A final example} with a realistic example of a joint spectral amplitude for a squeezed state generated in a ring resonator structure. Our conclusions and suggestions for future work are presented in section \ref{eq:conclusion}.

\section{Joint temporal amplitude}
\label{sec:Joint amplitude}
Besides the expression (\ref{eq:squeezed state}) for a squeezed state that is based on an integral over frequencies (or wavenumbers), it will be useful to have an expression based on integrals over time (or position). For simplicity we assume group velocity dispersion can be neglected and that light propagates with a velocity $v$; then putting 
\begin{align}
 & \overline{a}(t)=\int\frac{d\omega}{\sqrt{2\pi}}a(\omega)e^{-i\omega t},\label{eq:abar}
\end{align}
and with
\begin{align}
 & \overline{\gamma}(t_{1},t_{2})\equiv\int\frac{d\omega_{1}d\omega_{2}}{2\pi}\gamma(\omega_{1},\omega_{2})e^{-i\omega_{1}t_{1}}e^{-i\omega_{2}t_{2}}
\end{align}
identifying the ``joint temporal amplitude,'' we can write Eq. \eqref{eq:squeezed state}
as 
\begin{align}
 & \left|\Psi\right\rangle =e^{\frac{\beta}{2}\int dt_{1}dt_{2}\overline{\gamma}(t_{1},t_{2})\overline{a}^{\dagger}(t_{1})\overline{a}^{\dagger}(t_{2})-\text{h.c.}}\left|\text{vac}\right\rangle .\label{eq:PsiTime}
\end{align}

We take $\gamma(\omega_{1},\omega_{2})$ and $\overline{\gamma}(t_{1},t_{2})$
to identify the spectral and temporal representations of a ``joint
amplitude'' and refer to their absolute squares $|\gamma(\omega_{1},\omega_{2})|^2$ and $|\overline{\gamma}(t_{1},t_{2})|^2$ as the spectral and temporal representations of a ``joint intensity.'' While due to time-ordering corrections we would expect the joint amplitude to change as the pump intensity and thus $\beta$ is increased \cite{quesada2014effects}, here we neglect such effects for simplicity, and take the joint amplitude to be fixed when we consider varying $\beta$ below.

Of course, the ket $\left|\Psi\right\rangle $ (Eq. (\ref{eq:squeezed state})
or (\ref{eq:PsiTime})) is a Schrödinger ket at a particular time,
say $t=0.$ The variables $t_{1},t_{2}$ can be thought of as surrogates
for position, where $\overline{a}(t_{1})$ is identified with the
electric field at $z=-vt$$_{1}$; equivalently, if the ket $\left|\Psi\right\rangle $
were allowed to evolve in time, $\overline{a}(t)$ would identify
the field operator at $z=0$ at time $t$ (see Appendix \ref{sec:app:field operator}). 

Corresponding to the frequency correlation functions (Eq. \eqref{eq:gs}) we can also introduce time dependent first- and second-order correlation functions \cite{glauber1963quantum}
\begin{subequations}
\label{eq:gs-1}
    \begin{gather}
        \overline{G}^{(1)}(t_1,t_2)=\left\langle \Psi|\overline{a}^{\dagger}(t_1)\overline{a}(t_2)|\Psi\right\rangle ,\\
  \overline{G}^{(2)}(t_{1},t_{2})=\left\langle \Psi|\overline{a}^{\dagger}(t_{1})\overline{a}^{\dagger}(t_{2})\overline{a}(t_{2})\overline{a}(t_{1})|\Psi\right\rangle.
    \end{gather}
\end{subequations}
The ``equal-time'' first-order correlation function, given by $\overline G^{(1)}(t)\equiv \overline G^{(1)}(t,t)$, is the ``photon-density'' of the pulse of light, and is used to predict the counting rate of an ideal photo-detector; indeed, if we integrate over all time, then 
\begin{equation}
\label{eq:N_pulse}
    N_\text{pulse} = \int \overline G^{(1)}(t) dt
\end{equation}
is the expected photon number in the pulse. The second-order correlation function $\overline G^{(2)}(t_1,t_2)$ has a similar interpretation, and is used to predict the probability of detection coincidences at the indicated times.

\begin{figure}
    \centering
    \includegraphics[width = \linewidth]{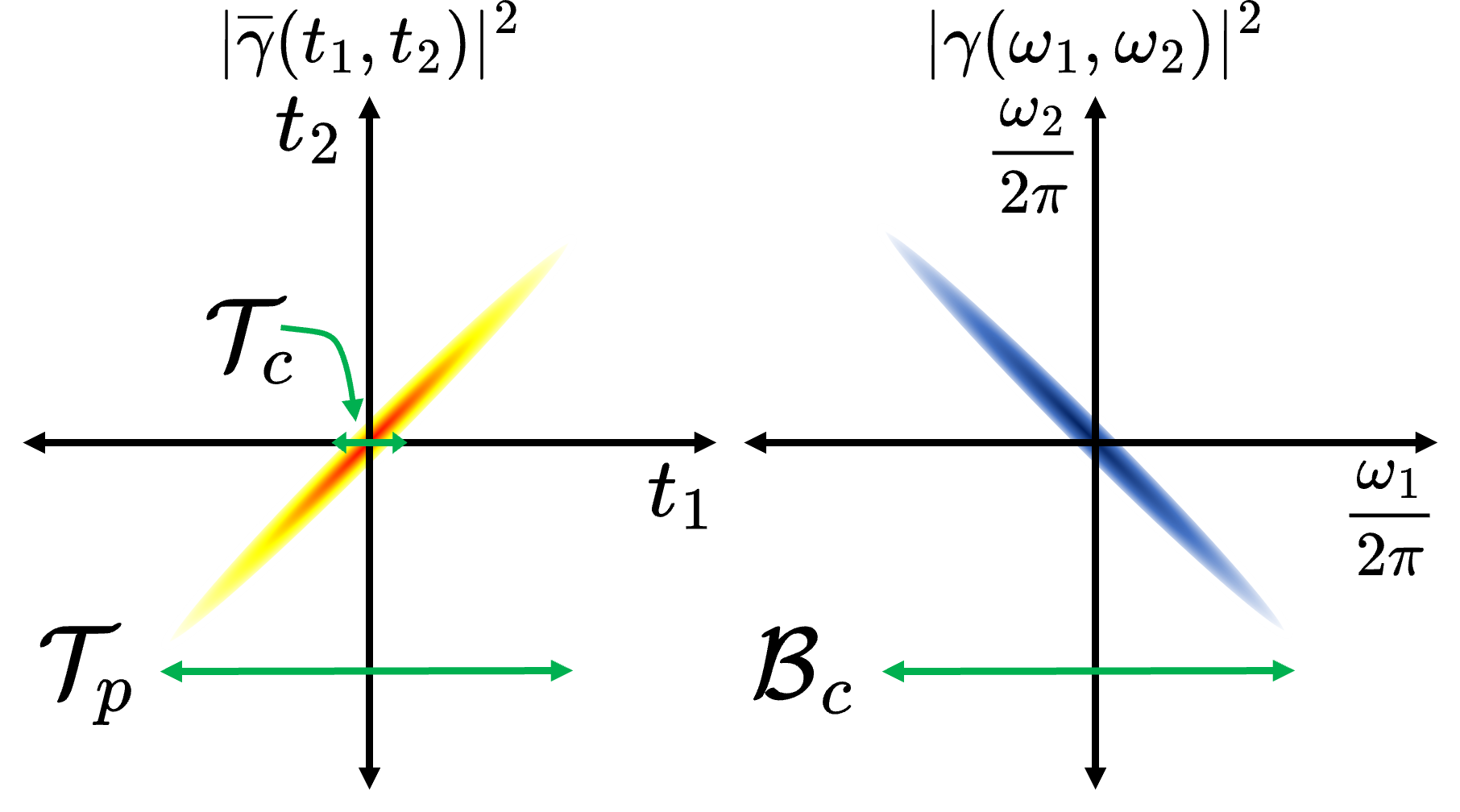}
    \caption{Schematic of a joint intensity represented in time on the left and frequency on the right. The horizontal width of the joint temporal (spectral) amplitude is denoted by $\mathcal{T}_p$, ($\mathcal{B}_c$, the bandwidth) is the effective pulse duration (bandwidth). The narrow horizontal width at $t_2 = 0$ is denoted by $\mathcal{T}_c = 1/\mathcal{B}_c$ and is the coherence time of photon pairs.}
    \label{fig:picture1}
\end{figure}

We will primarily be interested in joint intensities of the form shown schematically in Fig. \ref{fig:picture1}, where there are 
large spectral-temporal correlations; however, the formalism we introduce
is valid for a general joint amplitude. We characterize the joint intensity by two quantities, $\mathcal{T}_p$ and $\mathcal{B}_c$, indicated in Fig. \ref{fig:picture1}. 

The quantity $\mathcal{T}_p$ is an effective measure of the duration of the generated pulse of squeezed light; in practice this will depend on the pump pulse duration 
if a nonresonant structure is used to generate the light, or by the resonance time if a resonant structure is used.

The second quantity, $\mathcal{B}_c$, is an effective measure of the bandwidth of generated photons; in practice this is set by phase-matching constraints if a nonresonant structure is used to generate the light, or by the resonator linewidth if a resonant structure is used. Associated with this bandwidth we define a time $\mathcal{T}_c = 1/\mathcal{B}_c$, which is on the order of the narrow width of the joint temporal intensity; see Fig. \ref{fig:picture1}. The time $\mathcal{T}_c$ identifies the ``coherence time,'' the typical range of $\left|t_{2}-t_{1}\right|$ over which $\left|\overline{\gamma}(t_{1},t_{2})\right|^{2}$ is non-negligible.

\section{Schmidt modes}
\label{sec:Schmidt modes}
A natural approach to try to elucidate the physics of a squeezed state and calculate the correlation functions given by Eq. \eqref{eq:gs-1} is to employ a Schmidt decomposition of the joint amplitude, since the squeezed state can then be written as a direct product of squeezed states associated with the supermodes introduced via the Schmidt decomposition. That is the strategy we explore in this section. 

In writing Eq. \ref{eq:squeezed state} for the squeezed state we took the origin of $\gamma(\omega_{1},\omega_{2})$ to indicate the same frequency $\omega_o$ with respect to which both $\omega_1$ and $\omega_2$ are referenced -- this is the case of so-called ``degenerate" squeezing -- and here without loss of generality the joint amplitude can be taken as symmetric in its variables
($\gamma(\omega_{2},\omega_{1})=\gamma(\omega_{1},\omega_{2})$, or
equivalently $\overline{\gamma}(t_{2},t_{1})=\overline{\gamma}(t_{1},t_{2})$). From a Takagi factorization \cite{chebotarev2014singular} we can then construct the Schmidt decomposition,
\begin{equation}
    \label{eq:Schmidt}
    \begin{split}
        \gamma(\omega_{1},\omega_{2})&=\sum_{n}\sqrt{p_{n}}f_{n}(\omega_{1})f_{n}(\omega_{2}),\\
        \overline{\gamma}(t_{1},t_{2})&=\sum_{n}\sqrt{p_{n}}\overline{f}_{n}(t_{1})\overline{f}_{n}(t_{2}),
    \end{split}
\end{equation}
where the Schmidt weights $p_{n}\geq0$ sum to unity;
the elements of the sets $\left\{ f_{n}(\omega)\right\} $ and $\left\{ \overline{f}_{n}(t)\right\} $
of mutually orthogonal and normalized functions are related by 
\begin{align}
 & \overline{f}_{n}(t)=\int\frac{d\omega}{\sqrt{2\pi}}f_{n}(\omega)e^{-i\omega t}.\label{eq:fnbar}
\end{align}
As usual, we take the set $\left\{ f_{n}(\omega)\right\} $ of elements with $p_{n}\neq0$ to be expanded to be a complete set of functions, assigning $p_{n}=0$ to the added elements, and correspondingly for the $\left\{ \overline{f}_{n}(t)\right\} $. 

The Schmidt number $K$ of the expansion (\ref{eq:Schmidt}), which characterizes the effective number of spectral or temporal modes in the sum (\ref{eq:Schmidt}), is given by 
\begin{align}
 & K=\left(\sum_{n}p_{n}^{2}\right)^{-1}.\label{eq:SN}
\end{align}
It serves as an effective measure of the correlation in the dependence of the joint amplitude on its two variables; in the limit of a Schmidt number of unity the joint amplitude is simply a product of the same function of each variable, and there is no correlation at all in its dependence on those variables. 

We can introduce another measure of the correlation of the dependence of the joint amplitude on its two variables by 
\begin{equation}
    \mathcal{K} = \mathcal{T}_p\mathcal{B}_c = \frac{\mathcal{T}_p}{\mathcal{T}_c}.
\end{equation}
Clearly when $\mathcal{T}_p$ is much greater than $\mathcal{T}_c$ then the dependence of the joint amplitude on its two variables is highly correlated (see Fig. \ref{fig:picture1}), and then $\mathcal{K} \gg1$. Of course, at the moment we have only defined $\mathcal{T}_p$ and $\mathcal{B}_c$ in a ``rough-and-ready" way, and such then is our definition of $\mathcal{K}$; we will make the definition more precise below. Often similarly defined quantities are introduced and referred to as the ``time-bandwidth product.'' So to avoid confusion we henceforth refer to $\mathcal{K}$ as an ``effective Schmidt number.'' 

Confusions with the phrase ``time-bandwidth product" can arise because of different definitions used for ``time" and ``bandwidth." For example, Fedorov et al. calculate a time-bandwidth product by taking the width of the ``unconditional'' (``single-particle'') and ``conditional'' (``coincidence'') widths of the joint intensity; for a general double-Gaussian joint amplitude (considered below), they show that the time-bandwidth product is exactly equal to the Schmidt number \cite{fedorov2006short,mikhailova2008biphoton,fedorov2008spontaneous}.  
Alternatively, Brecht and Silberhorn calculate the time-bandwidth product taking the conditional and unconditional widths of the ``chronocyclic Wigner function,'' which for their double-Gaussian model is equal to the Schmidt number, even when a chirp is included \cite{brecht2013characterizing}. The agreement -- either exact or approximate -- between the ``time-bandwidth product'' and the Schmidt number suggest a deeper connection between the two quantities. 

However, there is an older and more rigorous meaning of the term ``time-bandwidth product" from classical information theory: For a one-dimensional bandlimited signal, it is the number of orthogonal functions optimally concentrated within a given timewidth needed to describe the signal \cite{landau1962prolate}, and in this context is referred to as the ``Shannon number'' \cite{simons2009slepian,freeden2010handbook}.  
Generalizations of the Shannon number exist for higher dimensional signals where one calculates the time-bandwidth product using the bandwidth and timewidth area, volume, etc. \cite{simons2009slepian,freeden2010handbook,miller2000communicating,pires2009direct}. Recent work has made the connection between the Shannon number from classical information theory to the Schmidt number in quantum information theory; see for example Pires et al. \cite{pires2009direct} or Pors et al. \cite{pors2008shannon} for both a theoretical and experimental investigation. 

In this spirit we define $\mathcal{K}$ more precisely by defining $\mathcal{T}_p$ and $\mathcal{B}_c$ more precisely. We assume that those quantities are chosen as small as possible but subject to the condition that, to within the level of approximation adopted in calculations,  they cover the range of the joint amplitude in time and frequency respectively. This means that the typical ways used to measure bandwidth or frequency -- such as the standard deviation, full-width-at-half-max etc. -- are too narrow. In particular, we choose $\mathcal{B}_c$ to be large enough that frequencies larger than $2\pi\mathcal{B}_c/2$ can be completely neglected, and we can treat the joint spectral amplitude as effectively bandwidth limited. For this reason the effective Schmidt number $\mathcal{K}$ will generally be larger than, but typically on the order of, other conventions used for the ``time-bandwidth product."  

In a later section we argue that generally the Schmidt number $K$ and effective Schmidt number $\mathcal{K}$ satisfy the inequality 
\begin{equation}
\label{eq:effectiveKandK}
    K\le \mathcal{K}.
\end{equation}
One might generally expect this to be true based on a physical argument: Suppose we have squeezed light propagating with an arbitrary joint amplitude with some Schmidt number $K$ and effective Schmidt number $\mathcal{K}$. Now if the squeezed light is sent through a dispersive medium, the joint spectral amplitude is multiplied by a complex but \emph{separable} phase that does \emph{not} change the Schmidt number. However, we know that in a dispersive medium the bandwidth remains constant but the pulse duration broadens, and so $\mathcal{K}$ will generally increase. Thus, in general one might indeed expect that $K\le \mathcal{K}$. Below we will discuss when the near exact equality holds.

We will find that we can use the effective Schmidt number $\mathcal{K}$ to introduce a ``weak squeezing'' regime characterized by the condition 
\begin{equation}
\label{weak squeezing}
    \frac{|\beta|}{\sqrt{\mathcal{K}}} \ll 1, 
\end{equation}
and a ``strong squeezing'' regime characterized by the condition
\begin{equation}
\label{strong squeezing}
    \frac{|\beta|}{\sqrt{\mathcal{K}}} \gg 1. 
\end{equation}
If neither of these conditions are satisfied we refer to the squeezing as ``moderate." 

Now as a first example of the use of the Schmidt decomposition to evaluate the correlation functions, consider a general normalized two-photon state, 
\begin{align}
 \left|\RN{2}\right\rangle& =\frac{1}{\sqrt{2}}\int dt_{1}dt_{2}\overline{\gamma}(t_{1},t_{2})\overline{a}^{\dagger}(t_{1})\overline{a}^{\dagger}(t_{2})\left|\text{vac}\right\rangle \label{eq:IIdef}\\
 & =\frac{1}{\sqrt{2}}\sum_{n}\sqrt{p_{n}}A_{n}^{\dagger}A_{n}^{\dagger}\left|\text{vac}\right\rangle ,\nonumber 
\end{align}
where we defined operators associated with the supermodes as
\begin{align}
\label{eq:Andef}
 & A_{n}^{\dagger}\equiv\int dt\overline{f}_{n}(t)\overline{a}^{\dagger}(t),
\end{align}
and so 
\begin{align}
\label{eq:a_expand}
 & \overline{a}(t)=\sum_{n}\overline{f}_{n}(t)A_{n}.
\end{align}
To write the inverted form (Eq. \eqref{eq:a_expand}) we have taken the adjoint of Eq. \eqref{eq:Andef} and used the completeness relation of the set of functions $\left\{ \overline{f}_{n}(t)\right\} $. The set of operators $\left\{ A_{n}\right\} $ and their adjoints satisfy the usual harmonic oscillator commutation relations. Just as the expressions (\ref{eq:Schmidt})
can be written in time or frequency form, so the first of (\ref{eq:IIdef})
and (\ref{eq:Andef}) can also be written involving integrals over
frequency of the corresponding quantities. We find 
\begin{equation}
\label{eq:II_terms1}
    \begin{split}
        \left.\overline{G}^{(1)}(t)\right|_{\ket{\RN{2}}}&\equiv\left\langle \RN{2}|\overline{a}^{\dagger}(t)\overline{a}(t)|\RN{2}\right\rangle\\
        &=2\sum_{n}p_{n}\left|\overline{f}_{n}(t)\right|^{2},
    \end{split}
\end{equation}
and
\begin{equation}
\label{eq:II_terms2}
    \begin{split}
        \left.\overline{G}^{(2)}(t_{1},t_{2})\right|_{\ket{\RN{2}}}&\equiv\left\langle \RN{2}|\overline{a}^{\dagger}(t_{1})\overline{a}^{\dagger}(t_{2})\overline{a}(t_{2})\overline{a}(t_{1})|\RN{2}\right\rangle\\
        &=2\left|\sum_{n}\sqrt{p_{n}}\overline{f}_{n}(t_{1})\overline{f}_{n}(t_{2})\right|^{2}.
    \end{split}
\end{equation}
In the first we have a contribution of $\left|\overline{f}_{n}(t)\right|^{2}$ from each Schmidt mode with a weight $p_{n}$, while in the second the amplitudes associated with each of the Schmidt modes add; the ``2" of course arises because we have pairs of photons. 

Moving to a squeezed state, in the limit $\left|\beta\right|\ll1$
we can write (\ref{eq:PsiTime}) as 
\begin{align}
 & \left|\Psi\right\rangle \rightarrow\left|\text{vac}\right\rangle +\frac{\beta}{\sqrt{2}}\left|\RN{2}\right\rangle +.....,
\end{align}
(cf. (\ref{eq:PsiOmega})), and we find 
\begin{align}
\label{eq:smallbeta_time}
 & \overline{G}^{(1)}(t)\rightarrow\sum_{n}\left|\beta_{n}\right|^{2}\left|\overline{f}_{n}(t)\right|^{2}=\left|\beta\right|^{2}\int\left|\overline{\gamma}(t,t')\right|^{2}dt'\\
 & \overline{G}^{(2)}(t_{1},t_{2})\rightarrow\left|\sum_{n}\beta_{n}\overline{f}_{n}(t_{1})\overline{f}_{n}(t_{2})\right|^{2}=\left|\beta\right|^{2}\left|\overline{\gamma}(t_{1},t_{2})\right|^{2},\nonumber 
\end{align}
where $\beta_{n}=\beta\sqrt{p_{n}}$ (cf. (\ref{eq:smallbeta_omega})). Treating $|\gamma(t_1,t_2)|^2$ as a normalized probability distribution, for $N_\text{pulse} = |\beta|^2\ll1$ we find  $\overline G^{(1)}(t)$ is  the probability distribution reduced by integrating over the second time variable, and $\overline{G}^{(2)}(t_{1},t_{2})$ is proportional to $|\overline{\gamma}(t_1,t_2)|^2$ itself, the norm squared of the joint temporal amplitude at the two corresponding times. 

More generally, using the Schmidt decomposition (\ref{eq:Schmidt})
and the supermode operators (\ref{eq:Andef}), which are all independent, we can write the squeezed
ket (\ref{eq:PsiTime}) as 
\begin{align}
\label{eq:schmidt product state}
 & \left|\Psi\right\rangle = \bigotimes_{n} S_{n}\vac_n ,
\end{align}
where $\vac_n$ is the vacuum state for the corresponding supermode and 
\begin{align}
 & S_{n}=e^{\frac{\beta_{n}}{2}A_{n}^{\dagger}A_{n}^{\dagger}-\text{h.c.}}.
\end{align}
With the standard result \cite{loudon2000quantum}
\begin{align}
 & S_{n}^{\dagger}A_{n}S_{n}=c_{n}A_{n}+e^{i\theta}s_{n}A_{n}^{\dagger},\label{eq:transformAn}
\end{align}
where we have put $\beta=\left|\beta\right|e^{i\theta}$ and 
\begin{align}
\label{eq:sncn}
 & c_{n}\equiv\cosh\left(\left|\beta_{n}\right|\right),\\
 & s_{n}\equiv\sinh\left(\left|\beta_{n}\right|\right),
\end{align}
using the expression (\ref{eq:a_expand}) to write $\overline{a}(t)$
in terms of the $\left\{ \overline{f}_{n}(t)\right\} $ -- and using
(\ref{eq:transformAn}) to evaluate $\left\langle \Psi|A_{n}^{\dagger}A_{m}|\Psi\right\rangle $
and $\left\langle \Psi|A_{n}^{\dagger}A_{m}^{\dagger}A_{p}A_{q}|\Psi\right\rangle $
-- from (\ref{eq:gs-1}) we have 
\begin{align}
 & \overline{G}^{(1)}(t)=\sum_{n}\left|\overline{f}_{n}(t)\right|^{2}s_{n}^{2},\label{eq:Gresults}\\
 & \overline{G}^{(2)}(t_{1},t_{2})=\overline{G}_\T{coh}^{(2)}(t_{1},t_{2})+\overline{G}_\T{incoh}^{(2)}(t_{1},t_{2}),\nonumber 
\end{align}
where 
\begin{subequations}
\label{eq:Gwork}
    \begin{gather}
        \overline{G}_\text{coh}^{(2)}(t_{1},t_{2})=\left|\sum_{n}s_{n}c_{n}\overline{f}_{n}(t_{2})\overline{f}_{n}(t_{1})\right|^{2}\\
        \hspace{-60mm}\overline{G}_\text{incoh}^{(2)}(t_{1},t_{2})\\
        \hspace{10mm}=\frac{1}{2}\sum_{n,m}\left|s_{n}s_{m}(\overline{f}_{n}(t_{1})\overline{f}_{m}(t_{2})+\overline{f}_{n}(t_{2})\overline{f}_{m}(t_{1}))\right|^{2},\nonumber
    \end{gather}
\end{subequations}
and the average photon number is 
\begin{align}
\label{eq:NpulseS}
 & N_\text{pulse} =\sum_{n}s_{n}^{2}.
\end{align}
Of the two contributions to $\overline{G}^{(2)}(t_{1},t_{2})$,
the ``coherent'' term $\overline{G}_\text{coh}^{(2)}(t_{1},t_{2})$ \cite{dayan2007theory},
which involves the square of a sum, is the generalization to a squeezed state of the term $\overline{G}^{(2)}(t_1,t_2)$ for the two-photon state (\ref{eq:II_terms2}), and the only term that survives in $\overline{G}^{(2)}(t_{1},t_{2})$
in the limit $\left|\beta\right|\ll1$, cf. (\ref{eq:smallbeta_time}). The ``incoherent'' term $\overline{G}_\text{incoh}^{(2)}(t_{1},t_{2})$ \cite{dayan2007theory} involves the sum of squares, and will only be significant at larger values of $\left|\beta\right|$. Note that the corresponding expressions for $G^{(1)}(\omega)$ and $G^{(2)}(\omega_{1},\omega_{2})$ take the same form as Eq. (\ref{eq:Gresults}), with $\overline{f}_{n}$ replaced by $f_{n},$ $t_{1}$ by $\omega_{1}$, etc. 

\subsection{Example 1: The double-Gaussian}
\label{sec:Example 1: The double  Gaussian}
\begin{figure*}[t]
    \centering
    \includegraphics[width = \linewidth]{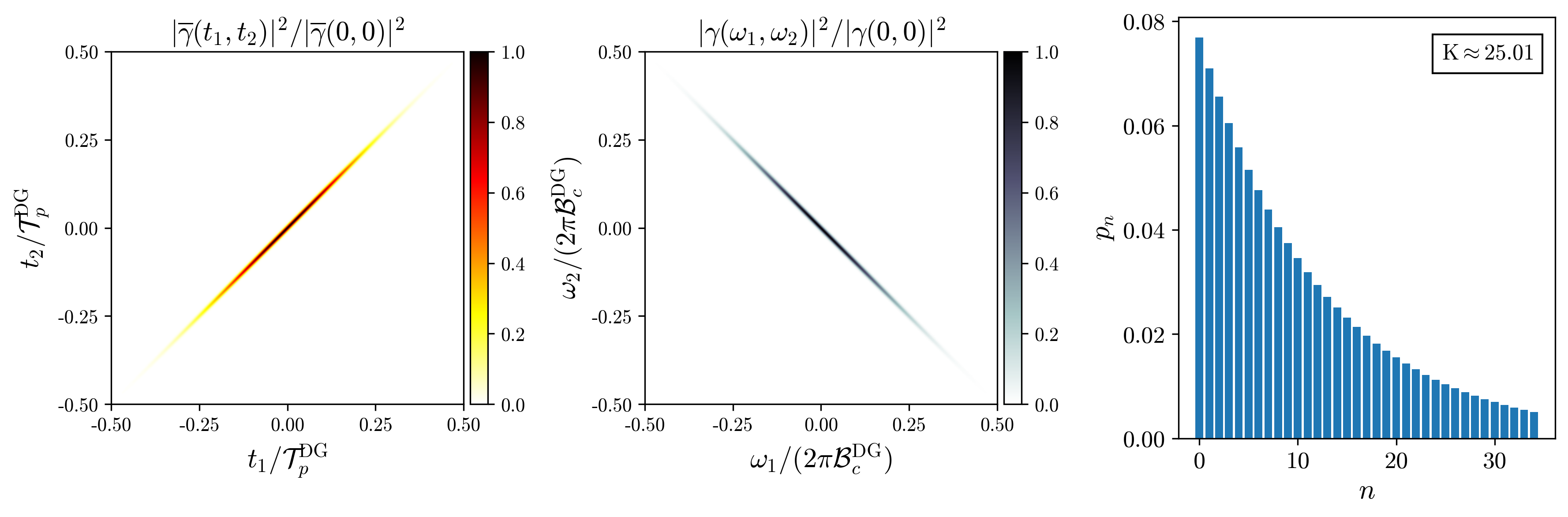}
    \caption{For the double-Gaussian, from left to right we plot the: joint temporal intensity divided by its maximum value with the axes normalized by $\mathcal{T}_p^\mathrm{DG}$; joint spectral intensity divided by its maximum value with the axes normalized by $2\pi \mathcal{B}_c^\mathrm{DG}$; Schmidt amplitudes $p_n$ up to $n = 34$.}
    \label{fig:1}
\end{figure*}
\begin{figure*}
    \centering
    \includegraphics[width = \linewidth]{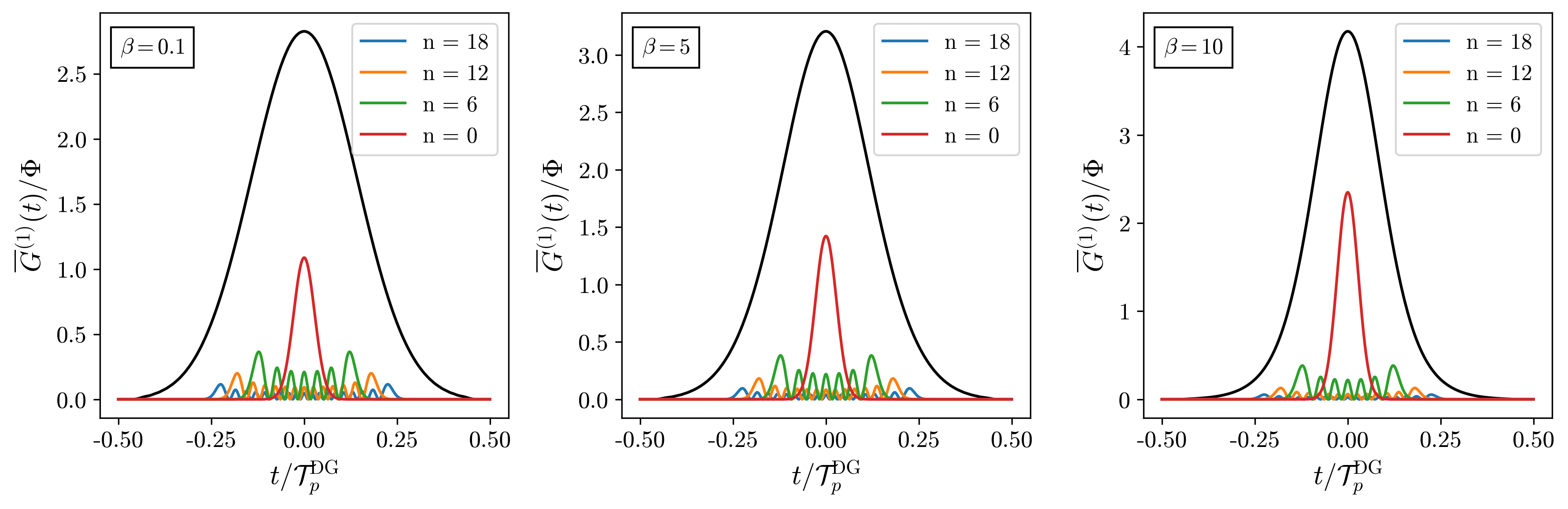}
    \caption{For the double-Gaussian, from left to right we plot $\overline{G}^{(1)}(t)/\Phi$ and a few contributions from different Schmidt modes in Eq. \eqref{eq:Gresults} with the horizontal axis normalized by $\mathcal{T}_p^\mathrm{DG}$ for $\beta = 0.1, 5,$ and $10$.}
    \label{fig:2}
\end{figure*}

A simple model for the joint amplitude is a double-Gaussian function,
\begin{subequations}
\label{double_gaussian}
    \begin{gather}
        \gamma(\omega_{1},\omega_{2})=\sqrt{\frac{1}{\pi\sigma_{p}\sigma_{c}}}e^{-\frac{(\omega_{1}-\omega_{2})^{2}}{4\sigma_{c}^{2}}}e^{-\frac{(\omega_{1}+\omega_{2})^{2}}{4\sigma_{p}^{2}}},\\
        \overline{\gamma}(t_{1},t_{2})=\sqrt{\frac{\sigma_{p}\sigma_{c}}{\pi}}e^{-\frac{\sigma_{c}^{2}(t_{1}-t_{2})^{2}}{4}}e^{-\frac{\sigma_{p}^{2}(t_{1}+t_{2})^{2}}{4}},
    \end{gather}
\end{subequations}
where $\sigma_{p}$ and $\sigma_{c}$ are the two parameters. The
Schmidt modes are harmonic oscillator wave functions. This can be
seen by noting that the reduced density operator of ``particle 1''
is equal to the density operator of a harmonic oscillator in thermal
equilibrium; its eigenfunctions are the Schmidt modes, and they are
obviously the harmonic oscillator wave functions. The details can
be worked out from this, or more mathematically from the Mehler kernal \cite{edrelyi1953higher, NIST:DLMF}. For $\sigma_{c}\geq\sigma_{p}$ , the Schmidt decompositions
are given by (\ref{eq:Schmidt}), with 
\begin{subequations}
    \begin{gather}
        \gamma(\omega_{1},\omega_{2})=\sum_{n\geq0}\sqrt{p_{n}}\mathcal{H}_{n}(\omega_{1})\mathcal{H}_{n}(\omega_{2}),\\
        \overline{\gamma}(t_{1},t_{2})=\sum_{n\geq0}\sqrt{p_{n}}\overline{\mathcal{H}}_{n}(t_{1})\overline{\mathcal{H}}_{n}(t_{2}),
    \end{gather}
\end{subequations}
and 
\begin{align}
 & p_{n}=\frac{4\sigma_{c}\sigma_{p}}{(\sigma_{c}+\sigma_{p})^{2}}\left(\frac{\sigma_{c}-\sigma_{p}}{\sigma_{c}+\sigma_{p}}\right)^{2n},
\end{align}
with a Schmidt number 
\begin{align}
 & K=\frac{\sigma_{c}^{2}+\sigma_{p}^{2}}{2\sigma_{c}\sigma_{p}}.\label{eq:SN_gaussian}
\end{align}
Here $\overline{f}_{n}(t)=\overline{\mathcal{H}}_{n}(t)$, where 
\begin{align}
\label{eq:H_n(t)mode}
 & \overline{\mathcal{H}}_{n}(t)=\frac{H_{n}(\frac{t}{t_{0}})e^{-t^{2}/(2t_{0}^{2})}}{\sqrt{2^{n}n!\pi^{1/2}t_{0}}},
\end{align}
with $H_{n}(x)$ the Hermite polynomials, is the standard coordinate-representation
harmonic oscillator energy eigenfunction, but with $t$ playing the
role of $x$ and 
\begin{align}
 & t_{0}\equiv\sqrt{\frac{1}{\sigma_{p}\sigma_{c}}}
\end{align}
playing the role of a reference length $x_{0}$ that is often introduced \cite{sakurai1995modern}; $f_{n}(\omega)=\mathcal{H}_{n}(\omega),$ where
\begin{align}
 & \mathcal{H}_{n}(\omega)=(-i)^{n}\sqrt{\frac{t_{0}}{2^{n}n!\pi^{1/2}}}H_{n}(\omega t_{0})e^{-\omega^{2}t_{0}^{2}/2}
\end{align}
is the standard momentum-representation harmonic oscillator energy
eigenfunction \cite{sakurai1995modern}, with $\omega$ playing the role of $p$. 

In Fig. \ref{fig:1} we plot the joint intensities $\left|\overline{\gamma}(t_{1},t_{2})\right|^{2}$ and $\left|\gamma(\omega_{1},\omega_{2})\right|^{2}$ for $\sigma_{c}/\sigma_{p}=50$, as well as the Schmidt weights $p_{n}$ as a function of $n$. For $\sigma_c\gg\sigma_p$, the Schmidt number \eqref{eq:SN_gaussian} is approximately given by $K\approx \sigma_c/(2\sigma_p) = 25$, indeed we find numerically that $K_\mathrm{DG} = 25.01$. The joint intensities in Fig. \ref{fig:1} vary over the widths $\mathcal{T}_p^\mathrm{DG}$ and $\mathcal{B}_c^\mathrm{DG}$ which we set to be
\begin{equation}
\label{eq:DGwidths}
    \mathcal{T}_p^\mathrm{DG} = \frac{a}{\sqrt{2}\sigma_p}, \hspace{5mm}\mathcal{B}_c = \frac{a}{2\pi} \frac{\sigma_c}{\sqrt{2}}, \hspace{5mm} \mathcal{T}_c^\mathrm{DG} = \frac{2\pi\sqrt{2}}{a\sigma_c},
\end{equation}
and choose $a = 2\sqrt{2\pi}$ for convenience. This choice of $a$ is large enough that the joint temporal and spectral amplitudes can essentially be taken to be confined within the ranges of $\mathcal{T}_p^\mathrm{DG}$ and $\mathcal{B}_c^\mathrm{DG}$ respectively, as can be gleaned from Fig. \ref{fig:1} and will in fact be confirmed by our calculations in later sections. Then the effective Schmidt number is
\begin{equation}
    \mathcal{K}_\mathrm{DG} = \frac{a^2}{2\pi}\frac{\sigma_c}{2\sigma_p} \approx 4K_\mathrm{DG} = 100.
\end{equation}

To illustrate the behaviour of the photon statistics for a large range of photon numbers, we consider the three values of $\beta = 0.1,5$, and $10$, corresponding to $|\beta|/\sqrt{\mathcal{K}_\mathrm{DG}} = 0.01,0.5,1$, and so ranging from weak to moderate squeezing; we dedicate section \ref{sec:The strongly squeezed limit} to 
the discussion of the strongly squeezed limit.

In Fig. \ref{fig:2} we plot $\overline{G}^{(1)}(t)$ for the three values of $\beta$. The expectation value of the number of photons is determined by (\ref{eq:NpulseS}), and for $\beta=0.1,$ $5$, and 
$10$ we have respectively $N_\text{pulse} \approx 0.01, 35$, and $383$; we take an effective photon flux (photons per unit time) to be given by $\Phi = N_\T{pulse}/$$\mathcal{T}^\mathrm{DG}_p$. We also plot the contribution to each $\overline{G}^{(1)}(t)$ from a number of the Schmidt modes (see (\ref{eq:Gresults})). For each value of $\beta$, the photon density at any particular time $t$ involves contributions from many Schmidt modes, and we cannot associate it with one or even a few Schmidt modes. Clearly as $\beta$ increases the contribution from the $n=0$ Schmidt mode increases and the shape of the photon density narrows. This occurs because the scaling of each contribution with $|\beta_n|$ is nonlinear and depends on the quantities $c_n$ and $s_n$ (\ref{eq:sncn}); since for the double-Gaussian the Schmidt amplitudes $p_n$ decrease as $n$ increases, $|\beta_0|$ has the largest contribution. This behavior suggests that in the strongly squeezed limit the photon statistics can be well approximated by the first few Schmidt modes, a point to which we return in section \ref{sec:The strongly squeezed limit}.

\begin{figure*}
    \centering
    \includegraphics[width = \linewidth]{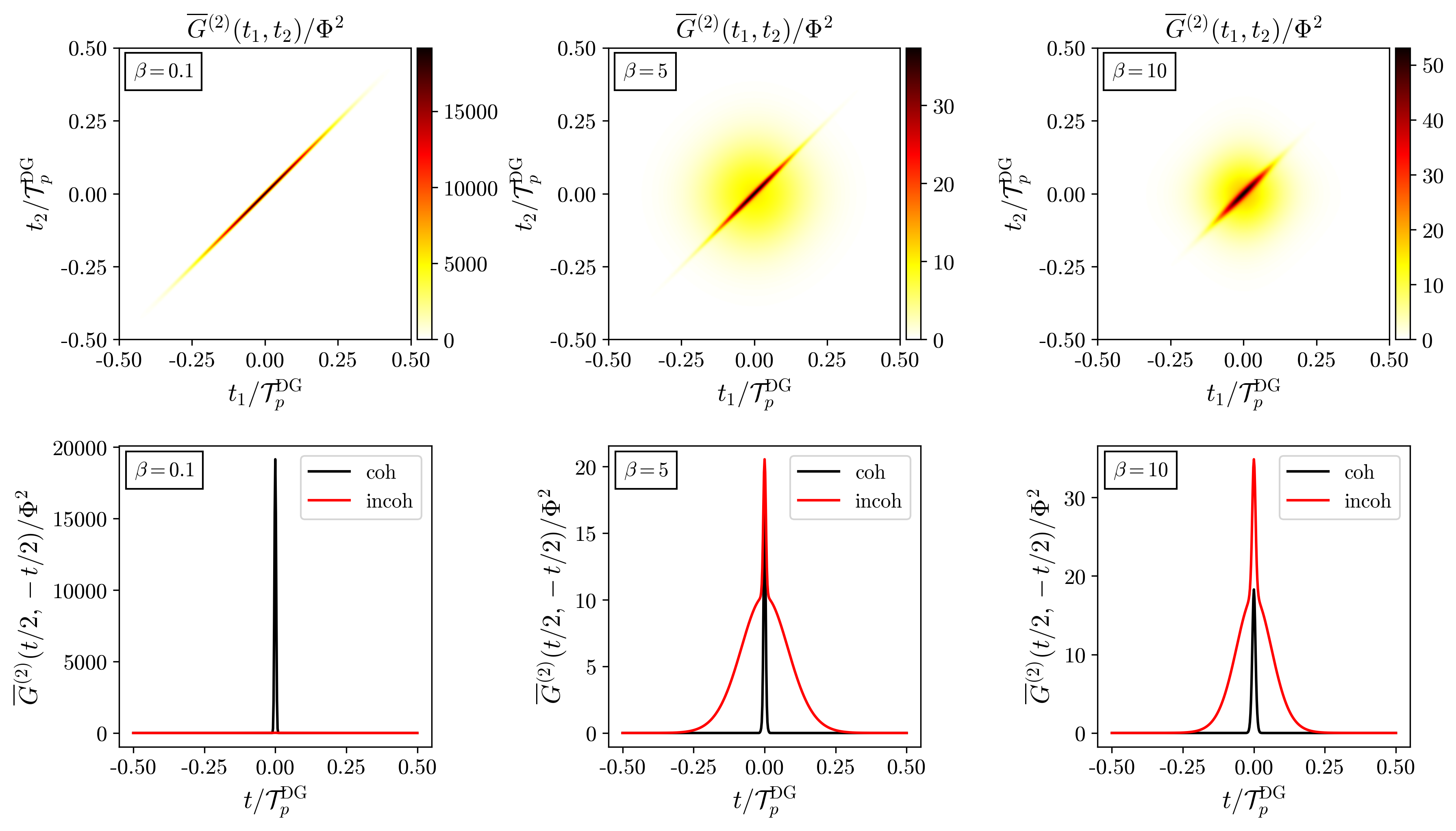}
    \caption{For the double-Gaussian, from left to right we plot $\overline{G}^{(2)}(t_1,t_2)/\Phi^2$ (top) and the coherent and incoherent contribution to $\overline{G}^{(2)}(t/2,-t/2)/\Phi^2$ (bottom) with the axes normalized by $\mathcal{T}_p^\mathrm{DG}$ for $\beta = 0.1, 5,$ and $10$.}
    \label{fig:3}
\end{figure*}

In Fig. \ref{fig:3} we turn to $\overline{G}^{(2)}(t_{1},t_{2})$ with again the three values of $\beta$ considered above. In the top row we show this function for the different values of $\beta$; at low $\beta$ the result is proportional to the square of the absolute value of the joint temporal amplitude (see Fig. \ref{fig:1}, Eq. (\ref{eq:smallbeta_time})), while for
larger $\beta$ there are significant corrections to this. In the bottom panel we plot $\overline{G}^{(2)}(t/2,-t/2)$, which corresponds to moving along a diagonal that runs from the upper-left to the lower-right of the plots in the first row; we give the coherent and incoherent contributions separately. 

The situation here is of course more complicated than that for $\overline{G}^{(1)}(t)$, because the expression (\ref{eq:Gresults}) for $\overline{G}^{(2)}(t_{1},t_{2})$ is more complicated than a simple sum over contributions from the individual Schmidt modes. But we see that at least in the weak squeezing regime the coherent contribution to $\overline{G}^{(2)}(t/2,-t/2)$ dominates, and it is nonzero only over a range of $t$ much less than the range of $t$ over which the individual Schmidt modes are nonzero; clearly the interference terms between the different Schmidt modes in (\ref{eq:Gresults}) play a critical role in the result for $\overline{G}^{(2)}(t_{1},t_{2})$. 

Further, the incoherent contribution at $\beta=5$ has a structure that consists partly of a broad background and partly of a contribution that mirrors the coherent contribution. Now note that the expression (\ref{eq:Gwork}) for $\overline{G}_{\text{incoh}}^{(2)}(t_1,t_2)$ can also be very generally written as
\begin{equation}
    \label{eq:G2_incoh_alt}
    \overline G_\text{incoh}^{(2)}(t_1,t_2) = \overline G^{(1)}(t_1)\overline G^{(1)}(t_2)  + |\overline G^{(1)}(t_1,t_2)|^2,
\end{equation}
and the broad background in $\overline{G}_{\text{incoh}}^{(2)}(t/2,-t/2)$ can be understood as arising from the first term on the right-hand-side. The contribution that mirrors the coherent contribution can be understood as arising from the second term, and in fact it is absent if we consider nondegenerate squeezed light, where signal and idler frequencies are well separated with different center  frequencies \cite{raymer2022theory}. The corresponding contribution to $G^{(2)}_\T{incoh}(\omega_1,\omega_2)$ is often discussed and referred to as the ``autocorrelation'' \cite{cutipa2022bright}. 
In any case, clearly the different features of $\overline{G}^{(2)}(t/2,-t/2)$ certainly do not follow in any simple way from the features of individual Schmidt modes, but arise from the interference of many of these modes.

As $\beta$ increases, the range of $\overline{G}^{(2)}(t,t)$ narrows, as does the range of the photon density. However, 
the coherent and incoherent contribution along $\overline{G}^{(2)}(t/2,-t/2)$ broaden. Thus as the squeezing parameter is increased, the photon statistics begin to look uncorrelated. This 
behavior matches that of the photon density, in that as $\beta$ increases fewer Schmidt modes are important in calculating the correlation functions. We also note that the incoherent contribution is approximately twice the coherent contribution, a point to which we return to in section \ref{sec:The strongly squeezed limit}.

While there are many features of interest here, we emphasize that
both the value of $\overline{G}^{(1)}(t)$ at a particular $t$, and that of $\overline{G}^{(2)}(t_{1},t_{2})$ at a particular $t_{1}$ and $t_{2}$, receive contributions from \emph{many} of the Schmidt modes. The same holds for the corresponding functions $G^{(1)}(\omega)$ and $G^{(2)}(\omega_{1},\omega_{2})$. The structure of the Schmidt modes themselves, in and of itself, does not help us understand the structure of the correlation functions.
\begin{figure*}
    \centering
    \includegraphics[width = \linewidth]{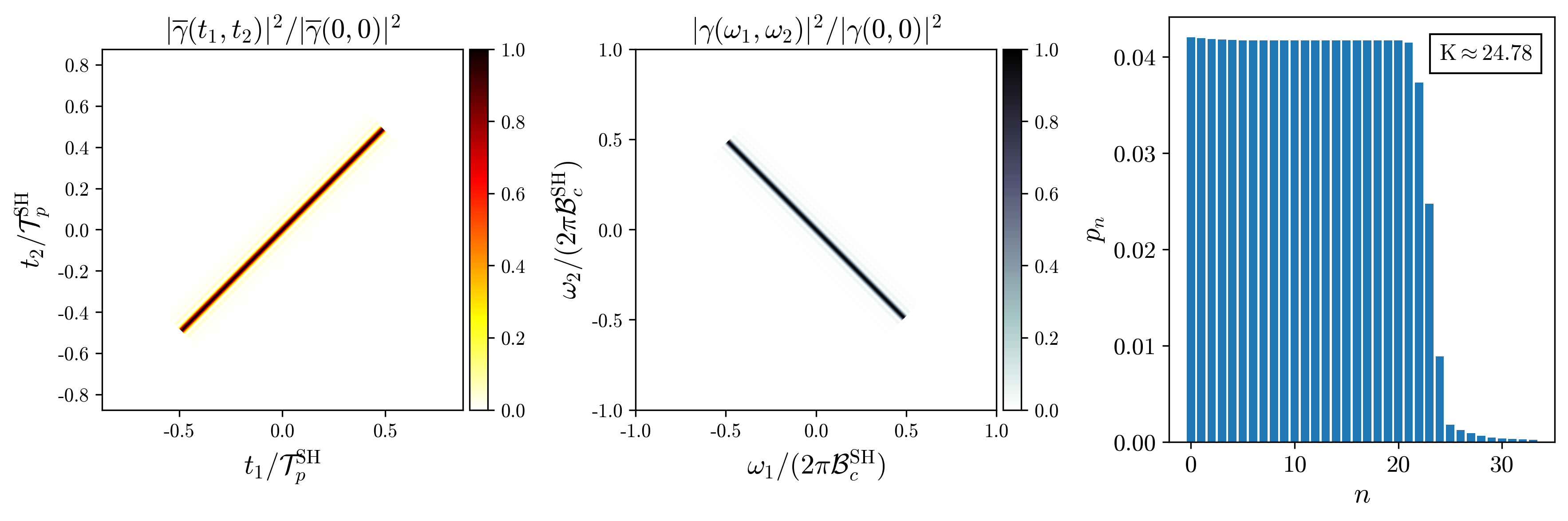}
    \caption{For the sinc-hat, from left to right we plot the: joint temporal intensity divided by its maximum value with the axes normalized by $\mathcal{T}^\mathrm{SH}_p$; joint spectral intensity divided by its maximum value with the axes normalized by $2\pi \mathcal{B}^\mathrm{SH}_c$; Schmidt amplitudes $p_n$ up to $n = 34$. }
    \label{fig:4}
\end{figure*}
\begin{figure*}
    \centering
    \includegraphics[width = \linewidth]{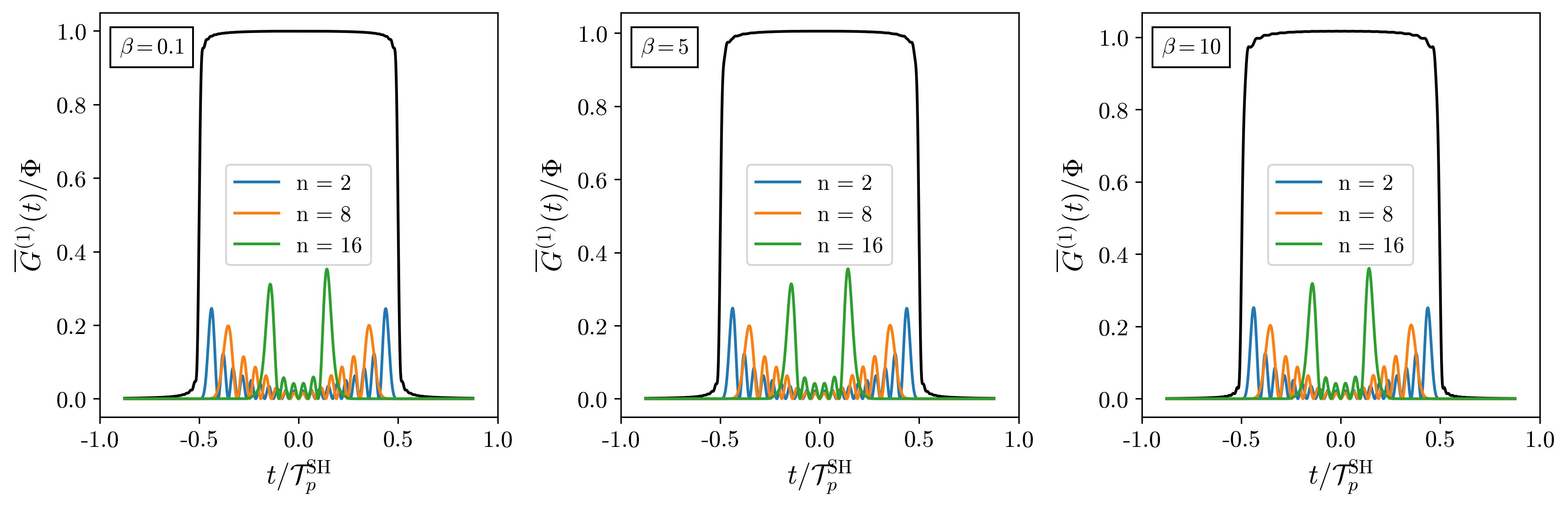}
    \caption{For the sinc-hat, from left to right we plot $\overline{G}^{(1)}(t)/\Phi$ and a few contributions from different Schmidt modes in Eq. \eqref{eq:Gresults} with the horizontal axis normalized by $\mathcal{T}^\mathrm{SH}_p$ for $\beta = 0.1, 5,$ and $10$.}
    \label{fig:5}
\end{figure*}

\subsection{Example 2: The sinc-hat}
\label{eq:Example 2: The sinc-hat}
This difference between the features of the Schmidt modes and the features of $\overline{G}^{(1)}(t)$ and $\overline{G}^{(2)}(t_{1},t_{2})$ is not specific
to the double-Gaussian joint amplitude. Consider another
form, 

\begin{align}
   \label{eq:sinc-hat}
   \gamma(\omega_{1},\omega_{2}) & =\alpha(\omega_{1}+\omega_{2})\phi\left(\frac{\omega_{1}-\omega_{2}}{2}\right),\\
   \label{eq:sinc-hat-time}
   \overline{\gamma}(t_{1},t_{2})&=\overline{\alpha}\left(\frac{t_{1}+t_{2}}{2}\right)\overline{\phi}(t_{1}-t_{2}),
\end{align}
where 
\begin{align}
    \label{eq:sinc-hat-parameters}
        \alpha(\omega)=\frac{1}{\sqrt{\Omega_{p}}}\text{sinc}\left(\frac{\pi\omega}{\Omega_{p}}\right),\\
        \phi(\omega)=\frac{1}{\sqrt{\Omega_{c}}}\text{ for } \ensuremath{-\frac{\Omega_{c}}{2}\leq\omega\leq\frac{\Omega_{c}}{2},}\\
        \hspace{-25mm}=0, \text{ otherwise,}\nonumber\\
        \bar{\alpha}(t)=\frac{1}{\sqrt{T_{p}}}\text{ for } \ensuremath{-\frac{T_{p}}{2}\leq t\leq\frac{T_{p}}{2}}, \\
        \hspace{-22mm}=0, \text{ otherwise,}\nonumber \\
        \overline{\phi}(t)=\frac{1}{\sqrt{T_{c}}}\text{sinc}\left(\frac{\pi t}{T_{c}}\right)
\end{align}
with $T_{p}=2\pi/\Omega_{p}$ and $T_{c}=2\pi/\Omega_{c}$, and as
usual $\alpha(\omega)$ and $\overline{\alpha}(t),$ and $\phi(\omega)$
and $\overline{\phi}(t)$, are Fourier transform pairs (cf. (\ref{eq:fnbar})). Since both $\gamma(\omega_1,\omega_2)$ and $\overline\gamma(t_1,t_2)$ are products of a ``sinc'' function and ``top-hat'' function we refer to this example as the ``sinc-hat'' joint amplitude. 

For the sinc-hat joint amplitude, the Schmidt modes must be found numerically. In Fig. \ref{fig:4} we plot the joint intensities $\left|\overline{\gamma}(t_{1},t_{2})\right|^{2}$
and $\left|\gamma(\omega_{1},\omega_{2})\right|^{2}$, as well as
the Schmidt weights $p_{n}$ as a function of $n$, for $T_{p}/T_{c}= 24$. We will see below that for large $T_{p}/T_{c}$ we have $K\approx T_p/T_c$, and indeed here we numerically find $K_\text{SH}=24.78$.

Evaluating $\overline\gamma(t_1,t_2)$ along the line $t_{1}=t_{2}$ those variables range from $-T_{p}/2$ to $T_{p}/2$, and similarly along $\omega_1=-\omega_2$, $\gamma(\omega_1,\omega_2)$ ranges from $-\Omega_{c}/2$ to $\Omega_{c}/2$. Na\"ively one would guess that we should set $\mathcal{T}_p^\text{SH}\to T_p$ and $2\pi\mathcal{B}_c^\mathrm{SH}\to \Omega_c$ or equivalently $\mathcal{T}_c^\text{SH}\to T_c$, however, this is only the range of $t$ along the \emph{diagonal} (or anti-diagonal in frequency) and the joint amplitude exists beyond it. In Appendix \ref{sec:app:Schematic of sinc-hat joint intensity} we show that  
\begin{equation}
    \mathcal{T}_p^\text{SH} = T_p +\frac{T_c}{2},
\end{equation}
and 
\begin{equation}
\label{eq:sinchatbandwidth}
    \mathcal{B}_c^\mathrm{SH} = \frac{\Omega_c}{2\pi} + \frac{1}{2}\frac{\Omega_p}{2\pi}, \hspace{5mm} \mathcal{T}_c^\mathrm{SH} = \frac{T_cT_p}{T_p + T_c/2}.
\end{equation}
This leads to an effective Schmidt number
\begin{equation}
    \mathcal{K}_\mathrm{SH} = \frac{(T_p + \frac{T_c}{2})^2}{T_pT_c} = 1 + \frac{T_p}{T_c}+ \frac{T_c}{4T_p},
\end{equation}
and for $T_p/T_c = 24$, $\mathcal{K}_\mathrm{SH}\approx 25$ to very good approximation. In section \ref{sec:The Whittaker-Shannon decomposition} we discuss this near equality.

\begin{figure*}
    \centering
    \includegraphics[width = \linewidth]{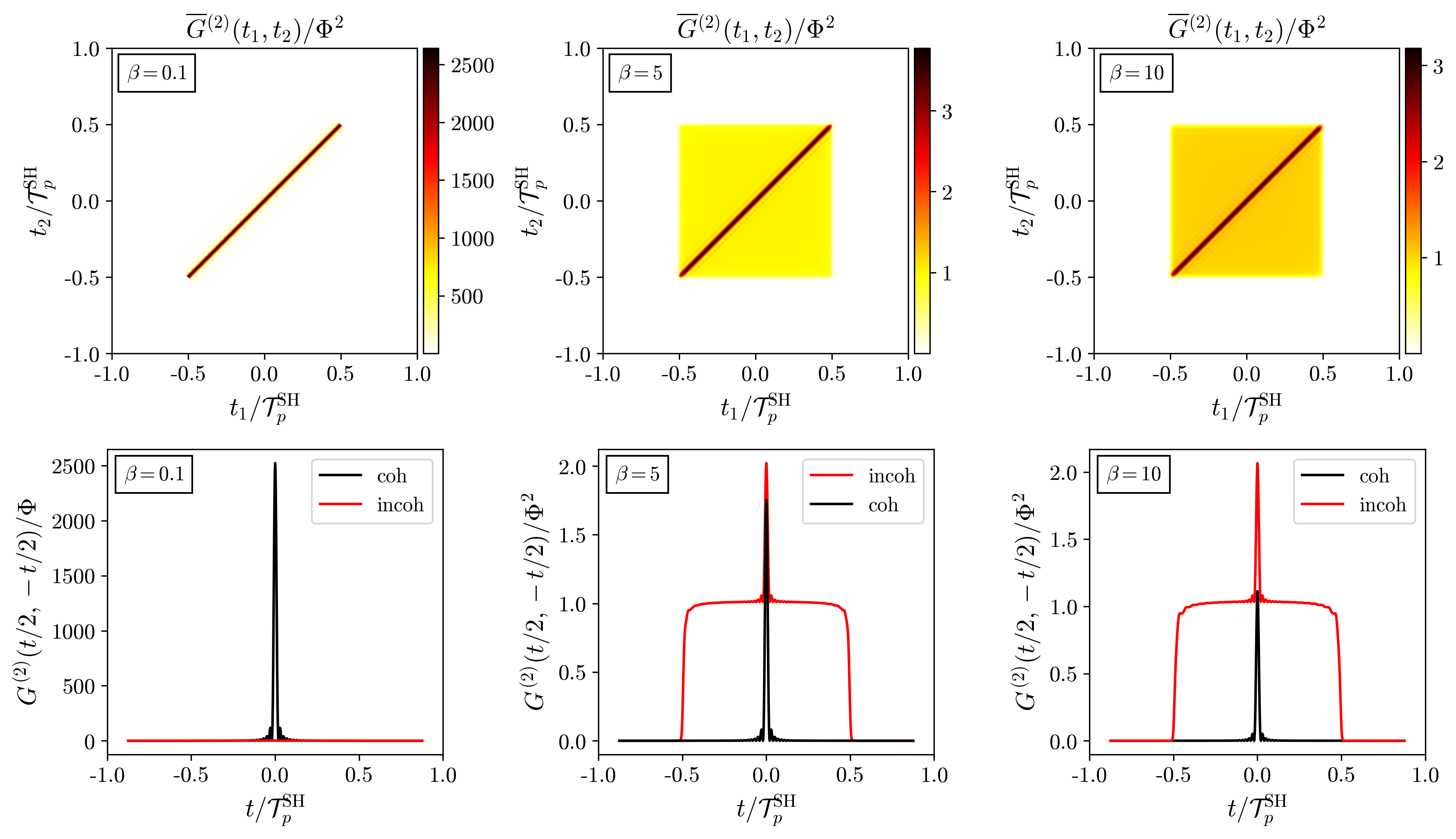}
    \caption{For the sinc-hat, from left to right we plot $\overline{G}^{(2)}(t_1,t_2)/\Phi^2$ (top) and the coherent and incoherent contribution to $\overline{G}^{(2)}(t/2,-t/2)/\Phi^2$ (bottom) with the axes normalized by $\mathcal{T}^\mathrm{SH}_p$ for $\beta = 0.1, 5,$ and $10$.}
    \label{fig:6}
\end{figure*}

In Fig. \ref{fig:5} we plot $\overline{G}^{(1)}(t)$ for the same three values of $\beta$ used in the example above, corresponding here to photon numbers $N_\text{pulse} \approx 0.01, 35$ and $335$; we take an effective photon flux to be given by $\Phi = N_\T{pulse}/\mathcal{T}_p^\mathrm{SH}$. Then for the three values of $\beta$ we have $|\beta|/\sqrt{\mathcal{K}_\mathrm{SH}}\approx 0.02, 1, 2$, which again corresponds to weak to moderate squeezing. In Fig. \ref{fig:5} we also include a few of the contributions from the Schmidt modes; we see that typically those contributions range over the whole duration of the pulse, analogous to what we saw for the double-Gaussian example.  In the top row of Fig. \ref{fig:6} we plot $\overline{G}^{(2)}(t_{1},t_{2})$
for the indicated values of $|\beta|$, and in the bottom row the coherent
and incoherent contributions to $\overline{G}^{(2)}(t/2,-t/2).$  Again, the range over which the individual Schmidt modes extend is much larger than these contributions, and so they must be understood as arising from a number of interfering Schmidt modes. 

So just as for squeezed states described by the double-Gaussian joint amplitude, the behavior of correlation functions of squeezed states described by the sinc-hat joint amplitude cannot be linked in a simple way to the behavior of the individual Schmidt modes. Quantitatively there are differences between the correlation functions resulting from those two joint amplitudes:  The relative amplitudes of the Schmidt modes of a given $n$ in Fig. \ref{fig:5} (sinc-hat joint amplitude) are roughly independent of $|\beta|$, while the relative amplitudes in of those in Fig. \ref{fig:2} (double-Gaussian) are not, and the two parts of the structure of $\overline{G}^{(2)}_{\text{incoh}}(t/2,-t/2)$ we noticed for the double-Gaussian joint amplitude are even more pronounced, and persist to larger $\beta$ than they did for that amplitude. These differences arise because the Schmidt weights of the sinc-hat joint amplitude are nearly identical in the $T_p/T_c \gg 1$ limit (see Fig. \ref{fig:4}), and thus each Schmidt mode contributes roughly equally to the resulting correlation functions even in the large $\beta$ limit.

To see how this near-degeneracy in the Schmidt weights arises, note that in general the Schmidt modes $\overline{f}_{n}(t)$ of a joint
temporal amplitude $\overline{\gamma}(t_{1},t_{2})$ are eigenfunctions
of the operator 
\begin{align}
 & M(t_{1},t_{2})\equiv\int\overline{\gamma}(t_{1},t)\overline{\gamma}^{*}(t,t_{2})dt,
\end{align}
with eigenvalue $p_{n}$,
\begin{align}
 & \int M(t_{1},t_{2})\overline{f}_{n}(t_{2})dt_{2}=p_{n}\overline{f}_{n}(t_{1}),\label{eq:Meigen}
\end{align}
which follows immediately from constructing $M(t_{1},t_{2})$ using
(\ref{eq:Schmidt}). Now for $T_{p}/T_{c}\gg1$ we can approximate the joint temporal amplitude (\ref{eq:sinc-hat-time}) as
\begin{align}
 \overline{\gamma}(t_{1},t_{2}) &=\overline{\alpha}\left(\frac{t_{1}+t_{2}}{2}\right)\overline{\phi}(t_{1}-t_{2})\\
 &\approx\overline{\alpha}(t_{1})\overline{\phi}(t_{1}-t_{2}),
\end{align}
and so 
\begin{align}
\label{eq:Mapprox}
 M(t_{1},t_{2})&\approx\int\overline{\alpha}(t_{1})\overline{\alpha}(t_{2})\overline{\phi}(t_{1}-t)\overline{\phi}(t-t_{2})dt\\
 &=\sqrt{T_{c}}\overline{\alpha}(t_{1})\overline{\alpha}(t_{2})\overline{\phi}(t_{1}-t_{2})\nonumber\\
 & =\frac{T_{c}}{T_{p}}\frac{\sin\left(\frac{\Omega_{c}}{2}(t_{1}-t_{2})\right)}{\pi(t_{1}-t_{2})},\text{ for }\ensuremath{-\frac{T_{p}}{2}\leq t_{1},t_{2}\leq\frac{T_{p}}{2},}\nonumber \\
 & \text{ }=0,\text{ otherwise.}\nonumber 
\end{align}
For specified $\Omega_{c}T_{p}/4$, the functions $\overline\psi_{n}(t')$
satisfying the eigenvalue equation 
\begin{align}
 & \int\limits_{-T_{p}/2}^{T_{p}/2}\frac{\sin\left(\frac{\Omega_{c}}{2}(t-t')\right)}{\pi(t-t')}\overline\psi_{n}(t')dt'=\lambda_{n}\overline\psi_{n}(t),
\end{align}
with the label $n=0,1,.....,$ are related to the angular prolate
spheroidal functions \cite{landau1962prolate,simons2009slepian,freeden2010handbook,slepian1961prolate,wang2017review}, and are defined 
with the normalization
\begin{align}
 & \int\limits_{-\infty}^{\infty}\left|\overline{\psi}_{n}(t)\right|^{2}dt=1.
\end{align}
The $\lambda_{n}$ are close to unity for small $n$, and fall off
quickly to zero for $n>T_{p}/T_{c} \equiv K_\text{app};$ we approximate them
as equal to unity for $n< K_\text{app}$ and zero for $n\ge K_\text{app}$. So within
the approximation (\ref{eq:Mapprox}) we have 
\begin{align}
 & \int M(t_{1},t_{2})\overline\psi_{n}(t_{2})dt_{2}\rightarrow\frac{T_{c}}{T_{p}}\lambda_{n}\overline\psi_{n}(t_{1}),
\end{align}
and comparing with (\ref{eq:Meigen}) we can identify 
\begin{align}
 & \overline{f}_{n}(t)\rightarrow\overline\psi_{n}(t),\label{eq:approx_pn}\\
 & p_{n}\rightarrow\sqrt{\frac{T_{c}}{T_{p}}} = \frac{1}{\sqrt{K_\text{app}}},\nonumber 
\end{align}
for $n<K_\text{app}$. That is, the Schmidt modes are approximately given
by the angular prolate spheroidal functions, and so 
\begin{align}
 & \overline{\gamma}(t_{1},t_{2})\rightarrow\sum_{n = 0}^{K_\text{app}-1}\sqrt{\frac{T_{c}}{T_{p}}}\overline\psi_{n}(t_{1})\overline\psi_{n}(t_{2}),\label{eq:approx_gammabar}
\end{align}
exhibiting a huge degeneracy of Schmidt mode amplitudes, with a approximate Schmidt
number (\ref{eq:SN}) $K_\text{app}=T_{p}/T_{c}\approx K_\text{SH}$, as expected. 

While (\ref{eq:approx_pn},\ref{eq:approx_gammabar}) are only approximate (cf. Fig. \ref{fig:4}), they do indicate that in the limit $T_{p}/T_{c}\gg1,$
which for the sinc-hat function we can characterize as the ``long pulse'' limit, a large near-degeneracy of Schmidt mode amplitudes can be expected. Like the exact Schmidt modes, the angular prolate spheroidal functions range over the whole duration $T_{p}$ associated
with the joint temporal amplitude. However, were the degeneracy exact it would imply that the Schmidt modes are not unique and that various superpositions of them could be constructed. While this freedom is
only approximate for near degeneracy, we will see below that we can use it to ``take apart" the joint amplitude in a different way by constructing approximate Schmidt modes that more explicitly reflect the properties of the light. 

\section{An approximate Schmidt decomposition}
\label{sec:An approximate Schmidt decomposition}
Focusing on the sinc-hat model (\ref{eq:sinc-hat},\ref{eq:sinc-hat-parameters})
in the limit where $T_{p}\gg T_{c}$ and $\mathcal{T}_p^\mathrm{SH}\to T_p$ and $\mathcal{T}_p^\mathrm{SH}\to T_c$, note that while in the second
row of Fig. \ref{fig:6} we have plotted $\overline{G}^{(2)}(\overline{t}+t/2,\overline{t}-t/2)$
for $\overline{t}=0,$ we would expect such plots to be similar for
values of $\overline{t}\equiv(t_{1}+t_{2})/2$ ranging over the pulse
duration, especially in the limit of small $|\beta|$. In that limit $\overline{G}^{(2)}(t_{1},t_{2})$
reflects the behavior of the joint temporal amplitude itself (see
(\ref{eq:smallbeta_time})), and generally $\overline{G}^{(2)}(\overline{t}+t/2,\overline{t}-t/2)$
will be nonzero for $t$ ranging on a time scale of the order of $T_c$; in our example of Fig. \ref{fig:6} that is $T_p/24$.   
This suggests that if we want to capture the behavior of the joint
temporal amplitude as a function of $(t_{1}-t_{2})$ in each of the
terms of an approximate Schmidt decomposition, rather than just when
they are all used together, we should look for approximate Schmidt
modes that vary over a range of $T_c$. Since, roughly speaking,
frequency components between $-\Omega_c/2$ and $\Omega_c/2$
are then available, one such function can easily be constructed by
taking 
\begin{equation}
\label{eq:etabar_def}
    \begin{split}
        \overline{\eta}(t)&\equiv\frac{1}{\sqrt{\Omega_c}}\int\limits_{-\Omega_c/2}^{\Omega_c/2}\frac{d\omega}{\sqrt{2\pi}}e^{-i\omega t}\\
        &=\frac{1}{\sqrt{T_c}}\text{sinc}\left(\frac{\pi t}{T_c}\right),
    \end{split}
\end{equation}
where the prefactor is chosen so the function is normalized (see (\ref{eq:chi_tilde_normalization})
below). However, we need a set of such functions that are orthonormal
to serve as approximate Schmidt functions; the way to do that is to
take the set of functions 
\begin{align}
 & \overline{\eta}_{n}(t)=\overline{\eta}(t-nT_c),
\end{align}
where $n$ is an integer, for then we have 
\begin{align}
 & \int\overline{\eta}_{n}^{*}(t)\overline{\eta}_{m}(t)dt=\delta_{nm}.\label{eq:chi_tilde_normalization}
\end{align}
Note that were time variables replaced by position variables, then
the $\left\{ \overline{\eta}_{n}(t)\right\} $ would correspond to
a set of Wannier functions of the lowest band in a one-dimensional
crystal of lattice spacing corresponding to $T_c$, when the potential
of the lattice is neglected (``empty-lattice approximation'') \cite{cohen2016fundamentals}. We
can then seek an approximate expression $\overline{\gamma}_\T{app}(t_{1},t_{2})$
for the sinc-hat joint temporal amplitude $\overline{\gamma}(t_{1},t_{2})$
of (\ref{eq:sinc-hat}) by writing 
\begin{align}
 & \overline{\gamma}_\T{app}(t_{1},t_{2})=T_{c}\sum_{n}u(nT_c)\overline{\eta}_{n}(t_{1})\overline{\eta}_{n}(t_{2}),\label{eq:gamma_bar_app}
\end{align}
which clearly takes the form of a Schmidt decomposition, with $u(nT_c)$
playing the role of an ``envelope function''; 
$\gamma_\mathrm{app}(t_1,t_2)$ is normalized as the exact function (\ref{eq:sinc-hat}) as long as
\begin{align}
 & T_c^{2}\sum_{n}\left|u(nT_c)\right|^{2}=1.
\end{align}
The introduction of the functions $\overline\eta_n(t)$ allows us to work with ``pseudo-Schmidt'' modes that are mutually orthogonal (like the real Schmidt modes), but are localized in time and range over
different center  times. We refer to the approximate Schmidt decomposition \eqref{eq:gamma_bar_app} we construct as the ``pseudo-Schmidt decomposition."
\begin{figure}
    \centering
    \includegraphics[width = \linewidth]{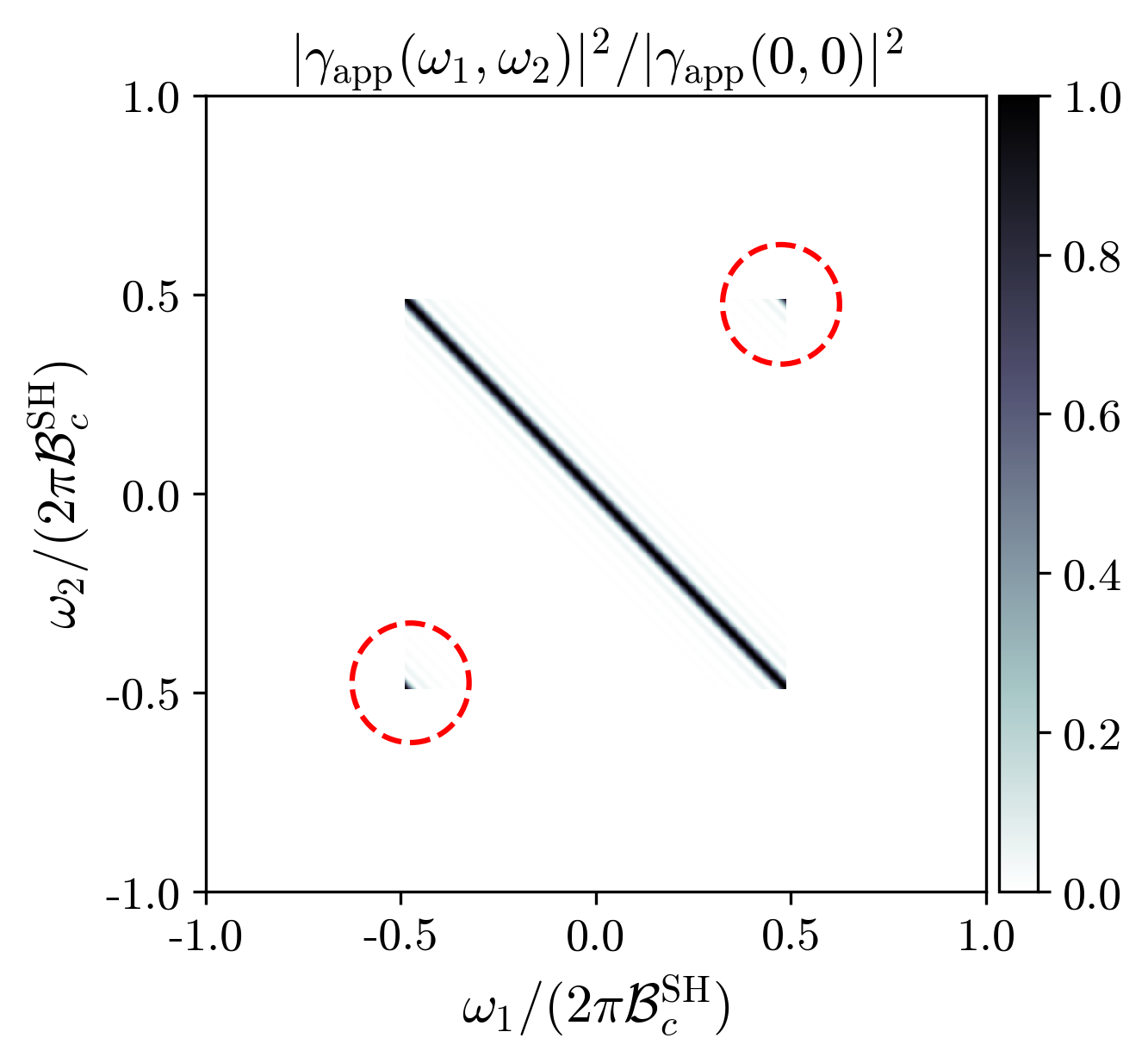}
    \caption{Plot of the approximate joint spectral intensity divided by its maximum value with the axes normalized by $2\pi\mathcal{B}^\mathrm{SH}_c$. The ``false'' contributions highlighted
    by the red dashed circles are due to the periodicity of the function $\widehat{u}(\omega)$ with a period $T_c$; see the discussion in the paragraph above Eq. \eqref{eq:ps1}.}
    \label{fig:7}
\end{figure}
\begin{figure*}
    \centering
    \includegraphics[width = \linewidth]{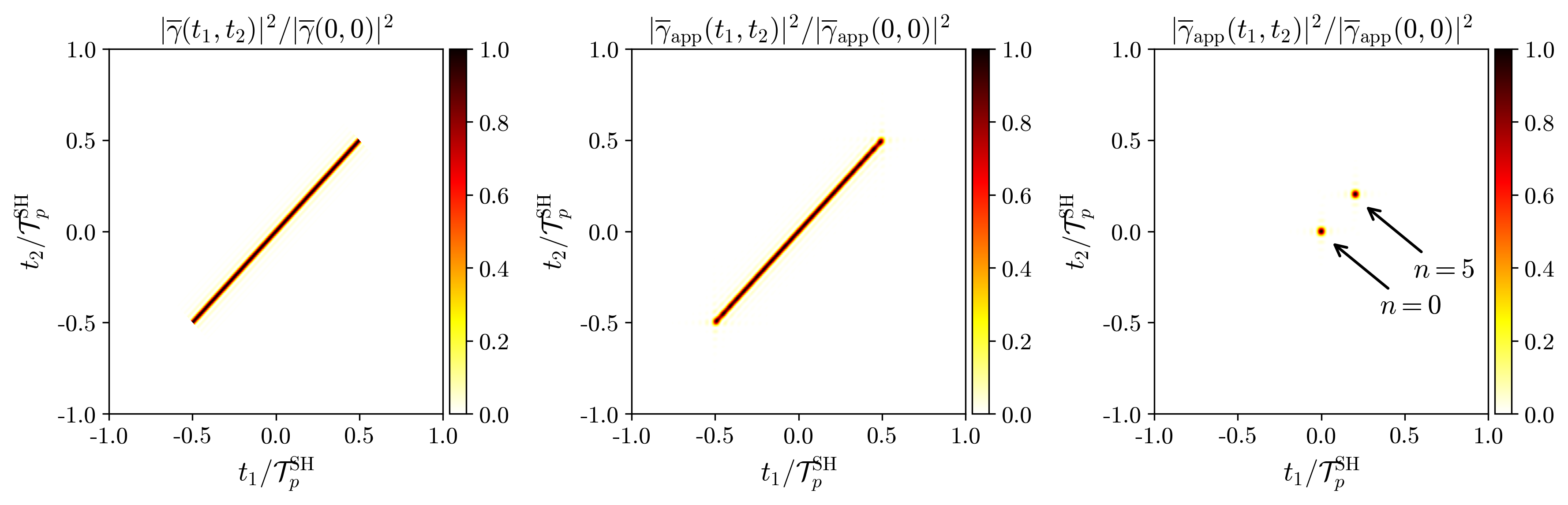}
    \caption{From left to right we plot the: sinc-hat joint temporal intensity divided by its maximum value; approximate joint temporal intensity divided by its maximum value; two contributions to the approximate joint temporal intensity divided by its maximum value when $n=0$ and $n=5$ with the axes all normalized by $\mathcal{T}^\mathrm{SH}_p$.}
    \label{fig:8}
\end{figure*}

Taking the Fourier transform of the approximate joint temporal amplitude
(\ref{eq:gamma_bar_app}) gives the approximate joint spectral amplitude
\begin{align}
  \gamma_\T{app}(\omega_{1},\omega_{2})&\equiv\int\frac{dt_{1}dt_{2}}{2\pi}\overline{\gamma}_\T{app}(t_{1},t_{2})e^{i\omega_{1}t_{1}}e^{i\omega_{2}t_{2}}\\
 & =\frac{T_c^2}{2\pi}\widehat{u}(\omega_{1}+\omega_{2})s(\omega_{1})s\left(\omega_{2}\right)\nonumber,
\end{align}
where 
\begin{align}
 & \widehat{u}(\omega)\equiv\sum_{n}u(nT_c)e^{in\omega T_c}\label{eq:f_hat_def}
\end{align}
and 
\begin{align}
  s(\omega)&\equiv\frac{1}{\sqrt{T_c}}\int_{-\infty}^{\infty}\overline{\eta}(t)e^{i\omega t}dt\label{eq:s_def}\\
 & =1,\nonumber \text{ for \ensuremath{-\frac{\Omega_c}{2}<\omega<\frac{\Omega_c}{2},}}\nonumber \\
 & =0, \text{ otherwise.}\nonumber 
\end{align}
We need to set $\widehat{u}(\omega),$ which satisfies $\widehat{u}(\omega+mT_c)=\widehat{u}(\omega)$
for any integer $m$. To ensure that $\overline{\gamma}_\T{app}(t_{1},t_{2})$
is a good approximation to $\overline{\gamma}(t_{1},t_{2})$ for the
sinc-hat model, we will want $u(nT_c)$ to be independent of $n$
over an appropriate range. Choosing $\mathcal{N}$ to be a large odd
integer, we put 
\begin{align}
  u(nT_c)&=\frac{1}{T_c\sqrt{\mathcal{N}}}, \text{ for \ensuremath{-\left(\frac{\mathcal{N}-1}{2}\right)\leq n\leq\left(\frac{\mathcal{N}-1}{2}\right),}}\nonumber\\
 & =0,  \text{ otherwise.}
\end{align}
Then from (\ref{eq:f_hat_def}) this leads to 
\begin{align}
 & \widehat{u}(\omega)=\frac{1}{T_c\sqrt{\mathcal{N}}}\frac{\sin\left(\frac{\mathcal{N}\omega T_c}{2}\right)}{\sin\left(\frac{\omega T_c}{2}\right)}.\label{eq:f_hat_work}
\end{align}
Now for the sinc-hat model we want the joint temporal amplitude $\overline{\gamma}(t_{1},t_{2})$
to be nonvanishing for $t_{1},t_{2}$ varying from $-T_p/2$ to
$T_p/2$. From the approximate form $\overline{\gamma}_\T{app}(t_{1},t_{2})$
that we are trying to construct (\ref{eq:gamma_bar_app}), this implies
that we should set 
\begin{align}
 & T_c\left(\frac{\mathcal{N}-1}{2}\right)=\frac{T_p}{2},
\end{align}
or 
\begin{align}
\label{eq:mathcalN def}
 & \mathcal{N}= 1 + \frac{T_p}{T_c} = \mathcal{K}_\mathrm{SH},
\end{align}
where the last equality holds to very good approximation when $T_p\gg T_c$. For $T_{p}\gg T_{c}$, which is the limit we consider here, we can take $\mathcal{N}$ to be either this or an odd integer close to it. The motivation for calling $\mathcal{K}$ 
the ``effective Schmidt number" is now clear, at least for this joint amplitude; $\mathcal{K}_\mathrm{SH}$ is the effective number of pseudo-Schmidt modes required for the decomposition.
We can then write (\ref{eq:f_hat_work}) as
\begin{align}
 & \widehat{u}(\omega)=\frac{1}{\sqrt{T_cT_p}}\frac{\sin\left(\frac{\omega}{2}(T_c+T_c)\right)}{\sin\left(\frac{\omega}{2}T_c\right)}.
\end{align}
Note that with this choice we find $\gamma_\T{app}(0,0)=\sqrt{T_c(T_p+T_c)}/(2\pi)$, while for the sinc-hat model we have exactly $\gamma(0,0)=\sqrt{T_{c}T_{p}}/(2\pi)$; so for $T_{p}/T_{c}\gg1$, $\gamma_\T{app}(0,0)$ is certainly equal
to $\gamma(0,0)$ to good approximation.

In Fig. \ref{fig:7} we plot $\left|\gamma_\T{app}(\omega_{1},\omega_{2})\right|^{2}$ for $T_{p}/T_{c}=24,$ taking $\mathcal{N}=\mathcal{K}_\mathrm{SH}=25$.
Comparing to the exact plot (middle plot in Fig. \ref{fig:4}) we see that there is indeed generally good agreement, with two main
differences: First, $\gamma_\T{app}(\omega_{1},\omega_{2})$ is only
nonzero for $(\omega_{1},\omega_{2})$ satisfying the bandwidth limiting
conditions $-\Omega_c/2\leq\omega_{1}\leq\Omega_c/2$ and $-\Omega_c/2\leq\omega_{2}\leq\Omega_c/2$,
while $\gamma(\omega_{1},\omega_{2})$ extends beyond that; this is
not so apparent in Fig. \ref{fig:4}, because outside the bandwidth limiting
conditions the true $\gamma(\omega_{1},\omega_{2})$ is very small. Second,
$\gamma_\T{app}(\omega_{1},\omega_{2})$ contains ``false contributions''
near $(\omega_{1},\omega_{2})=(\Omega_c/2,\Omega_c/2)$ and $(-\Omega_c/2,-\Omega_c/2$), see the contributions highlighted
by the red dashed circles in Fig. \ref{fig:7}.
This is related to the first difference, and arises because the postulated approximate form (\ref{eq:gamma_bar_app}) of $\overline{\gamma}_\T{app}(t_{1},t_{2})$
involves the function $\widehat{u}(\omega)$ that is periodic in $\omega$
with a period $1/T_c$.

We now look at the correlation functions that follow for a squeezed
state with the approximate joint temporal amplitude identified here,

\begin{align}
\label{eq:ps1}
 & \overline{\gamma}_\T{app}(t_{1},t_{2})={\displaystyle \sum_{n=-\frac{\mathcal{N}-1}{2}}^{\frac{\mathcal{N}-1}{2}}\sqrt{p_{n}}\overline{\eta}_{n}(t_{1})\overline{\eta}_{n}(t_{2}),}
\end{align}
where
\begin{align}
\label{eq:app_p_n}
 p_{n}&=\frac{1}{\mathcal{N}}, \text{ for}\;\ensuremath{-\left(\frac{\mathcal{N} - 1}{2}\right)\leq n\leq}\left(\frac{\mathcal{N} - 1}{2}\right)\nonumber\\
 & =0\;\text{otherwise.}
\end{align}
From this point forward we will denote the sums over $n$ leaving the bounds implicit. We plot $|\overline{\gamma}_\T{app}(t_{1},t_{2})|^2$ for the sinc-hat model
with $T_{p}/T_{c}=24$ in the middle diagram of Fig. \ref{fig:8}, repeating
the exact $|\overline{\gamma}(t_{1},t_{2})|^2$ for this model in the
left-most diagram; we can see the general level of agreement that
might be expected from the results shown in Fig. \ref{fig:4} and  Fig. \ref{fig:7}. In the right-most
plot we show $\overline{\eta}_{n}(t_{1})\overline{\eta}_{n}(t_{2})$
for $n=0$ and $n=5$; all such functions are of course well-localized, and
from (\ref{eq:ps1}) we see that the contribution to $\overline{\gamma}_\T{app}(t_{1},t_{2})$
from each of these is their product with the (pseudo-) Schmidt weight $p_{n}$. 

\begin{figure*}
    \centering
    \includegraphics[width = \linewidth]{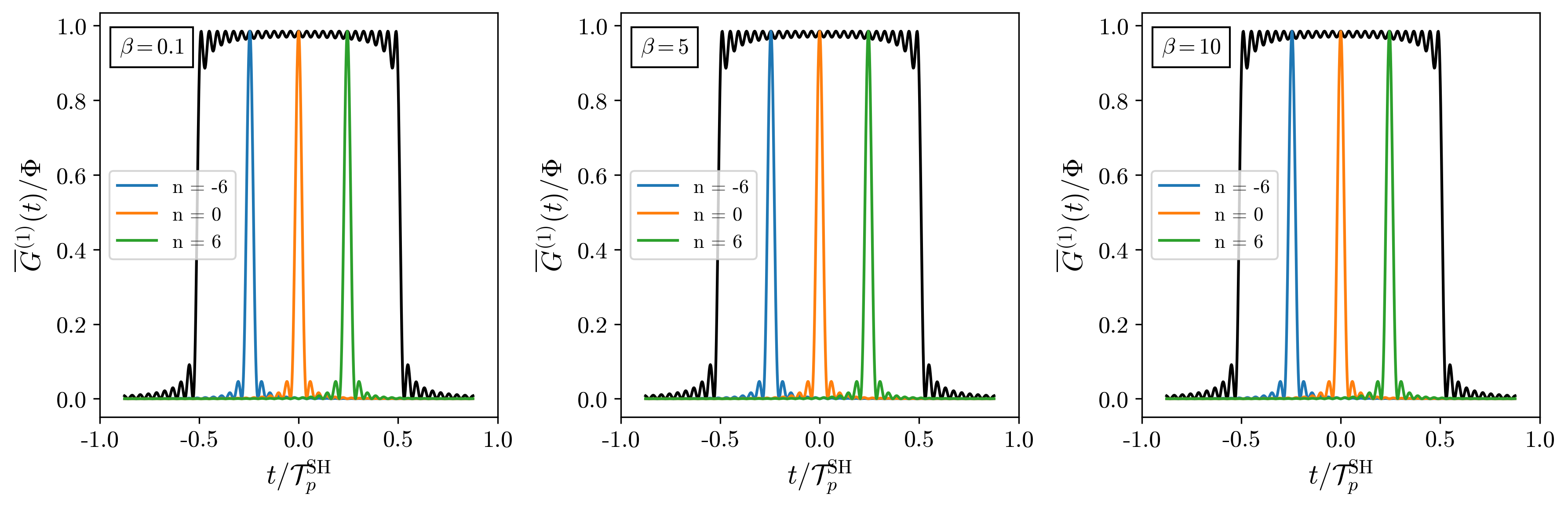}
    \caption{For the sinc-hat calculated from the pseudo-Schmidt decomposition, from left to right we plot $\overline{G}^{(1)}(t)/\Phi$ and a few contributions from different pseudo-Schmidt modes in Eq. \eqref{eq:Gresults_approx} with the horizontal axis normalized by $\mathcal{T}^\mathrm{SH}_p$ for $\beta = 0.1, 5,$ and $10$.}
    \label{fig:9}
\end{figure*}

Using the
general expressions (\ref{eq:Gresults},\ref{eq:Gwork}) for $\overline{G}^{(1)}(t)$
and $\overline{G}^{(2)}(t_{1},t_{2})$, we find that for our approximate
model (\ref{eq:ps1}) we have 
\begin{align}
\label{eq:Gresults_approx}
 & \overline{G}^{(1)}(t)=s^{2}\sum_{n}\left|\overline{\eta}_{n}(t)\right|^{2},\\
 & \overline{G}^{(2)}(t_{1},t_{2})=\overline{G}_\T{coh}^{(2)}(t_{1},t_{2})+\overline{G}_\T{incoh}^{(2)}(t_{1},t_{2}).\nonumber 
\end{align}
where 
\begin{align}
\label{eq:Gwork_approx}
 \overline{G}_\T{coh}^{(2)}(t_{1},t_{2})&=s^{2}c^{2}\left|\sum_{p}\overline{\eta}_{p}(t_{2})\overline{\eta}_{p}(t_{1})\right|^{2},\\
  \overline{G}_\T{incoh}^{(2)}(t_{1},t_{2})&=\frac{1}{2}s^{4}\sum_{n,m}\left|\overline{\eta}_{n}(t_{1})\overline{\eta}_{m}(t_{2})+\overline{\eta}_{n}(t_{2})\overline{\eta}_{m}(t_{1})\right|^{2}.\nonumber 
\end{align}
with 
\begin{equation}
\label{eq:PSsc}
    s=\sinh\left(\frac{\left|\beta\right|}{\sqrt{\mathcal{N}}}\right),\hspace{5mm}c=\cosh\left(\frac{\left|\beta\right|}{\sqrt{\mathcal{N}}}\right).
\end{equation}
These are of course the same formulas as for the exact Schmidt modes, except that here $s$ and $c$ are the same for each pseudo-Schmidt mode. But since the pseudo-Schmidt modes are localized over a time of order  $T_c$, we see that this  decomposition of $\overline{G}^{(1)}(t)$ identifies contributions from each pseudo-Schmidt mode that are localized in time windows much less than the width of $\overline{G}^{(1)}(t)$, which is on the order of $T_p$. We show some of these contributions in Fig. \ref{fig:9}. Here the range over which $n$ varies indicates the overall range of $\overline{G}^{(1)}(t)$ to very good approximation, and at any given time $t$ the contributions to $\overline{G}^{(1)}(t)$ come from at most a very few of the functions $\overline{\eta}_{n}(t)$ with $nT_c$ close to $t$.

\begin{figure*}
    \centering
    \includegraphics[width = \linewidth]{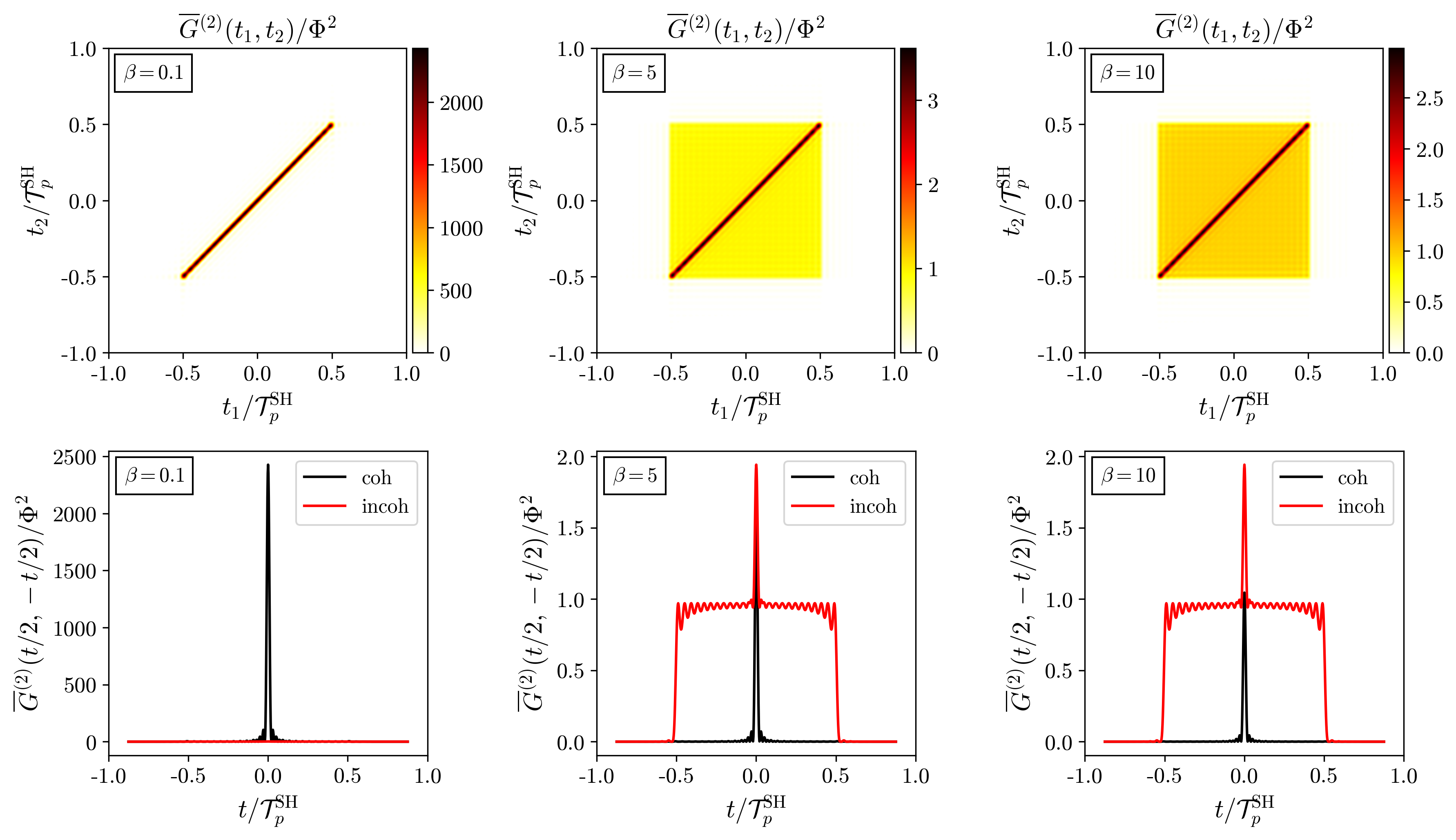}
    \caption{For the sinc-hat calculated from the pseudo-Schmidt decomposition, from left to right we plot $\overline{G}^{(2)}(t_1,t_2)/\Phi^2$ (top) and the coherent and incoherent contribution to $\overline{G}^{(2)}(t/2,-t/2)/\Phi^2$ (bottom) with the axes normalized by $\mathcal{T}^\mathrm{SH}_p$ for $\beta = 0.1, 5,$ and $10$.}
    \label{fig:10}
\end{figure*}

At least within the approximate pseudo-Schmidt decomposition we can now justify the use of the condition (\ref{weak squeezing}) to identify the weak squeezing regime.  Since in our example here $\mathcal{N} = \mathcal{K}_\text{SH}$, the weak squeezing limit can be written as 
\begin{equation}
\label{weak squeezing 2}
    \frac{|\beta|}{\sqrt{\mathcal{N}}} \ll 1,
\end{equation}
the expectation value of the number of photons in any pseudo-Schmidt mode is given by 
\begin{align}
\label{eq:Nmode}
 & N_\text{mode} =s^2 = \sinh^{2}\left(\frac{\left|\beta\right|}{\sqrt{\mathcal{N}}}\right),
\end{align}
while the expectation value of the number of photons in the full pulse is
\begin{align}
\label{eq:Npulse}
 & N_\text{pulse} =\mathcal{N}\sinh^{2}\left(\frac{\left|\beta\right|}{\sqrt{\mathcal{N}}}\right).
\end{align}
So in the weak squeezing limit (\ref{weak squeezing 2}) we have $N_{\text{mode}} \ll 1$; the expected number of photons in any pseudo-Schmidt mode is much less than unity.  However, the expected number of photons in the pulse, $N_{\text{pulse}}$ can be arbitrarily large; indeed, in the limit of weak squeezing we have $N_{\text{pulse}} \rightarrow |\beta|^2.$  Very generally, regardless of the level of squeezing, we can understand the CW limit as $|\beta|^2 \rightarrow \infty$ and $\mathcal{N} \rightarrow \infty$ such that $|\beta|^2/\mathcal{N}$ is a constant. 

In Fig. \ref{fig:10} we plot the approximate $\overline{G}^{(2)}(t_{1},t_{2})$
given by (\ref{eq:Gresults_approx},\ref{eq:Gwork_approx}), and compare
with the exact $\overline{G}^{(2)}(t_{1},t_{2})$ in Fig. \ref{fig:6} for the
sinc-hat model; we see very good agreement. Importantly, the decomposition
(\ref{eq:Gresults_approx},\ref{eq:Gwork_approx}) of $\overline{G}^{(2)}(t_{1},t_{2})$
into pseudo-Schmidt modes immediately illustrates the behavior of
that function in a way that the decomposition into the exact Schmidt
modes does not.

Considering the first plot when $\beta = 0.1$ ($|\beta|/\sqrt{\mathcal{N}}$ = 0.02), the dominance of $\overline{G}^{(2)}_{\text{coh}}(t/2,-t/2)$ over $\overline{G}^{(2)}_{\text{incoh}}(t/2,-t/2)$ follows immediately from the prefactors $s^2c^2$ and $s^4$ in their expressions (\ref{eq:Gwork_approx}). And in the sum over $p$ in the expression for $\overline{G}^{(2)}_{\text{coh}}(t/2,-t/2)$, of the terms $\overline{\eta}_{p}(t/2)\overline{\eta}_{p}(-t/2)$ 
both $\overline{\eta}_{p}(t/2)$ and $\overline{\eta}_{p}(-t/2)$ must be significant for a contribution to be made, which requires $p$ to be close to $0$ and $|t| \lesssim T_c$. The same behavior extends for the larger values of $\beta$ as in the exact sinc-hat Schmidt decomposition. Thus the largest contribution to $\overline{G}^{(2)}(t/2,-t/2)$ when $|t|\lesssim T_c$ comes from only a few terms in the pseudo-Schmidt decomposition, while it involves many terms in the Schmidt decomposition, and their interference.

The expression for $\overline{G}^{(2)}_{\text{incoh}}(t/2,-t/2)$, which in the pseudo-Schmidt decomposition (\ref{eq:Gwork_approx}) involves sums over two indices $n$ and $m$, contains two types of contributions. In the first, with $m=n$, we get terms that will only be significant if $m=n$ is close to $0$ and $|t| \lesssim T_c$, as in the expression for $\overline{G}^{(2)}_{\text{coh}}(t/2,-t/2)$; this gives the contribution to $\overline{G}^{(2)}_{\text{incoh}}(t/2,-t/2)$ that mirrors the form of $\overline{G}^{(2)}_{\text{coh}}(t/2,-t/2)$. This contribution to $\overline{G}^{(2)}_{\text{incoh}}(t/2,-t/2)$, and the term $\overline{G}^{(2)}_{\text{coh}}(t/2,-t/2)$, can thus be seen to arise from pairs of photons, each photon in a pair associated with the same pseudo-Schmidt mode. But the terms with $m \neq n$ can give contributions for $t$ on the order of $T_p$; they give rise to the broad background, which can be understood as arising from pairs of photons, with the photons in a pair associated with different pseudo-Schmidt modes. 

In a similar way one can understand the behavior of $\overline{G}^{(2)}_{\text{coh}}(\overline{t}+t/2,\overline{t}-t/2)$ and $\overline{G}^{(2)}_{\text{incoh}}(\overline{t}+t/2,\overline{t}-t/2)$ for $\overline{t} \neq 0$. Unlike in the decomposition of the joint temporal amplitude in terms of Schmidt modes, the decomposition in terms of pseudo-Schmidt modes immediately reveals the structure of those correlation functions.

\begin{figure*}
    \centering
    \includegraphics[width = \linewidth]{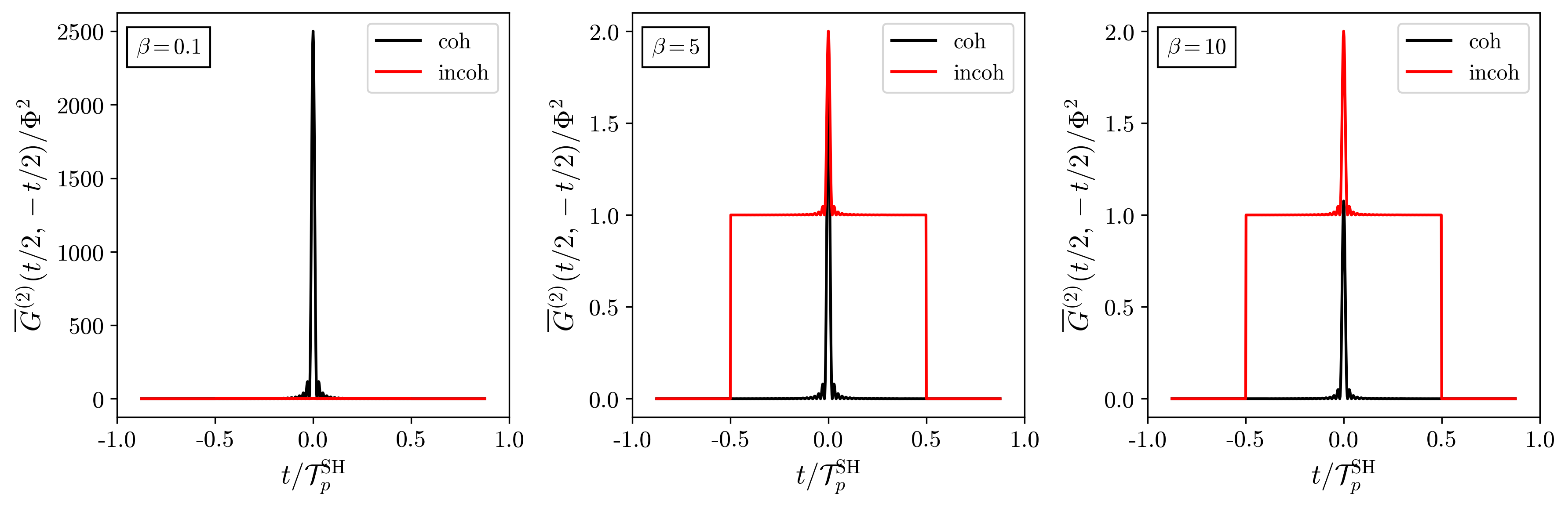}
    \caption{Using the analytical result (\ref{eq:G2analyticlCWlimit}),  from left to right we plot $\overline{G}^{(2)}(t_1,t_2)/\Phi^2$ (top) and the coherent and incoherent contribution to $\overline{G}^{(2)}(t/2,-t/2)/\Phi^2$ (bottom) with the axes normalized by $\mathcal{T}^\mathrm{SH}_p$ for $\beta = 0.1, 5,$ and $10$.}
    \label{fig:11}
\end{figure*}

And in fact, we can derive an analytic expression for
the approximate (\ref{eq:Gresults_approx},\ref{eq:Gwork_approx})
$\overline{G}^{(1)}(t)$ and $\overline{G}^{(2)}(t_{1},t_{2})$ in
the CW limit. Noting that 
\begin{align}
\label{eq:approx_comp}
 & \sum_{n=-\infty}^{\infty}\overline{\eta}_{n}(t_{2})\overline{\eta}_{n}(t_{1})=\frac{1}{T_c}\text{sinc}\left(\frac{\pi\:\Delta t}{T_c}\right),
\end{align}
where $\Delta t\equiv t_{2}-t_{1}$, for $t_{1}$ and $t_{2}$ in
the center  of a pulse of duration $T_p$, when $T_p\rightarrow\infty$
we can write 
\begin{align}
 & \overline{G}_\T{coh}^{(2)}(t_{1},t_{2})\rightarrow\frac{s^{2}c^{2}}{T_c^{2}}\text{sinc}^{2}\left(\frac{\pi\:\Delta t}{T_c}\right),
\end{align}
while 
\begin{align}
 & \overline{G}^{(1)}(t)\rightarrow\frac{s^{2}}{T_c}.
\end{align}
Finally, noting that
\begin{equation}
\label{eq:G1(t1,t2)PS}
    \begin{split}
    \overline{G}^{(1)}(t_1,t_2)& = s^2\sum_n\overline{\eta}_{n}(t_{2})\overline{\eta}_{n}(t_{1})\\
    &\rightarrow \frac{s^2}{T_c}\text{sinc}\left(\frac{\pi\:\Delta t}{T_c}\right),
    \end{split}
\end{equation}
then with the alternate form of $\overline{G}_\T{incoh}^{(2)}$ (see Eq. \eqref{eq:G2_incoh_alt}) we have
\begin{align}
\label{eq:G_incoh_PS}
 & \overline{G}_\T{incoh}^{(2)}(t_{1},t_{2})\rightarrow\frac{s^{4}}{T_c^{2}}\left(1+\text{sinc}^{2}\left(\frac{\pi\:\Delta t}{T_c}\right)\right),
\end{align}
and so 
\begin{align}
\label{eq:G2analyticlCWlimit}
 \overline{G}^{(2)}(t_{1},t_{2})&\rightarrow\overline{G}^{(2)}(\Delta t) = \frac{s^{2}c^{2}}{T_c^{2}}\text{sinc}^{2}\left(\frac{\pi\:\Delta t}{T_c}\right)\\
 &\hspace{15mm}+\frac{s^{4}}{T_c^{2}}\left(1+\text{sinc}^{2}\left(\frac{\pi\:\Delta t}{T_c}\right)\right)\nonumber.
\end{align}
We plot this in Fig. \ref{fig:11} (cf. Fig. \ref{fig:6} and \ref{fig:10}.)

From the above discussion, we are motivated to think of $\overline{G}^{(2)}(\overline{t} + t/2, \overline{t} - t/2)$ for times $|t|\lesssim T_c$ when $n=m$ as arising from one pseudo-Schmidt mode at a time, as it were, on a ``mode-by-mode'' basis. Since the pseudo-Schmidt decomposition is valid only for joint amplitudes such as the sinc-hat joint amplitude, where there is a high degeneracy in the Schmidt weights, we give this calculation in Appendix \ref{sec:app:Mode-by-mode calculation}. 

\section{The Whittaker-Shannon decomposition}
\label{sec:The Whittaker-Shannon decomposition}
The ability to construct approximate pseudo-Schmidt modes for the sinc-hat joint amplitude that
are ``localized'' in time ($\overline{\eta}(t)$, see (\ref{eq:etabar_def}))
relied on the near-degeneracy of the Schmidt modes. But this cannot be generally expected; see, e.g., Fig. \ref{fig:1} for the Schmidt
amplitudes of the double-Gaussian joint amplitude, where there is no
such near-degeneracy. Nonetheless, the recognition used above that
only a finite frequency range -- there from $-\Omega_c/2$ to $\Omega_c/2$
-- is important can be employed to construct an extension of a Schmidt decomposition more generally; it shares the feature of the pseudo-Schmidt decomposition above that the functions involved are localized in time.

We consider a joint spectral amplitude $\gamma(\omega_{1},\omega_{2})$
to be bandwidth limited, in that at
least approximately it can be taken to be nonzero only for $-\Omega/2\leq\omega_{1},\omega_{2}\leq\Omega/2$,
where $\Omega$ is a positive frequency. Then we can construct a Whittaker-Shannon
decomposition of the joint amplitude based on the Whittaker-Shannon
Sampling Theorem \cite{whittaker1915xviii,shannon1949communication,butzer1992sampling}. That theorem states that for a function
$\overline{g}(t)$ that is bandwidth limited in the sense defined
above -- i.e., involving only constituent frequencies in the range
$-\Omega/2\leq\omega\leq\Omega/2$ -- we can write 
\begin{align}
 & \overline{g}(t)=\sum_{n}\overline{g}(n\tau)\text{sinc}\left(\frac{(t-n\tau)\pi}{\tau}\right),\label{eq:gbar_expand}
\end{align}
where $n$ ranges over the integers and 
\begin{align}
 & \tau=\frac{2\pi}{\Omega}.
\end{align}
Defining 
\begin{align}
 & \overline{\chi}(t)=\frac{1}{\sqrt{\tau}}\text{sinc}\left(\frac{\pi t}{\tau}\right)
\end{align}
and putting 
\begin{align}
 & \overline{\chi}_{n}(t)=\overline{\chi}(t-n\tau),
\end{align}
we have 
\begin{align}
 & \int\overline{\chi}_{n}^{*}(t)\overline{\chi}_{m}(t)dt=\delta_{nm},
\end{align}
and we can write (\ref{eq:gbar_expand}) as 
\begin{align}
 & \overline{g}(t)=\sqrt{\tau}\sum_{n}\overline{g}(n\tau)\overline{\chi}_{n}(t).
\end{align}
The Fourier transform of $\overline{g}(t)$ is 
\begin{align}
 & g(\omega)=\int\frac{dt}{\sqrt{2\pi}}\overline{g}(t)e^{i\omega t}=\sqrt{\tau}\sum_{n}\overline{g}(n\tau)\chi_{n}(\omega),
\end{align}
where 
\begin{align}
    \chi_{n}(\omega)&=\frac{e^{i\omega n\tau}}{\sqrt{\Omega}}\text{ for \ensuremath{-\frac{\Omega}{2}\leq\omega\leq\frac{\Omega}{2}}}\\
 & =0 \text{ otherwise,}
\end{align}
is the Fourier transform of $\overline{\chi}_{n}(t)$;
we have
\begin{align}
 & \int\chi_{n}^{*}(\omega)\chi_{m}(\omega)d\omega=\delta_{nm}.
\end{align}
For frequencies within the band limited region the set of functions $\{\chi_n(\omega)\}$ and $\{\chi_n(t)\}$ form an approximately complete set of functions and we refer to them as the ``Whittaker-Shannon modes.''

The ``Whittaker-Shannon decomposition" of the joint temporal amplitude
$\overline{\gamma}(t_{1},t_{2})$, and its Fourier transform $\gamma(\omega_{1},\omega_{2})$,
follow immediately from this if $\gamma(\omega_{1},\omega_{2})$
is bandwidth limited, with frequency bandwidth $\Omega.$ Using the
sampling theorem for both variables, we have 
\begin{align}
 \overline{\gamma}(t_{1},t_{2})&=\tau\sum_{n,m}\overline{\gamma}(n\tau,m\tau)\overline{\chi}_{n}(t_{1})\overline{\chi}_{m}(t_{2}),\label{eq:WSequations}\\
 \gamma(\omega_{1},\omega_{2})&=\tau\sum_{n,m}\overline{\gamma}(n\tau,m\tau)\chi_{n}(\omega_{1})\chi_{m}(\omega_{2}),\nonumber 
\end{align}
Unlike a Schmidt (or pseudo-Schmidt) decomposition, this involves a double sum. But the Whittaker-Shannon supermodes involved, with raising operators given by 
\begin{align}
 & B_{n}^{\dagger}=\int d\omega\chi_{n}(\omega)a^{\dagger}(\omega)=\int dt\overline{\chi}_{n}(t)\overline{a}^{\dagger}(t),
\end{align}
are associated with functions that are localized in time ($\overline{\chi}_{n}(t))$ and can be inverted so that
\begin{equation}
\label{eq:inverseB_n}
    \overline{a}(t) = \sum_n \overline{\chi}_n(t)B_n.
\end{equation} 
It will be useful to define
\begin{equation}
    \beta_{nm} = \beta\tau\overline\gamma(n\tau,m\tau),
\end{equation}
which is generally complex but symmetric, $\beta_{nm}=\beta_{mn}$; then we can write the squeezed state $\left|\Psi\right\rangle $ (\ref{eq:squeezed state},\ref{eq:PsiTime}) as 
\begin{align}
\label{eq:sq_state_WS}
 & \left|\Psi\right\rangle =e^{\frac{1}{2}\sum\limits_{n,m}\beta_{nm}B_{n}^{\dagger}B_{m}^{\dagger}-\text{h.c.}}\vac .
\end{align}
Although non-diagonal terms of $\beta_{nm}$ will be important, since
$\overline{\gamma}(t_{1},t_{2})$ is only significant for $\left|t_{1}-t_{2}\right|$
on the order of the coherence time, we can expect $\beta_{nm}$ to be significant only for $m$ ``reasonably close'' to $n$; we will examine this in more detail below. Note that our
examples of the double-Gaussian and sinc-hat joint spectral amplitudes
illustrate that imposing the approximation that those amplitudes are
bandwidth limited requires a choice of $\Omega\ge 2\pi\mathcal{B}_c$ that corresponds to $\tau\le \mathcal{T}_c$, that is, $\tau$ is smaller than, but on the order of, the coherence time.

As a first example, we apply 
the Whittaker-Shannon decomposition to the sinc-hat joint amplitude (\ref{eq:sinc-hat},\ref{eq:sinc-hat-parameters}), and 
show how the approximate pseudo-Schmidt decomposition (\ref{eq:ps1},\ref{eq:app_p_n}) arises as the 
limiting case when $T_p\gg T_c$. To apply the Whittaker-Shannon decomposition we must choose a bandlimit, and for the sinc-hat example the joint spectral amplitude ranges mostly within the ``box'' set by the width $\Omega_c$; a natural choice for the bandlimit $\Omega$ is then to set $\Omega \to \Omega_c$ and it follows that $\tau \to T_c$. However, the sinc-hat joint spectral amplitude will never fit exactly inside a box of width $\Omega_c$  because it is only exactly bandlimited in the diagonal direction set by the range of $\phi((\omega_1 - \omega_2)/2)$, see Eq. \eqref{eq:sinc-hat}, \eqref{eq:sinc-hat-parameters}, so the corners will always exist outside the box, see the discussion surrounding Eq. \eqref{eq:sinchatbandwidth} and Appendix \ref{sec:app:Schematic of sinc-hat joint intensity}.  When we work in the limit that $T_p\gg T_c$, 
where the contributions that exist outside the boundary of the box are very small, then to good approximation we can treat the sinc-hat joint spectral amplitude as bandlimited with bandwidth $\Omega = \Omega_c$ ($\tau = T_c$).

With this choice of $\Omega$ and the sinc-hat joint temporal amplitude (Eq. \eqref{eq:sinc-hat}),  we evaluate
\begin{equation}
    \begin{split}
        \tau\:\overline\gamma(n\tau, m\tau) &= \delta_{nm}\sqrt{\frac{T_c}{T_p}},\text{for} -\frac{T_p}{2}\le nT_c \le \frac{T_p}{2}\\
        & = 0,  \text{otherwise},
     \end{split}
\end{equation}
where we have used $\text{sinc}((n-m)\pi) = \delta_{nm}$. We find that the range of Whittaker-Shannon modes is then diagonal, and only ranges over the pulse duration set by $\overline{\alpha}(nT_c)$; identifying that $T_p/T_c = \mathcal{N} - 1$ \eqref{eq:mathcalN def} and $p_n$ \eqref{eq:app_p_n} as in section \ref{sec:An approximate Schmidt decomposition},  it immediately follows that
\begin{align}
 \overline{\gamma}(t_{1},t_{2}) \rightarrow\sum_{n=-\left(\frac{\mathcal{N}-1}{2}\right)}^{\left(\frac{\mathcal{N}-1}{2}\right)}\sqrt{p_{n}}\overline{\chi}_{n}(t_{1})\overline{\chi}_{n}(t_{2}),
\end{align}
where $\overline{\chi}_{n}(t) = \overline{\eta}_{n}(t)$.

For the sinc-hat model (see Fig. \ref{fig:4}), the reduction of the double sum to its diagonal contributions is only possible because the joint spectrum is exactly bandlimited in the $(\omega_1 - \omega_2)$ direction (see \eqref{eq:sinc-hat} and \eqref{eq:sinc-hat-parameters}); however, the resulting Whittaker-Shannon decomposition will only be accurate as a single sum in the $T_p\gg T_c$ limit for which the contributions that exist outside the box become very small and can safely be neglected.

In general, this reduction will not be possible and we will need to include some -- but typically only a few -- 
off diagonal contributions to properly approximate the joint amplitude with the Whittaker-Shannon decomposition. As we will see below, the single sum and product state that follows if only diagonal terms are included (see Eq. \eqref{eq:schmidt product state}) is much easier and more intuitive to work with, so we always want to be as close to that limit as we can be. This means that for a not-exactly-bandlimited joint spectral amplitude, the choice of a bandwidth $\Omega=2\pi/\tau$ always involves a trade-off: We want $\tau$ as large as possible so that each $\overline{\gamma}(n\tau,n\tau)$ covers most of the $|t_1 - t_2|$ behavior; however, if we choose $\tau$ too large then $\Omega$ becomes too small to even approximately cover the bandwidth of the joint spectral amplitude. 

To see how this plays out in practice, consider the double-Gaussian joint amplitude (\ref{double_gaussian}). Although it is not strictly bandlimited, if 
we set $\Omega/2\pi = \mathcal{B}_c^\mathrm{DG} = a\sigma_c/(2\pi\sqrt{2})$ (see Eq. \ref{eq:DGwidths}), to very good approximation we can neglect the high frequency contributions for a reasonable choice of $a$. Then $\tau = 2\pi\sqrt{2}/(a\sigma_c)$, and
\begin{equation}
    \label{eq:gammpropto}
    \overline{\gamma}(n\tau,m\tau)\propto e^{-\frac{\sigma_{c}^{2}\tau^2(n-m)^{2}}{4}}
\end{equation}
sets the off diagonal range; for our choice of $\tau$ we have $\sigma_c^2\tau^2 = 8\pi^2/a^2$ and as $a$ increases the bandlimit gets larger, but so does the range over which $|n-m|$ is significant. Setting $a = 2\sqrt{2\pi}$ as in Section \ref{sec:Schmidt modes}, we have $\sigma_c^2\tau^2/4 = \pi$ and the $1/e$ drop-off in Eq. \eqref{eq:gammpropto} occurs when $|n-m| = 1/\sqrt{\pi} \approx 0.5$.  In this example, and more generally for only approximately band limited joint spectral amplitudes, one could investigate optimizing $\tau$ constrained by a specified error tolerance on the Whittaker-Shannon interpolation.

Clearly the Whittaker-Shannon decomposition depends on the two index parameter $\beta_{nm}$, but to make a comparison to the pseudo-Schmidt decomposition we focus on $\mathring{\beta}$, which we define to be the value of $\beta_{nm}$ at the $n$ and $m$ for which $|\overline\gamma(n\tau,m\tau)|$  takes its maximum value.  Denoting the value of $|\overline\gamma(n\tau,m\tau)|$ at this $n$ and $m$ by
$\overline{\gamma}_\text{max}$, we have
\begin{equation}
\label{eq:mathringbeta}
\mathring\beta = \tau\beta\overline\gamma_\text{max},
\end{equation}
and
\begin{equation}
\label{eq:betabardef}
    \beta_{nm} = \mathring\beta r_{nm},
\end{equation}
with $r_{nm} = \overline{\gamma}(n\tau,m\tau)/\overline{\gamma}_\text{max}$.  The range of $|n-m|$ over which $r_{nm}$ is significant identifies the range over which $\beta_{nm}$ varies.  

Again using the double-Gaussian joint amplitude as an example, which achieves its maximum at $t_1=t_2=0$, we find
\begin{equation}
\label{eq:beta_0_DG}
    |\mathring\beta| =\sqrt{2}|\beta|\frac{\tau}{\sqrt{\mathcal{T}^\text{DG}_p\mathcal{T}^\text{DG}_c}},
\end{equation}
where we have used $\mathcal{T}^\text{DG}_p\mathcal{T}^\text{DG}_c = 2\pi/\sigma_c\sigma_p$. From the discussion above, we set $\tau = \mathcal{T}_c^\text{DG}$, and obtain
\begin{equation}
    |\mathring\beta| = \sqrt{2}|\beta| \sqrt{\frac{\mathcal{T}_c^\text{DG}}{\mathcal{T}_p^\text{DG}}} = \sqrt{2}\frac{|\beta|}{\sqrt{\mathcal{K}_\text{DG}}},
\end{equation}
so the effective Schmidt number naturally arises and -- aside from the benign factor of $\sqrt{2}$ -- identifies the weakly or strongly squeezed limit as $|\mathring\beta|\ll1$ or $|\mathring\beta|\gg 1$ respectively, justifying the definition in section \ref{sec:Schmidt modes}. Since each $\overline{\chi}_n(t)$ has a width $\tau=\mathcal{T}_c^\text{DG}$, the effective Schmidt number $\mathcal{K}_\text{DG}$ roughly identifies the number of Whittaker-Shannon modes that are relevant along $r_{nn}$. Further, following the discussion around Eq. \eqref{weak squeezing 2}, for a very long pulse such that $\mathcal{K}_\text{DG}\gg1$ the squeezing parameter $|\beta|$ can be quite large but $|\mathring\beta|$ remains finite.

In Fig. \ref{fig:13} we plot $|\overline \gamma(t_1,t_2)|^2$ and  $|\gamma(\omega_1,\omega_2)|^2$, reconstructed from using the Whittaker-Shannon decomposition for the double-Gaussian joint amplitude of Fig. \ref{fig:1}, where we have chosen $\Omega = 2\pi\mathcal{B}_c$, and plot $r_{nm}$. Comparing to Fig. \ref{fig:1}, we see very good agreement, and from the zoomed in plot of $r_{nm}$ in the lower right corner it is clear that only a few neighbouring Whittaker-Shannon modes in the $|n-m|$ direction are relevant.

The argument that even though $|\beta|$ can be quite large $|\mathring\beta|$ remains finite holds true for \emph{any} joint amplitude, because as long as it is square normalized \eqref{eq:jtanorm} it will carry pre-factors on its behavior in the two directions in the plane (of either $(t_1,t_2)$ or $(\omega_1,\omega_2)$) over which it is defined. In Appendix \ref{sec:app:Joint temporal amplitude scaling}, we show that given a joint temporal amplitude characterized by $\mathcal{T}_p$ and $\mathcal{T}_c$, its maximum value $\overline{\gamma}_\text{max}$ is on the order of
\begin{equation}
    \overline{\gamma}_\text{max} \sim \frac{1}{\sqrt{\mathcal{T}_p\mathcal{T}_c}}
\end{equation}
To apply the Whittaker-Shannon decomposition we set $\Omega = 2\pi \mathcal{B}_c$ so that $\tau= \mathcal{T}_c$. Then one immediately finds 

\begin{equation}
\label{eq:effectiveK}
    \begin{split}
        |\mathring\beta| = \tau|\beta|\overline\gamma_\text{max} \sim \frac{|\beta|}{\sqrt{\mathcal{K}}}.
    \end{split}  
\end{equation}

What remains to be shown is the relation between the Schmidt number $K$ and the effective Schmidt number $\mathcal{K}$. In Appendix \ref{sec:app:relation between K and Delta}, we show that for a general joint temporal amplitude characterized by widths $\mathcal{T}_p$, $\mathcal{T}_c$ indicated schematically in Fig. \ref{fig:picture1}, and a given $\tau$ which we set to be equal to $\mathcal{T}_c$, the inequality 
\begin{equation}
    K \le \frac{\mathcal{T}_p}{\tau} = \frac{\mathcal{T}_p}{\mathcal{T}_c} = \mathcal{K},
\end{equation}
generally holds; of course this is conditioned on $\tau$ being sufficiently small that the Whittaker-Shannon decomposition can be accurately used. Thus by applying the Whittaker-Shannon interpolation formula -- which previously has not been linked to squeezed light or the Schmidt decomposition -- we are able to show that the Schmidt number and effective Schmidt number are intimately related. Further, the Whittaker-Shannon decomposition may be an important stepping stone in formally linking the Schmidt number to other measures of the correlation, like the time-bandwidth product and its generalizations, in the same way that the Whittaker-Shannon interpolation formula is intimately linked to the Shannon number for classical signals and the information content of images \cite{fedorov2006short,mikhailova2008biphoton,fedorov2008spontaneous,brecht2013characterizing,landau1962prolate,simons2009slepian,freeden2010handbook,miller2000communicating,pires2009direct,pors2008shannon,slepian1961prolate,wang2017review,whittaker1915xviii,shannon1949communication,butzer1992sampling,slepian1983some}.  

Finally, we are now in a position to explain why the effective Schmidt number for the sinc-hat example is nearly equal to the exact Schmidt number. First we emphasize that the sinc-hat is an idealization with a sharp cutoff in time (frequency) and is constant along $t_1=t_2$ $(\omega_1=-\omega_2)$. This means that in the limit when $T_p\gg T_c$, to very good approximation it is exactly bandlimited with $\Omega = \Omega_c$ ($\tau = T_c$). Mathematically, the sharp cutoff in frequency results in an exact diagonalization of $\overline{\gamma}(n\tau,m\tau)$, and since $\overline{\gamma}(n\tau,n\tau)$ is identical for all $n$ where it is not zero (see Eq. (\ref{eq:ps1})) the sum in Eq. ({\ref{eq:SN})} can be carried out exactly resulting in $K \approx T_p/T_c$, as we found in section \ref{eq:Example 2: The sinc-hat}; we then expect
\begin{equation}
    \frac{\mathcal{K}_\mathrm{SH}}{K_\mathrm{SH}} = 1 + \frac{T_c}{T_p} \to 1,
\end{equation}
for $T_p/T_c\to \infty$. Physically, the Schmidt amplitudes are near-degenerate, so there is no unique set of Schmidt modes.

\begin{figure*}
    \centering
    \includegraphics[width = \linewidth]{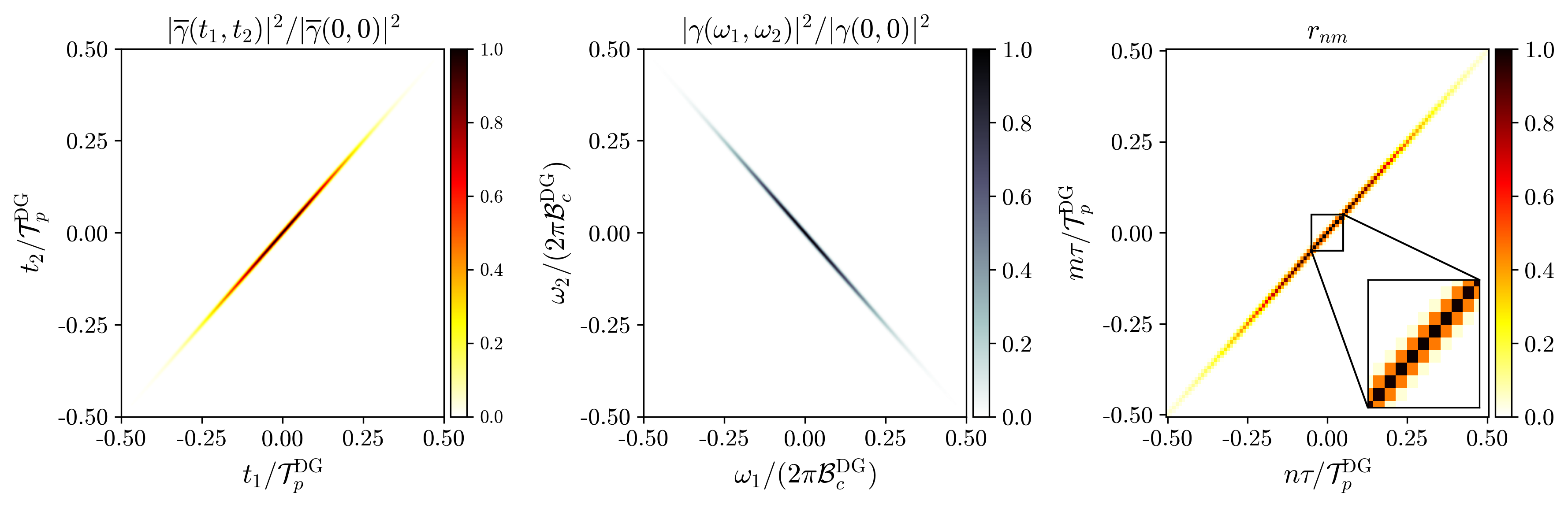}
    \caption{For the double-Gaussian, from left to right we plot the: joint temporal intensity divided by its maximum value with the axes normalized by $\mathcal{T}_p^\mathrm{DG}$; joint spectral intensity divided by its maximum value with the axes normalized by $2\pi \mathcal{B}_c^\mathrm{DG}$; amplitudes $r_{nm}$.}
    \label{fig:12}
\end{figure*}

\section{Employing the Whittaker-Shannon decomposition}
\label{sec:Employing the Whittaker-Shannon decomposition}
In using the pseudo-Schmidt decomposition to ``take apart" the sinc-hat joint temporal amplitude, we showed that the expressions for the correlation functions $\overline{G}^{(1)}(t)$ and $\overline{G}^{(2)}(t_1,t_2)$ could be easily understood in terms of the properties of the supermodes involved in the decomposition. In this section we look at the corresponding expressions for the correlation functions when we ``take apart" the joint amplitudes using the Whittaker-Shannon decomposition instead. The results are more complicated, but again the behavior of the correlation functions can be understood in terms of the properties of the Whittaker-Shannon supermodes in an intuitive way. And since the Whittaker-Shannon decomposition can be much more widely applied than a pseudo-Schmidt decomposition, the results here are much more general.

To simplify the notation we write the squeezed state in Eq. \eqref{eq:sq_state_WS} as $\ket{\Psi} = \tilde{S}\vac$, where
\begin{equation}
\label{eq:sqoperatordisentangled}
    \Tilde{S} = e^{\frac{1}{2}\sum\limits_{n,m}\beta_{nm}B_{n}^{\dagger}B_{m}^{\dagger}-\text{h.c.}},
\end{equation}
is the squeezing operator. To calculate the correlation functions analogously to what was done with the Schmidt and pseudo-Schmidt decompositions, we use the inverse relation (Eq. \eqref{eq:inverseB_n}) and the transformation \cite{lo1993generalized},
\begin{subequations}
\label{eq:transformBn}
    \begin{gather}
        \Tilde{S}^\dagger B_r \Tilde{S} = \mu_{rs}B_s + \nu_{rs}B_s^\dagger ,\\
        \Tilde{S}^\dagger B_r^\dagger \Tilde{S} = \mu_{rs}^*B_s^\dagger  + \nu_{rs}^*B_s,
    \end{gather}
\end{subequations}
where we adopt the convention that repeated indices are to be summed over, and
\begin{subequations}
\label{eq:CSdef}
    \begin{gather}
        \mu_{rs} = \delta_{rs} + \frac{1}{2!}\beta_{ra}\beta_{as}^* + \frac{1}{4!}\beta_{ra}\beta_{ab}^*\beta_{bc}\beta_{cs}^* + ...,\\
        \nu_{rs} = \beta_{rs} + \frac{1}{3!}\beta_{ra}\beta^*_{ab}\beta_{bs} + ...\,.
    \end{gather}
\end{subequations}
Note that from the symmetric property of $\beta_{nm}$ it follows that $\mu_{rs}$  is Hermitian and $\nu_{rs}$ is symmetric. 

For the Schmidt or pseudo-Schmidt decomposition the transformation used always involves a single supermode
(see Eq. \eqref{eq:transformAn}); for the general Whittaker-Shannon decomposition the squeezing transformation 
is more complicated. The structure of the squeezing transformation \eqref{eq:transformBn}, and the form of $\mu_{rs}$ and $\nu_{rs}$, motivates the use of matrix multiplication; $\beta_{nm}$ is now treated as a complex square symmetric matrix which we denote by $\m{\beta}$ ($\m{\beta}^* = \m{\beta}^\dagger$). To implement the squeezing transformation it is convenient to use the ``left'' polar decomposition of $\m{\beta}$ (valid for \emph{any} complex square matrix), given by
\begin{equation}
\label{eq:polardecomp}
    \m{\beta} = \mathbf{UP}, 
\end{equation}
where $\mathbf{P} = \sqrt{\m{\beta^\dagger\beta}}$ is Hermitian and $\mathbf{U}$ is unitary. Equivalently, if we set $\mathbf{Q} = \mathbf{UPU}^\dagger$ then we have the ``right'' polar decomposition $\m{\beta} = \mathbf{QU}$; since $\m{\beta}$ is symmetric, $\mathbf{P}^T = \mathbf{Q}$ and $\mathbf{U}^T = \mathbf{U}$. Using the polar decomposition and its properties, the matrices $\m{\mu}$ and $\m{\nu}$ are given by \cite{ma1990multimode}
\begin{subequations}
    \begin{gather}
    \m{\mu} = \T{cosh}(\mathbf{UPU^\dagger}) = \T{cosh}\mathbf{Q}, \\
    \m{\nu} = \mathbf{U}\T{sinh}\mathbf{P} = (\T{sinh}\mathbf{Q})\mathbf{U}.
    \end{gather}
\end{subequations}
The form of $\m{\mu}$ and $\m{\nu}$ guarantees that the transformation in Eq. \eqref{eq:transformBn} preserves the commutation relations of the $B_n$ operators, as expected since the transformation is unitary.

In Appendix \ref{sec:app:Correlation functions} we calculate the correlation functions (Eq. \eqref{eq:gs-1}); we find
\begin{equation}
    \begin{split}
    \label{eq:WSG1}
        \overline{G}^{(1)}(t_1,t_2)& = \m{\overline{\chi}}\dg(t_1)(\T{sinh}^2\mathbf{P})\m{\overline{\chi}}(t_2),
    \end{split}
\end{equation}
and
\begin{equation}
\label{eq:WSG2}
    \begin{split}
        \overline{G}^{(2)}(t_1,t_2)&=|\m{\overline{\chi}}^T(t_1)\mathbf{U}(\T{sinh}\mathbf{P})(\T{cosh}\mathbf{P})\m{\overline{\chi}}(t_2)|^2\\
        &+ \overline{G}^{(1)}(t_1)\overline{G}^{(1)}(t_2) + |\overline{G}^{(1)}(t_1,t_2)|^2
    \end{split}
\end{equation}
where $\m{\overline{\chi}}(t) = (..., \overline{\chi}_{-1}(t), \overline{\chi}_0(t), \overline{\chi}_1(t),...)^T$ is the column vector formed from the set $\{\overline{\chi}_n(t)\}$ for a given $t$. 

Since the functions $\overline{\chi}_n(t)$ are similar to the pseudo-Schmidt modes $\overline{\eta}_n(t)$, Eqs. \ref{eq:WSG1} and \ref{eq:WSG2} for the correlation functions calculated using the Whittaker-Shannon decomposition are the generalization of the pseudo-Schmidt results (\ref{eq:Gresults_approx}, \ref{eq:Gwork_approx}), valid for an approximately bandlimited, but otherwise general joint amplitude. Both the pseudo-Schmidt and Whittaker-Shannon mode functions are 
localized, so for short time differences the structure of Eq. \ref{eq:WSG2} reduces to that of the pseudo-Schmidt decomposition, and we can again think of the correlation function on a ``mode-by-mode'' basis; however, since this correspondence 
is only approximate, we discuss it in Appendix  \ref{sec:app:Mode-by-mode calculation}.

\subsection{Packet expansion}
\label{sec:Packet expansion}
From Eq. \eqref{eq:WSG1}, we can immediately write the photon density as 
\begin{equation}
\label{eq:WSG1-2}
\begin{split}
    \overline{G}^{(1)}(t)& = \sum_n \Gamma_n^2 |\overline{\rho}_n(t)|^2,
\end{split}
\end{equation}
where $\overline{\rho}_n(t)$ is a normalized function set by
\begin{equation}
\label{eq:rhodef}
    \overline{\rho}_n(t) = \frac{1}{\Gamma_n}\sum_m 
    (\T{sinh} \mathbf{P})_{nm}
    \overline{\chi}_m(t),
\end{equation}
and 
\begin{equation}
\label{eq:gamdef}
    \Gamma_n = \sqrt{(\T{sinh}^2\mathbf{P})_{nn}},
\end{equation}
which is real. The expression \eqref{eq:WSG1-2} for $\overline{G}^{(1)}(t)$  is the generalization of Eq. \eqref{eq:Gresults_approx} in the pseudo-Schmidt decomposition, and clearly has the same form; indeed, if we were to set $\beta_{nm}$ to be diagonal and independent of $n$ for the $n$ for which it does not vanish, we would have  $\overline{\rho}_n(t)\to \overline{\chi}_n(t)$. Even more generally, the expression \eqref{eq:WSG1-2} mirrors the form of the expansion (\ref{eq:Gresults}) of $\overline{G}^{(1)}(t)$ in terms of Schmidt modes, with $\Gamma_{n}$ here taking the role of $s_{n}$ there. But the $\overline{\rho}_{n}(t)$ cannot be identified as a supermode; while the functions in the set $\{\overline{\chi}_n(t)\}$ 
are mutually orthogonal, the functions in the set $\{\overline{\rho}_n(t)\}$ are not, because in general $\T{sinh}\mathbf{P}$ is not unitary. Nonetheless, the functions in the latter set are generally localized compared to the duration of the pulse, especially for weak squeezing. We refer to the $\overline{\rho}_n(t)$ simply as ``packets," and to the expansion \eqref{eq:WSG1-2} for $\overline{G}^{(1)}(t)$ as its ``packet expansion;" we will see packet expansions of other correlation functions below.

The expected number of photons in the pulse is given by integrating $\overline{G}^{(1)}(t)$ over all time; we find
\begin{equation}
\label{eq:NavgWS}
    N_\T{pulse} = \T{Tr}(\T{sinh}^2\mathbf{P}) = \sum _n  \Gamma_n^2,
\end{equation}
where $\T{Tr}(\cdot)$ denotes the trace. This
is reminiscent of the corresponding expressions \eqref{eq:NpulseS} and \eqref{eq:Npulse} for the Schmidt and pseudo-Schmidt expansions respectively. From Eq. \eqref{eq:NavgWS} it is clear that $\Gamma_n^2$ is the number of photons in each packet, and summing over all packets gives the total number of photons.

\begin{figure*}
    \centering
    \includegraphics[width = \linewidth]{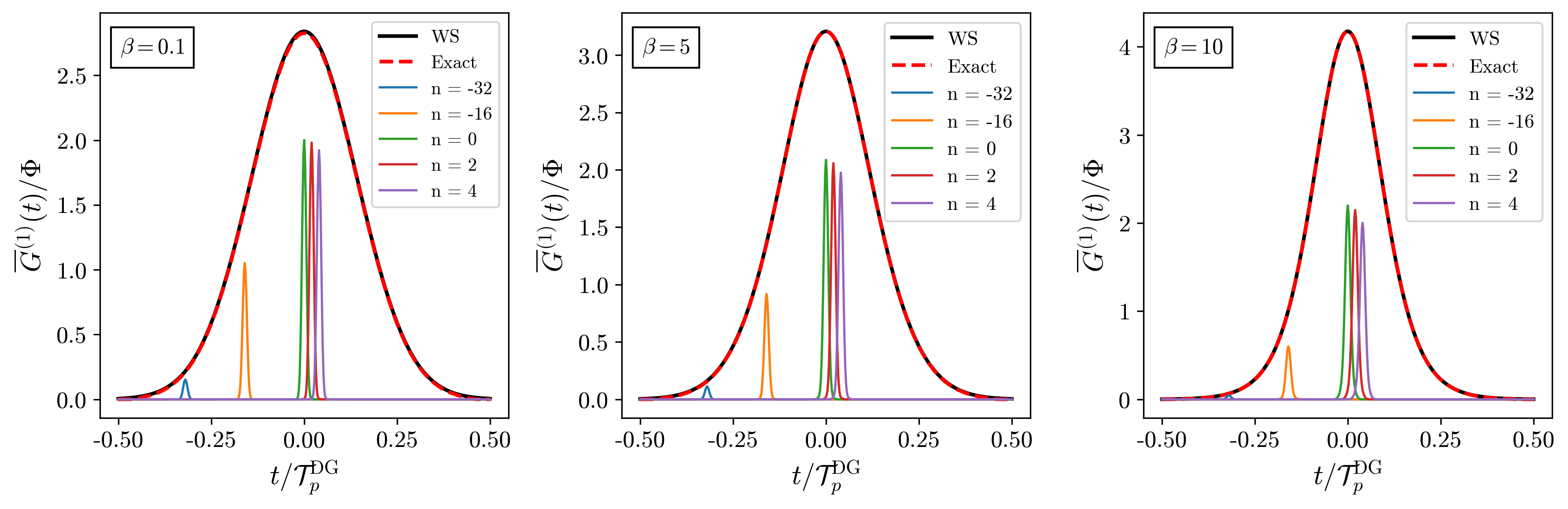}
    \caption{For the double-Gaussian, from left to right we plot $\overline{G}^{(1)}(t)/\Phi$ calculated using the Whittaker-Shannon decomposition and a few contributions from different packets in Eq. \eqref{eq:WSG1-2} compared with the exact calculation (\ref{eq:Gresults}), with the horizontal axis normalized by $\mathcal{T}_p^\mathrm{DG}$ for $\beta = 0.1, 5,$ and $10$.}
    \label{fig:13}
\end{figure*}

For the double-Gaussian and three values of $\beta$ chosen in section \ref{sec:Example 1: The double  Gaussian}, we have $|\mathring\beta| \approx 0.014, 0.7,$ and $1.4$ which is on the order of the three values $|\beta|/\sqrt{\mathcal{K}}$, in agreement with the discussion surrounding Eq.  \eqref{eq:effectiveK}. In Fig. \ref{fig:13} we plot $G^{(1)}(t)$ calculated using Eq. \eqref{eq:WSG1}, as well as the contributions given by Eq. \eqref{eq:WSG1-2} for a few values of $n$, together with the exact $\overline{G}^{(1)}(t)$ for the three chosen values of $\beta$, which correspond to $N_\T{pulse} \approx  0.01, 35, 383$; we find excellent agreement between the exact and Whittaker-Shannon decomposition. From Fig. \ref{fig:13} we see that each $\overline{\rho}_n(t)$ is clearly localized compared to the duration of the pulse, and so using the packets we can ``take apart" the squeezed light and provide a simple description of the photon density (Eq. \eqref{eq:WSG1-2}); this extends our understanding from the pseudo-Schmidt decomposition valid for the sinc-hat joint amplitude to more general joint amplitudes, such as the double-Gaussian, where a Whittaker-Shannon decomposition is necessary. 

\begin{figure*}
    \centering
    \includegraphics[width = \linewidth]{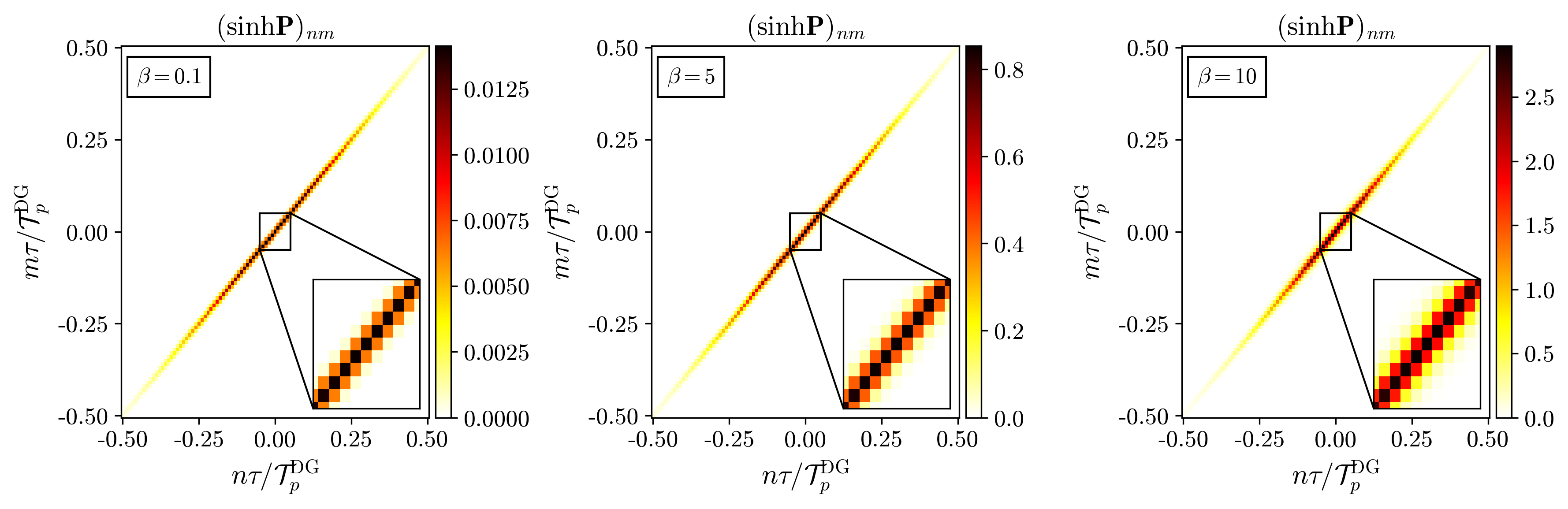}
    \caption{For the double-Gaussian, from left to right we plot $(\T{sinh}\mathbf{P})_{nm}$ for $\beta = 0.1, 5,$ and $10$ with the horizontal axis normalized by $\mathcal{T}_p^\mathrm{DG}$.}
    \label{fig:14}
\end{figure*}

Notice that as $|\beta|$ ($|\mathring\beta|$) increases so does the width of each $\overline{\rho}_n(t)$. Referring back to Eq. \eqref{eq:rhodef}, this occurs because elements of $\T{sinh}\mathbf{P}$ that are further off-diagonal become more important as $|\beta|$ increases. 
And this is a consequence of the fact that more powers of $\mathbf{P}$ become important in the expansion of $\T{sinh}\mathbf{P}$ as $|\beta|$\ increases, since $\mathbf{P}$ depends on $\m{\beta}$ (see Eq. \eqref{eq:polardecomp}). Thus 
the off-diagonality of $\T{sinh}\mathbf{P}$ is extended beyond that of $\m{\beta}$, and elements of $\T{sinh}\mathbf{P}$ further from the diagonal become larger as $|\beta|$ increases; see Fig. \ref{fig:14} for plots of $\T{sinh}\mathbf{P}$ with increasing $|\beta|$, which demonstrates this effect.


We can now construct the expansions for $\overline{G}^{(2)}_{\text{coh}}(t_1,t_2)$ and $\overline{G}^{(2)}_{\text{incoh}}(t_1,t_2)$. For the first of these, comparing the expression \eqref{eq:WSG2} for the full $\overline{G}^{(2)}(t_1,t_2)$ with our earlier general expression \eqref{eq:G2_incoh_alt} for $\overline{G}^{(2)}_{\text{incoh}}(t_1,t_2)$, we can identify
\begin{equation}
    \begin{split}
    \label{eq:WScoh}
    \overline{G}^{(2)}_{\text{coh}}(t_1,t_2)&=|\m{\overline{\chi}}^T(t_1)\mathbf{U}(\T{sinh}\mathbf{P})(\T{cosh}\mathbf{P})\m{\overline{\chi}}(t_2)|^2\\
    &=\left|\sum_{n,m}(\mathbf{U}(\T{sinh}\mathbf{P})(\T{cosh}\mathbf{P}))_{nm}\overline{\chi}_n(t_1)\overline{\chi}_m(t_2)\right|^2
    \end{split}.
\end{equation}
This can be compared with the corresponding expressions \eqref{eq:Gwork} and \eqref{eq:Gwork_approx} for the Schmidt and pseudo-Schmidt decompositions respectively. The appearance of terms with $m \neq n$ here, as opposed to the single summation that appears in the Schmidt and pseudo-Schmidt expansion, is expected given that the squeezed state written using the Whittaker-Shannon decomposition involves a double sum (\ref{eq:sqoperatordisentangled})

We show in Appendix \ref{sec:app:D} that the expression \eqref{eq:WScoh} for $\overline{G}^{(2)}_\T{coh}(t_1,t_2)$ can also be written using the set of packet functions $\{\overline{\rho}_n(t)\}$,
\begin{equation}
\label{eq:WScohcont}
    \overline{G}^{(2)}_\T{coh}(t_1,t_2) = \left|\sum_{n,m}\Gamma_n\Gamma_m(\mathbf{U}\T{coth}\mathbf{P})_{mn}\overline{\rho}_m(t_1)\overline{\rho}_n(t_2)\right|^2,
\end{equation}
but this is not a convenient expression to use in practice:  For if $\m{\beta}$ is close to diagonal some functions of $\m{\beta}$, such as $\text{tanh}\mathbf{P}$, will be as well, but not $\text{coth}\mathbf{P}$. Further, the weakly squeezed limit is not directly apparent from the form of Eq. (\ref{eq:WScohcont}), so it seems preferable to write $\overline{G}_\T{coh}^{(2)}(t_1,t_2)$ using $\{\overline{\chi}_n(t)\}$ instead of $\{\overline{\rho}_n(t)\}$.  Perhaps this is not surprising, for earlier we found that when $|\beta|\ll 1$, $\overline{G}^{(2)}(t_1,t_2)\to |\beta|^2|\overline{\gamma}(t_1,t_2)|^2$ \eqref{eq:smallbeta_time}, which directly depends on the correlations contained in the joint temporal amplitude; in same limit, using the Whittaker–Shannon decomposition, we expect it to depend on the analogous quantity $\beta_{nm}$. So unlike $G^{(1)}(t)$, where the photon density at a particular time involves contributions from all possible pairs and is written in terms of $\{\overline{\rho}_n(t)\}$, $\overline{G}_\T{coh}^{(2)}(t_1,t_2)$ -- at least in the weakly squeezed limit -- should directly depend on the temporal correlations, and so it is more suitable to write $\overline{G}_\T{coh}^{(2)}(t_1,t_2)$ in terms of $\{\overline{\chi}_n(t)\}$, as is done in Eq. \eqref{eq:WScoh}.

Turning finally to the general expression \eqref{eq:G2_incoh_alt} for $\overline{G}^{(2)}_{\text{incoh}}(t_1,t_2)$, using the definition of $\overline{\rho}_n(t)$ and $\Gamma_n$ (\ref{eq:rhodef}, \ref{eq:gamdef}) we can write a packet expansion for the incoherent contribution as
\begin{equation}
\begin{split}
\label{eq:eq:WSG2incoh-2}
    \overline{G}^{(2)}_\T{incoh}(t_1,t_2) \\
    &\hspace{-10mm}= \frac{1}{2}\sum_{n,m}\left|\Gamma_n\Gamma_m(\overline{\rho}_n(t_1)\overline{\rho}_m(t_2) + \overline{\rho}_n(t_2)\overline{\rho}_m(t_1))\right|^2,
\end{split}
\end{equation}
which has the same form as the Schmidt \eqref{eq:Gwork} and pseudo-Schmidt \eqref{eq:Gwork_approx} decompositions, but is in terms of the set of packets $\{\overline{\rho}_n(t)\}$.

\begin{figure*}
    \centering
    \includegraphics[width = \linewidth]{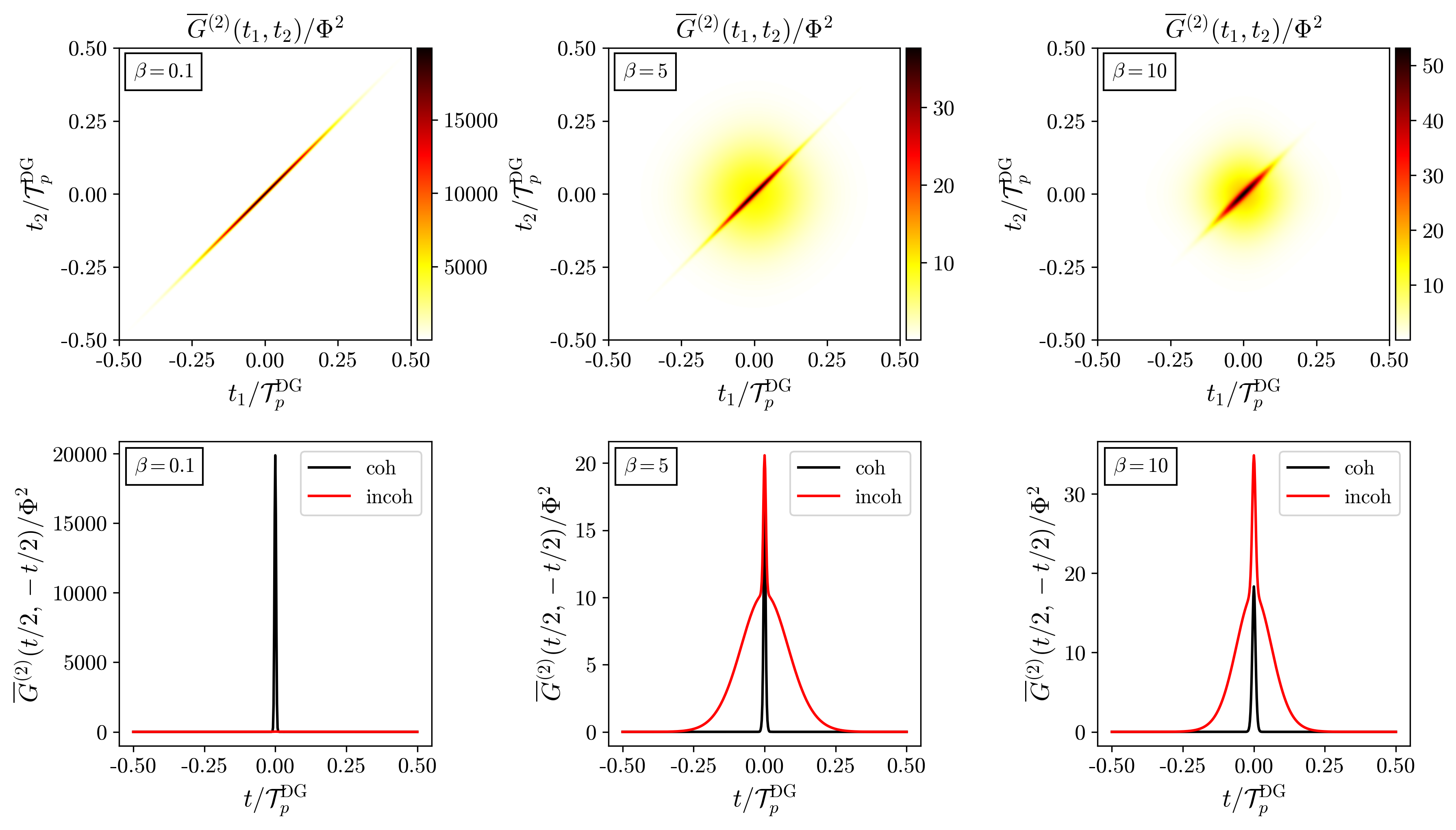}
    \caption{For the double-Gaussian and using Whittaker–Shannon decompositions, from left to right we plot $\overline{G}^{(2)}(t_1,t_2)/\Phi^2$ (top) and the coherent and incoherent contribution to $\overline{G}^{(2)}(t/2,-t/2)/\Phi^2$ (bottom) with the axes normalized by $\mathcal{T}_p^\mathrm{DG}$ for $\beta = 0.1, 5,$ and $10$.}
    \label{fig:15}
\end{figure*}

In Fig. \ref{fig:15} we plot $G^{(2)}(t_1,t_2)$ calculated using Eq. \eqref{eq:WSG2} for the double-Gaussian joint amplitude, as well as the coherent and incoherent contributions; comparing to Fig. \ref{fig:3} for the exact calculation we find excellent agreement between the two. Although the Whittaker-Shannon decomposition does not allow the simple factorization of the ket into product kets associated with each Schmidt or pseudo-Schmidt mode, correlation functions can still be evaluated. And, 
since the off-diagonal elements of the squeezing matrix $\m{\beta}$ typically drop off quickly away from the diagonal, the correlation functions can be written in a form involving either the set of functions $\{\overline{\chi}_n(t)\}$ or the set of functions $\{\overline{\rho}_n(t)\}$; all these functions are localized compared to the Schmidt modes. 

\subsection{Correlation functions in the weakly squeezed limit}
In this section we identify approximate expressions for the correlation functions valid in the weakly squeezed limit when $|\mathring\beta|\ll 1$. The correlation functions for the Whittaker–Shannon decomposition involve the matrix $\mathbf{P}$, which using the form of  Eq. \eqref{eq:betabardef} is given by $\mathbf{P} = |\mathring\beta|^2 \sqrt{\mathbf{r\dg r}}$. Taking the weakly squeezed limit we approximate 
\begin{equation}
    \begin{split}
        &\T{sinh}\mathbf{P}\to \mathbf{P}\\
         &\T{cosh}\mathbf{P}\to \mathbf{1},
    \end{split}
\end{equation}
where $\mathbf{1}$ is the identity matrix. Then $\Gamma_n\to \sqrt{(\mathbf{P}^2)_{nn}}$,
\begin{equation}
\label{eq:rhoapprox}
        \overline{\rho}_n(t)\to \frac{1}{\Gamma_n} \sum_m P_{nm}\overline{\chi}_m(t),\\
\end{equation}
and for $r_{nm}$ nonzero only for $|n-m|$ less than a small integer, the set \{$\overline{\rho}_n(t)\}$ will be localized.  The expression \eqref{eq:WSG1-2} then gives 
\begin{equation}
\begin{split}
    \overline{G}^{(1)}(t)& \to \sum_{n,m} |P_{nm}\overline{\chi}_m(t)|^2.
\end{split}
\end{equation}

Turning to $\overline{G}^{(2)}(t_1,t_2)$, for the coherent contribution $\overline{G}_\T{coh}^{(2)}(t_1,t_2)$, the general expression \eqref{eq:WScoh}, using the result
\begin{equation}
    \mathbf{U}(\T{sinh}\mathbf{P})(\T{cosh}\mathbf{P})\to \mathbf{U}\mathbf{P} = \m{\beta},
\end{equation}
to find
\begin{equation}
\label{eq:G2_WS_weakly pumped}
    \begin{split}
        \overline{G}_\T{coh}^{(2)}(t_1,t_2)&\to |\m{\overline{\chi}}^T(t_1)\m{\beta}\m{\overline{\chi}}(t_2)|^2\\
        &=\left|\sum_{n,m}\overline{\chi}_n(t_1)\beta_{nm} \overline{\chi}_m(t_2)\right|^2,
    \end{split}
\end{equation}
which clearly shows that the resulting photon statistics depends on $\beta_{nm}$. Finally, in this limit the expression \eqref{eq:eq:WSG2incoh-2} for $\overline{G}^{(2)}_\T{incoh}(t_1,t_2)$ gives
\begin{equation}
    \begin{split}
    \overline{G}^{(2)}_\T{incoh}(t_1,t_2) \to \\
    &\hspace{-10mm} \frac{1}{2}\sum_{n,m}\left|\sum_{u,v} P_{nu}P_{mv}(\overline{\chi}_{u}(t_1)\overline{\chi}_{v}(t_2)+\overline{\chi}_u(t_2)\overline{\chi}_v(t_1))\right|^2.
\end{split}
\end{equation}

\section{Local states and correlation functions}
\label{sec:Local states and calculation of correlation functions}
The general equations we derived for $\overline{G}^{(1)}(t)$ \eqref{eq:WSG1-2} and $\overline{G}^{(2)}(t_1,t_2)$ \eqref{eq:WSG2}, and their weakly squeezed approximations, are valid for any times $t, t_1, t_2$. In the discussion surrounding Fig. \ref{fig:13} for the photon density, it was noted that since each packet is localized we only need a few to properly represent the photon density at any particular time. This suggests that for some calculations, including some more general than the correlation functions considered above, we can rely on an approximate form of the ket itself.

Suppose we are interested in features of the state around a small neighbourhood centered at the time $t_\RN{1}$. Now for a correlated joint temporal amplitude $\beta_{nm}$ will typically only be nonzero for $|n-m|$ ranging up to a small integer, so for times around a small neighbourhood of $t_\RN{1}$ only a few Whittaker-Shannon modes centered around $n_\RN{1} = [t_\RN{1}/\tau]$ will be relevant; here we use $[\cdot]$ to denote the closest integer. We identify the range of $m$ around $n_\RN{1}$ for which $\beta_{n_{\RN{1}}m}$ will be non-negligible by the odd integer $d$, assuming that $\beta_{n_\RN{1} m}$ is sufficiently small for $|n_\RN{1} - m|> (d-1)/2$ that for those values of $m$ it can be neglected. 

\begin{figure}
    \centering
    \includegraphics[width = 0.9\linewidth]{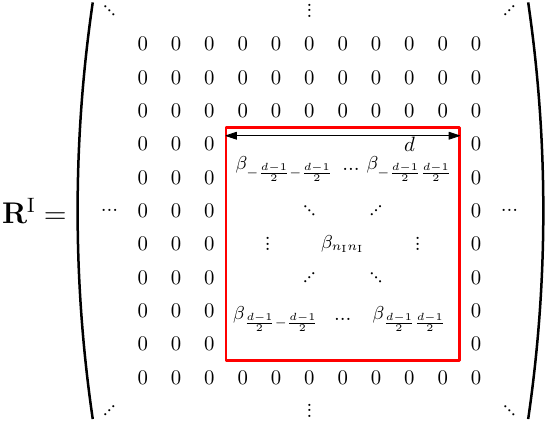}
    \caption{Schematic of the matrix $\mathbf{R}^\RN{1}$ which has nonzero entries centered at $\beta_{n_\RN{1}n_\RN{1}}$ with a width $d$ and zeros everywhere else. The matrix $\mathbf{K} = \m{\beta} - \mathbf{R}^\RN{1}$ and consists of every nonzero element that we set to zero in $\mathbf{R}^\RN{1}$.}
    \label{fig:16}
\end{figure}

We then split the matrix $\m{\beta}$ into two contributions
\begin{equation}
\label{eq:RKdef}
    \m{\beta} = \mathbf{R}^\RN{1} + \mathbf{K},
\end{equation}
where $\mathbf{R}^\RN{1}$ is a symmetric matrix with nonzero entries centered at $\mathrm{R}^\RN{1}_{n_\RN{1}n_\RN{1}}$. It contains the elements of $\m{\beta}$ as the row and column indices range over $(d-1)/2$ in all directions from the center at (${n_\RN{1}},{n_\RN{1}}$), and all its other elements are set to zero. The matrix $\mathbf{R}^\RN{1}$ is shown schematically in Fig. \ref{fig:16} with the nonzero contributions existing inside the red ``box'' containing $d^2$ elements; all the other elements of $\m{\beta}$ are contained in $\mathbf{K}$. For times of interest we assume that $d$ is chosen large enough so that significant contributions to the quantities of interest, such as correlation functions involving times near $t_\RN{1}$, only involve the elements of $\m\beta$ contained in $\mathbf{R}^\RN{1}$.

\begin{figure*}
    \centering
    \includegraphics[width = \linewidth]{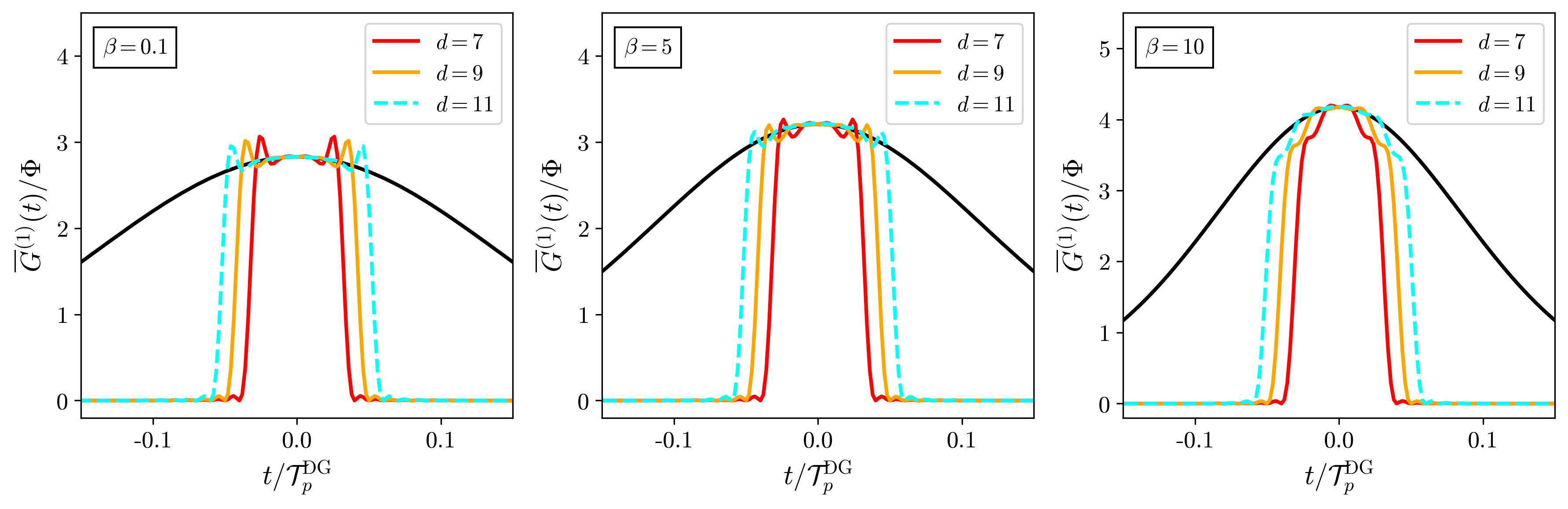}
    \caption{For the double-Gaussian, from left to right we plot $\overline{G}^{(1)}(t)/\Phi$ calculated using the Whittaker-Shannon decomposition with the full $\beta_{nm}$ compared to the approximate calculation using $R^\RN{1}$ near $t_\RN{1} = 0$ for $d=7,9$ and $11$, with the horizontal axis normalized by $\mathcal{T}_p^\mathrm{DG}$ for $\beta = 0.1, 5,$ and $10$.}
    \label{fig:17}
\end{figure*}
\begin{figure}
    \centering
    \includegraphics[width = 0.9\linewidth]{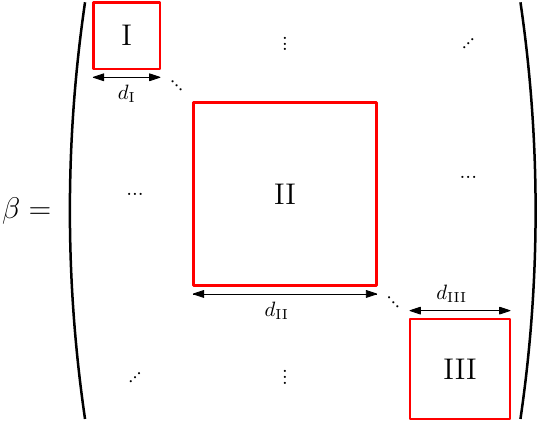}
    \caption{Schematic of the matrix $\m{\beta}$ partitioned into a set of non-overlapping matrices $\mathbf{R}^J$, each with nonzero values centered at $\beta_{n_Jn_J}$ of size $d_J$ denoted by the red squares. The matrix $\mathbf{K} = \m{\beta} - \mathbf{R}^\RN{1} -  \mathbf{R}^\RN{2} - ...$, and consists of every other non zero element contained in $\m{\beta}$. }
    \label{fig:18}
\end{figure}

The squeezing operator \eqref{eq:sqoperatordisentangled} is then approximated by
\begin{equation}
    \begin{split}
        \tilde{S} & = e^{\frac{1}{2}\sum\limits_{n,m}(R^\RN{1} _{nm} + K_{nm})B_{n}^{\dagger}B_{m}^{\dagger}-\text{h.c.}}\\
        &\approx  e^{\frac{1}{2}\sum\limits_{n,m}R^\RN{1} _{nm}B_{n}^{\dagger}B_{m}^{\dagger}-\text{h.c.}},
    \end{split}
\end{equation}
where in the second line we dropped all the terms from $\mathbf{K}$.  The dimension of $\mathbf{R}^\RN{1}$
can be quite large if the pulse is long, but since most of its entries are zero and the only nonzero entries have a width given by $d$, we use a ``prime''
symbol below to indicate that we can restrict the sum to be only over the nonzero elements of $\mathbf{R}^\RN{1}$.
In practice, this can drastically increase the efficiency of numerical computations involving only a limited region of time. The state is then taken to be $\ket{\Psi_\RN{1}} = \tilde{S}_\RN{1}\vac_\RN{1}$, where $\vac_\RN{1}$ is the vacuum state corresponding to the modes associated with the nonzero elements in $\mathbf{R}^\RN{1}$, and we set
\begin{equation}
    \tilde{S}_\RN{1} = e^{\frac{1}{2}\sum^\prime\limits_{n,m}R^\RN{1} _{nm}B_{n}^{\dagger}B_{m}^{\dagger}-\text{h.c.}}.
\end{equation}
From the approximate state $\ket{\Psi_\RN{1}}$ we can apply the same steps as above to calculate the correlation function, but instead of using the full matrix $\m\beta$ for the transformation and polar decomposition we use the reduced matrix $\mathbf{R}^\RN{1}$. Then for 
times of interest we have (see Eq. \eqref{eq:inverseB_n})
\begin{equation}
\label{eq:approxinverse}
    \begin{split}
        \overline{a}(\overline t_\RN{1})& = \sum_n \overline{\chi}_n(\overline t_\RN{1})B_n\\
        &\approx \sum_n^\prime \overline{\chi}_n(\overline t_\RN{1})B_n,
    \end{split}
\end{equation} 
where again we restrict the sum to be over the modes associated with the nonzero elements of $\mathbf{R}^\RN{1}$, as in the approximate state $\ket{\Psi_\RN{1}}$. The equations for the correlation functions (\ref{eq:WSG1}, \ref{eq:WSG2}) can be applied with  $\m\beta$ replaced by $\mathbf{R}^\RN{1}$ for times near $ t_\RN{1}$, with the appropriate restriction of the sums. 

In Fig. \ref{fig:17} we plot a zoomed-in version of $\overline G^{(1)}(t)$ calculated from the full state $\ket{\Psi}$ and the approximate state $\ket{\Psi_\RN{1}}$ around the time $t_\RN{1} = 0$ for the double-Gaussian, the three values of $\beta$ ($\mathring\beta$), and $d = 7,9$ and $11$. We see that the choice of $d$ generally determines over how wide a neighbourhood around $t_\RN{1}$ the contributions from the full state are well approximated by the contributions from $\ket{\psi}_\RN{1}$. For $\beta = 0.1$ ($\mathring\beta = 0.014$), which is well within the weakly squeezed limit, only three Whittaker-Shannon modes on each side of the center mode are relevant to accurately determine the photon density at $t=0$; this is much less then the total number of modes along $r_{nn}$ which is set by $\mathcal{K}$, and for this example is $\mathcal{K}_\mathrm{DG} = 100$. Thus, the state $\ket{\Psi_\RN{1}}$ provides a ``local'' description of the photon density around $t=t_{\RN{1}}$. With the state $\ket{\Psi_\RN{1}}$ one can also calculate $G^{(2)}(t_1,t_2)$ as long as both times are near the time $t_\RN{1}$. We find similar agreement and trends with $\beta$ and $d$ as for the photon density.

As $|\beta|$ ($|\mathring\beta|$) increases more neighbouring Whittaker-Shannon modes are required to accurately reproduce the photon statistics at a given time and we need to increase the size of the nonzero box of elements in $\mathbf{R}^\RN{1}$. For as $|\beta|$ increases there is a larger amplitude for photons to be described by different modes spread further apart from each other; see the discussion in the paragraph after Eq. \eqref{eq:NavgWS}. In the same way that $(\T{sinh}\mathbf{P})_{nm}$ spreads in the $|n-m|$ direction as $|\beta|$ increases (see Fig. \ref{fig:14}) we need to choose a larger box to capture all the possible contributions near a given $t_\RN{1}$. We return to this point below.

Suppose now we are interested in the properties of the state associated with two or more times $t_\RN{1}, t_\RN{2}, t_\RN{3}, ...$, ``sufficiently far apart'' from one another. Then we can split $\m{\beta}$ into a \emph{set} of contributions given by
\begin{equation}
    \m{\beta} = \mathbf{R}^\RN{1} + \mathbf{R}^\RN{2} + \mathbf{R}^\RN{3} + ... + \mathbf{K},
\end{equation}
with $\mathbf{R}^\RN{1}$ associated with $t_{\RN{1}}$, $\mathbf{R}^\RN{2}$ associated with $t_{\RN{2}}$, etc., and 
where by ``sufficiently far apart'' we mean that the corresponding boxes of sizes $d_\RN{1}, d_\RN{2}$, etc.
associated with the regions of $\mathbf{R}^\RN{1},\mathbf{R}^\RN{2}$, etc. that contain nonzero elements do not overlap; this is shown schematically in Fig. \ref{fig:18}, where we indicate the regions of $\mathbf{R}^\RN{1},\mathbf{R}^\RN{2}$, etc. that contain nonzero elements by I,II, etc. Again, the matrix $\mathbf{K}$ contains the remaining contributions to $\m\beta$ not in any of the nonzero regions of the $\mathbf{R}^J$ matrices, $J = \RN{1}, \RN{2}$, etc. Then since each $\mathbf{R}^J$ has nonzero elements only in the region where the others do not, each matrix in $\{\mathbf{R}^J\}$ commutes with the rest and each contribution to the squeezing operator can be split apart. So for the times of interest the state is given by 
\begin{equation}
\label{eq:fullketproductstate}
    \ket{\Psi} \approx \bigotimes_J \ket{\Psi_J} = \bigotimes_J \tilde{S}_J \ket{\T{vac}}_J,
\end{equation}
where $\ket{\Psi_J}$, $\tilde{S}_J$, and $\ket{\T{vac}}_J$ are the obvious generalization of $\ket{\Psi_\RN{1}}$, $\tilde{S}_\RN{1}$ and $\ket{\T{vac}}_\RN{1}$. Using equation \eqref{eq:NavgWS} for the average photon number we similarly calculate that each time region $t_J$ has 
\begin{equation}
\label{eq:N_J}
    N_J = \T{Tr}(\T{sinh}^2\mathbf{P}^J)
\end{equation}
photons, where $\mathbf{P}^J$ is the matrix calculated from doing a polar decomposition of the corresponding $\mathbf{R}^J$.

To calculate the correlation functions in the neighbourhood of a time $t_J$ we again use Eqs. \ref{eq:WSG1} and \ref{eq:WSG2} with the replacement of $\beta_{nm}$ with the appropriate $\mathbf{R}^J$ as discussed above. If instead we want to calculate $\overline{G}^{(2)}(t_J, t_{J^\prime})$ for $J\neq J^\prime$ then the corresponding operators $a(t_J)$ and $a^\dagger(t_{J^\prime})$ in Eq. \eqref{eq:gs-1} commute and the resulting second-order correlation function is
\begin{equation}
    \overline{G}^{(2)}(t_J, t_{J^\prime}) =   \overline{G}^{(1)}(t_J)\overline{G}^{(1)}(t_{J^\prime}),
\end{equation}
where the 
photon densities evaluated at the times $t_J, t_{J^\prime}$ are evaluated using the respective contributions from $\ket{\Psi_J}$ and $\ket{\Psi_{J^\prime}}$.

\subsection{Disentangling the squeezing operator}
While the approximation of the state $\ket{\Psi}$ into the set of states \{$\ket{\Psi_J}$\}, takes apart the squeezed state and provides a local calculation of the correlation functions, it does not really give us intuition of the state itself. To gain insight into that, we make use of the general ``disentangling formula'' \cite{ma1990multimode} applied to each squeezing operator $\tilde{S}_J$,

\begin{equation}
    \tilde{S}_J = |\mathbf{W}^J|^\frac{1}{2}e^{\frac{1}{2}\hspace{-1mm}\sum\limits_{n,m}^\prime T_{nm}^J B_n\dg B_m\dg}e^{\sum\limits_{n,m}^\prime L_{nm}^JB_n\dg B_m}e^{-\frac{1}{2}\sum\limits_{n,m}^\prime T^{J,*}_{nm}B_nB_m},
\end{equation}
where
\begin{subequations}
    \begin{gather}
    |\mathbf{W}^J|^\frac{1}{2} =\sqrt{\T{det}(\T{sech}(\mathbf{Q}^J))}\\
        \mathbf{T}^J = (\mathbf{T}^J)^T = \T{tanh}(\mathbf{Q}^J)\mathbf{U}^J = \mathbf{U}^J\T{tanh}(\mathbf{P}^J)\\
        \mathbf{L}^J = \text{ln}(\T{sech}(\mathbf{Q}^J)),
    \end{gather}
\end{subequations}
and with $\mathbf{Q}^J$, $\mathbf{U}^J$ and $\mathbf{P}^J$ the same as before \eqref{eq:polardecomp}, but calculated from the reduced matrix $\mathbf{R}^J$. Then acting the disentangled squeezing operator on the vacuum state we have for each {$\ket{\Psi_J}$}
\begin{equation}
\label{eq:disentangledform}
    \begin{split}
    \ket{\Psi_J} &= |\mathbf{W}^J|^\frac{1}{2} e^{\frac{1}{2}\sum\limits_{n,m}^\prime T^J_{nm} B_n\dg B_m\dg}\vac_J\\
        &\equiv \overline{S}_J\vac_J,
    \end{split}
\end{equation}
where $\overline{S}_J$ is the disentangled squeezing operator after acting on the vacuum state. We point out that one could also apply the disentangling formula to the whole state valid at all times, but this is not very illuminating because one can already calculate the full correlation functions; further, had we first applied the disentangling formula and then reduced the state by getting rid of the terms that are negligible, the resulting state would not be normalized, whereas $\ket{\Psi_J}$, as given in Eq. \eqref{eq:disentangledform} always is.

\begin{figure}
    \centering
    \includegraphics[width = 0.8\linewidth]{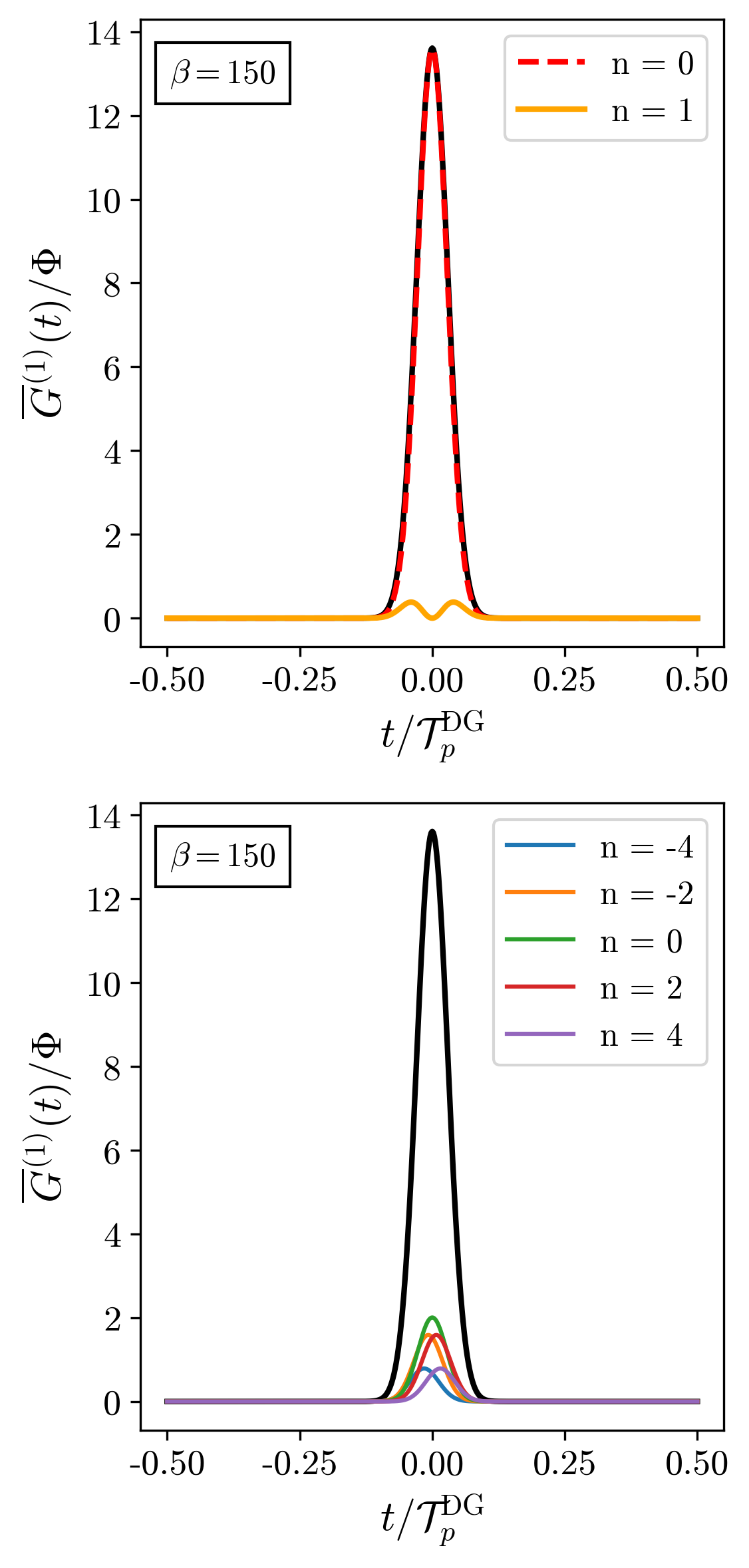}
    \caption{For the double-Gaussian joint amplitude, we plot $\overline{G}^{(1)}(t)/\Phi$ calculated using the Schmidt decomposition (top) and Whittaker-Shannon decomposition (bottom) as well as a few contributions from each calculation, for $\beta = 150$ with the horizontal axis normalized by $\mathcal{T}_p^\mathrm{DG}$.}
    \label{fig:19}
\end{figure}
\begin{figure}
    \centering
    \includegraphics[width = \linewidth]{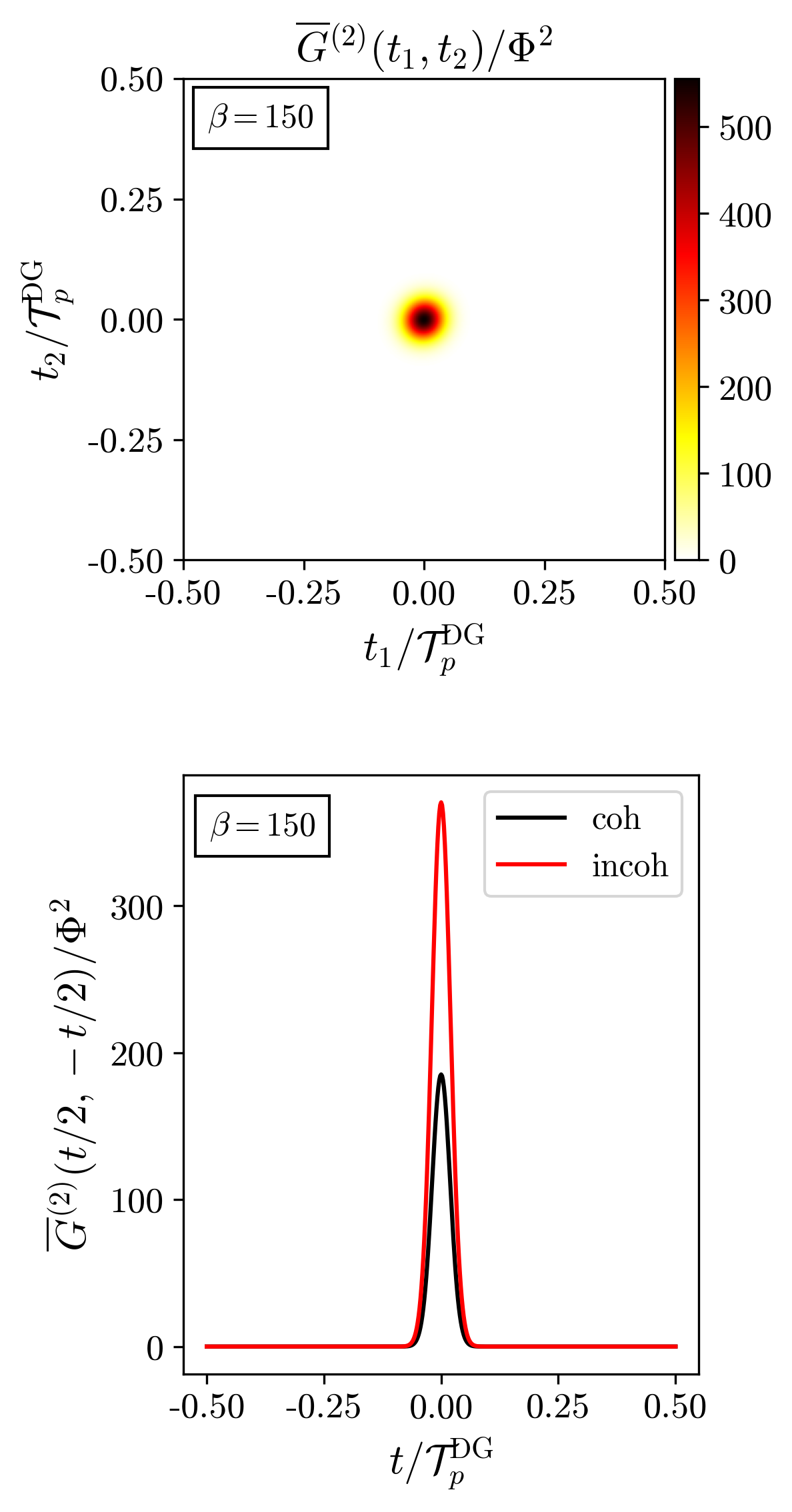}
    \caption{For the double-Gaussian joint amplitude, we plot $\overline{G}^{(2)}(t_1,t_2)/\Phi^2$ (top), and the coherent and incoherent contribution to $\overline{G}^{(2)}(t/2,-t/2)/\Phi^2$ (bottom) calculated using the Schmidt decomposition, for $\beta = 150$ with the horizontal and vertical axis normalized by $\mathcal{T}_p^\mathrm{DG}$.}
    \label{fig:20}
\end{figure}

Unlike the squeezing operator in its ``entangled form'' \eqref{eq:sqoperatordisentangled}, all the operators in the exponent of Eq. \eqref{eq:disentangledform} are creation operators which always commute so we can equivalently write the state for each $J$ as
\begin{equation}
    \ket{\Psi_J} = |\mathbf{W}^J|^\frac{1}{2} \bigotimes_{n,m} e^{\frac{1}{2}T^J_{nm} B_n\dg B_m\dg}\vac_J.
\end{equation}
Unfortunately this form of the state is not as intuitive as the \emph{single} product state in the pseudo-Schmidt decomposition because for a given $n$ we must include contributions from every other $m$ in the range of $T^J_{nm}$. 

\subsection{The ket in the weakly squeezed limit}
Instead, in situations where $|\beta|$ can be quite large but $|\mathring{\beta}|$ is sufficiently small, as in a long pulse, we can take advantage of the fact that it is $|\mathring{\beta}|$ that sets the magnitude of the matrix $T^J_{nm}$. Given that the sum in Eq. \eqref{eq:disentangledform} is only over the modes of interest, and not the whole joint amplitude, in the weakly squeezed limit when $|\mathring\beta|\ll1$ we can Taylor expand the exponential so that
\begin{equation}
\label{eq:expandedstate}
    \begin{split}
        \ket{\Psi_J} & \approx |\mathbf{W}^J|^\frac{1}{2}\left(\vac_J + \sqrt{\frac{N_J}{2}}\ket{\RN{2}}_J + ...\right),
    \end{split}
\end{equation}
where we have used
$(\mathbf{T}^J)^{\dagger} \mathbf{T}^J\to (\mathbf{R}^{J} )^\dagger \mathbf{R}$, Eq. \eqref{eq:N_J} reduces to 
\begin{equation}
    N_J\to \T{Tr}((\mathbf{R}^{J} )^\dagger \mathbf{R}^J),
\end{equation}
and we have introduced the normalized two-photon state for each $J$ by
\begin{equation}
    \ket{\RN{2}}_J = \frac{1}{\sqrt{2}} \sum_{nm}\frac{T_{nm}^J}{\sqrt{N_J}}B_n^\dagger B_m^\dagger \vac_J.
\end{equation}
The state $\ket{\Psi_J}$ has $N_J\ll 1$ photons and the prefactor $|\mathbf{W}^J|$ in the weakly squeezed limit is
\begin{equation}
    |\mathbf{W}^J| \approx 1 - \frac{N_J}{2},
\end{equation}
so the state remains normalized to order $N_J^2$, as expected. The two-photon state $\ket{\RN{2}}_J$ is a superposition of all the ways in which pairs of photons can be associated with the same Whittaker-Shannon supermode, or different supermodes, within a neighbourhood of the time $t_J$; one can easily extend the state in Eq. \eqref{eq:expandedstate} to higher order in which two pairs, three pairs, etc. are considered. For a set of times, $\{t_J\}$, in the weakly squeezed limit the full state in equation \eqref{eq:sq_state_WS} can be expanded as 
\begin{equation}
\label{eq:fullket_approx}
    \ket{\Psi}\approx \bigotimes_J |\mathbf{W}^J|^\frac{1}{2}\left(\vac_J + \sqrt{\frac{N_J}{2}}\ket{\RN{2}}_J + ...\right),
\end{equation}
providing a localized description of squeezing light for correlated but otherwise arbitrary joint amplitudes. In the long pulse limit, despite the fact that $|\beta|\to \infty$, $|\mathring\beta|$ remains finite and we can describe the light in the weakly squeezed limit as being composed of approximately two-photons within a neighbourhood around each time $t_J$.

\section{The strongly squeezed limit}
\label{sec:The strongly squeezed limit}
In this section we consider the strongly squeezed limit, where $|\beta|/\sqrt{\mathcal{K}}\gg1$, or equivalently $|\mathring\beta|\gg 1$. The results we derived in section \ref{sec:The Whittaker-Shannon decomposition} and \ref{sec:Employing the Whittaker-Shannon decomposition} are valid for any approximately bandlimited joint amplitude and for any squeezing parameter $\beta$. However, following the discussion around Fig. \ref{fig:2}, for the double-Gaussian with $\mathcal{K}_\mathrm{DG} = 100$ we found that as $|\beta|$ increased fewer Schmidt modes were required to calculate the correlation functions, although many were required to calculate the joint amplitude. 

To explore this further, consider the double-Gaussian joint amplitude (Fig. \ref{fig:1}, with $\mathcal{K}_\mathrm{DG} = 100$) but for $\beta = 150$, corresponding to $|\beta|/\sqrt{\mathcal{K}_\mathrm{DG}} \approx 15$ or $|\mathring\beta| \approx 21$, well in the strongly squeezed regime. In Fig. \ref{fig:19} we plot $\overline{G}^{(1)}(t)$ calculated using the Schmidt and Whittaker-Shannon decompositions. Clearly the first Schmidt mode is sufficient to produce an accurate $\overline{G}^{(1)}(t)$, despite the fact that all Schmidt modes are required to correctly calculate the joint amplitude. This is because the correlation functions depend on $s_n$ and $c_n$ (recall Eqs. \eqref{eq:Gresults},\eqref{eq:Gwork}), but when $|\beta|$ is large these scale exponentially; since the Schmidt modes drop off as $n$ increases, $s_0\gg s_1$ and the sums in Eq. \eqref{eq:Gresults} for the correlation functions are well approximated by the $n=0$ term. For the Whittaker-Shannon decomposition we still find very good agreement with the Schmidt calculation. However, as in Fig. \ref{fig:13}, we find each packet is significantly broadened. In the strongly squeezed limit, the amplitude that many photon pairs will contribute is large, and so the contribution of photons corresponding to two Whittaker-Shannon modes for which $|n-m|\gg 1$ is significant; since the ${\overline{\rho}_n}(t)$ include contributions from all other $m$ for a given $n$, they are necessarily broader. More mathematically, in the strongly squeezed limit many matrix multiplications are involved in calculating $\T{sinh}\mathbf{P}$, and so (following the discussion in section \ref{sec:Packet expansion} and surrounding Fig. \ref{fig:14}, \ref{fig:17},) the width of each packet is significantly broadened.

Since packets can then significantly overlap with a number of their neighbours, the local description of the photon statistics and resulting state breaks down. This is not surprising, given that the second-order correlation function, calculated using the Schmidt decomposition and plotted in Fig. \ref{fig:20},  is completely uncorrelated, a local description to identify the photon correlations is not necessary. Note that here we do not include the plot of $\overline{G}^{(2)}(t_1,t_2)$ calculated using the Whittaker-Shannon decomposition, because it is essentially identical to Fig. \ref{fig:20}.

Finally, we point out that in Fig. \ref{fig:20} the incoherent contribution is approximately twice the coherent contribution. If we restrict the sum in Eq. \eqref{eq:Gresults} and \eqref{eq:Gwork} to the first term we have
\begin{subequations}
    \begin{gather}
        \overline{G}^{(1)}(t)\to s_0^2 |\overline f_0(t)|^2,\\
        N_\T{pulse}\to s_0^2,\\
         \overline{G}_\T{coh}^{(2)}(t_{1},t_{2})\to  N_\T{pulse}^2|\overline f_0(t_1)|^2|\overline f_0(t_2)|^2,\\
         \overline{G}_\T{incoh}^{(2)}(t_{1},t_{2})\to 2N_\T{pulse}^2|\overline f_0(t_1)|^2|\overline f_0(t_2)|^2,
    \end{gather}
\end{subequations}
where we have used the fact that for large $|\beta|$, $c_n\approx s_n$; it is clear from these expressions that $\overline{G}_\T{incoh}^{(2)}(t_{1},t_{2})\approx 2\overline{G}_\T{coh}^{(2)}(t_{1},t_{2})$. Then putting them together we have
\begin{equation}
    \overline{G}^{(2)}(t_{1},t_{2}) = 3N_\T{pulse}^2|\overline f_0(t_1)|^2|\overline f_0(t_2)|^2.
\end{equation}
Since the calculation in Appendix \ref{sec:app:Mode-by-mode calculation} was done for a pseudo-Schmidt mode, it can be applied to this instance of a single Schmidt mode,
so we identify $3N_\T{pulse}^2$ as the expectation value of the number of ways of ``picking" two photons in the large $|\beta|$ limit from the first Schmidt mode.

\begin{figure*}
    \centering
    \includegraphics[width = 0.9\linewidth]{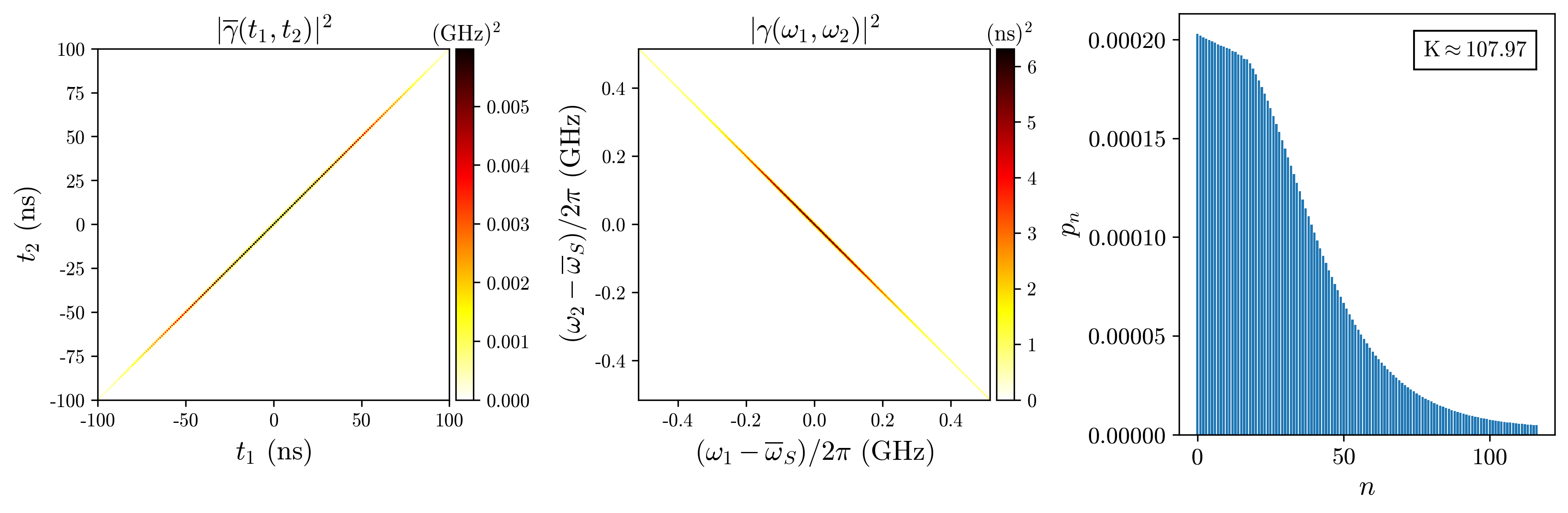}
    \caption{From left to right we plot the: joint temporal intensity; joint spectral intensity; and Schmidt amplitudes generated from a dual-pump spontaneous four-wave mixing process. The two pump functions are centered at the wavelengths $\overline\lambda_\T{P1} = 1.556$ $\mu$m and $\overline\lambda_\T{P2} = 1.547$ $\mu$m and each have temporal FWHM of $100$ \T{ns} and an energy of $10^3$ \T{pJ}. The generated photons are centered at $\overline\lambda_\T{S} = 1.552$ $\mu$m ($\overline\omega_\T{S}/2\pi = 193.164$ \T{THz}) and have a bandwidth on the order of a \T{GHz}. The ring resonator has quality factors $Q_\T{P1} = 1529378$, $Q_\T{P2} = 3844257$, and $Q_\T{S} = 2704405$ for the three modes and a nonlinear coupling $\Lambda = 5$ \T{THz} \cite{quesada2022beyond}. }
    \label{fig:X1}
\end{figure*}

\section{A final example}
\label{sec:A final example}
\begin{figure}
    \centering
    \includegraphics[width = 0.8\linewidth]{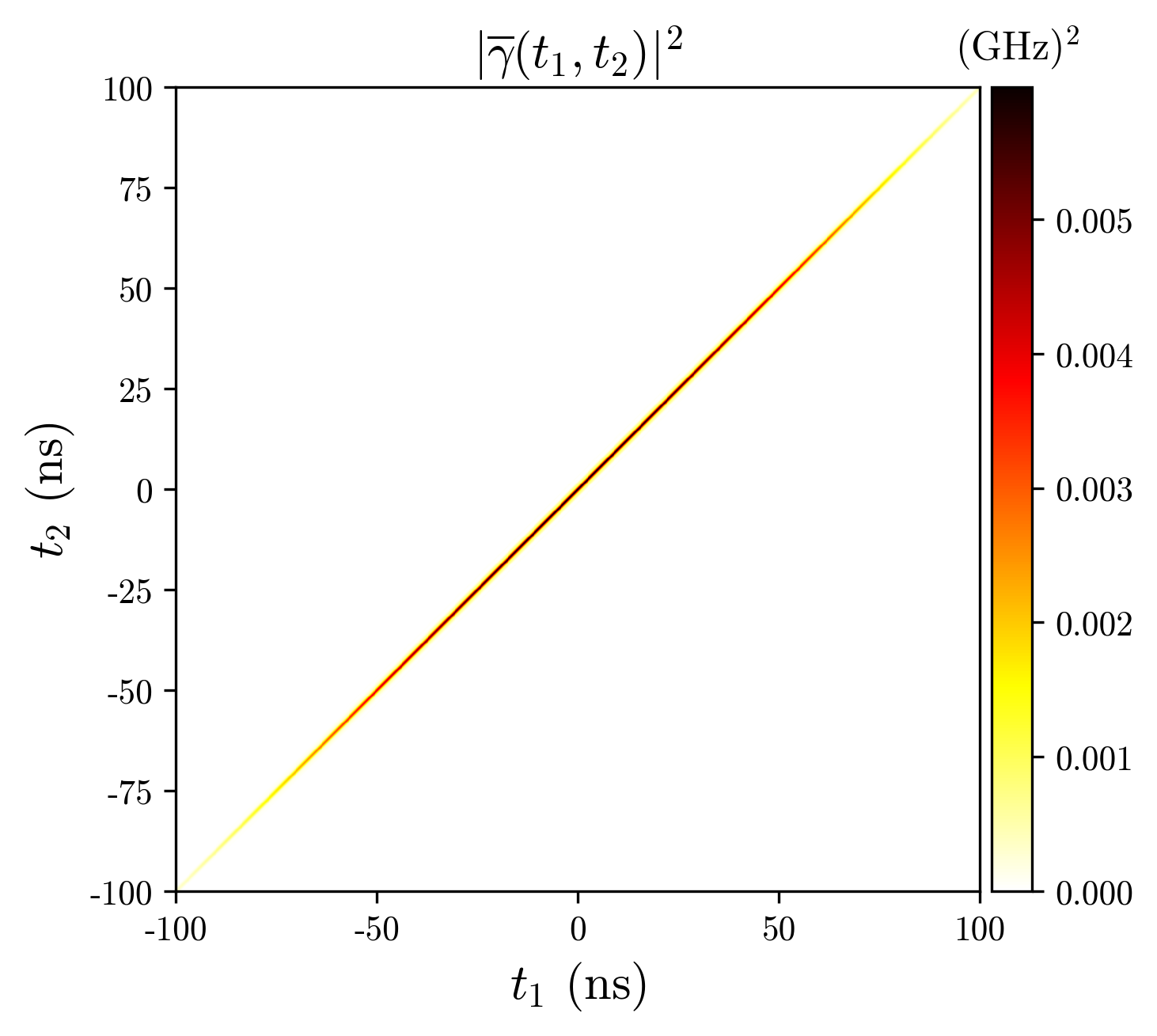}
    \caption{Joint temporal intensity calculated using the Whittaker-Shannon decomposition.}
    \label{fig:X2}
\end{figure}

In this section we consider a realistic joint amplitude generated from a dual-pump spontaneous four-wave mixing process, in a ring resonator system, when time ordering effects and self- and cross-phase modulation are included \cite{quesada2022beyond}. In Fig. \ref{fig:X1} we plot the joint intensity and the Schmidt amplitudes. The joint intensity has widths $\mathcal{T}_p^\mathrm{R} = 200 \text{ ns}$ and $\mathcal{B}_c^\text{R} \approx 1.03 \text{ GHz}$ ($\mathcal{T}_c^\text{R}\approx 0.97 \text{ ns}$) corresponding to an effective Schmidt number $\mathcal{K}_\text{R} \approx 206$, and the joint amplitude has a Schmidt number $K_\text{R} \approx 108$, where we use ``R'' to identify this ``ring'' calculation. The squeezing parameter for the generation is $\beta = 3.72$, corresponding to $N_\T{pulse}\approx 11$ photons; other system parameters, such as the center wavelengths and pump duration, are given in the caption of Fig. \ref{fig:X1}. Comparing Fig. \ref{fig:X1} with Fig. \ref{fig:X2} for the joint temporal intensity calculated using the Whittaker-Shannon decomposition, with $\Omega/2\pi = \mathcal{B}_c^\text{R}$ ($\tau = \mathcal{T}_c^\text{R}$), we see excellent agreement.

\begin{figure}
    \centering
    \includegraphics[width = 0.8\linewidth]{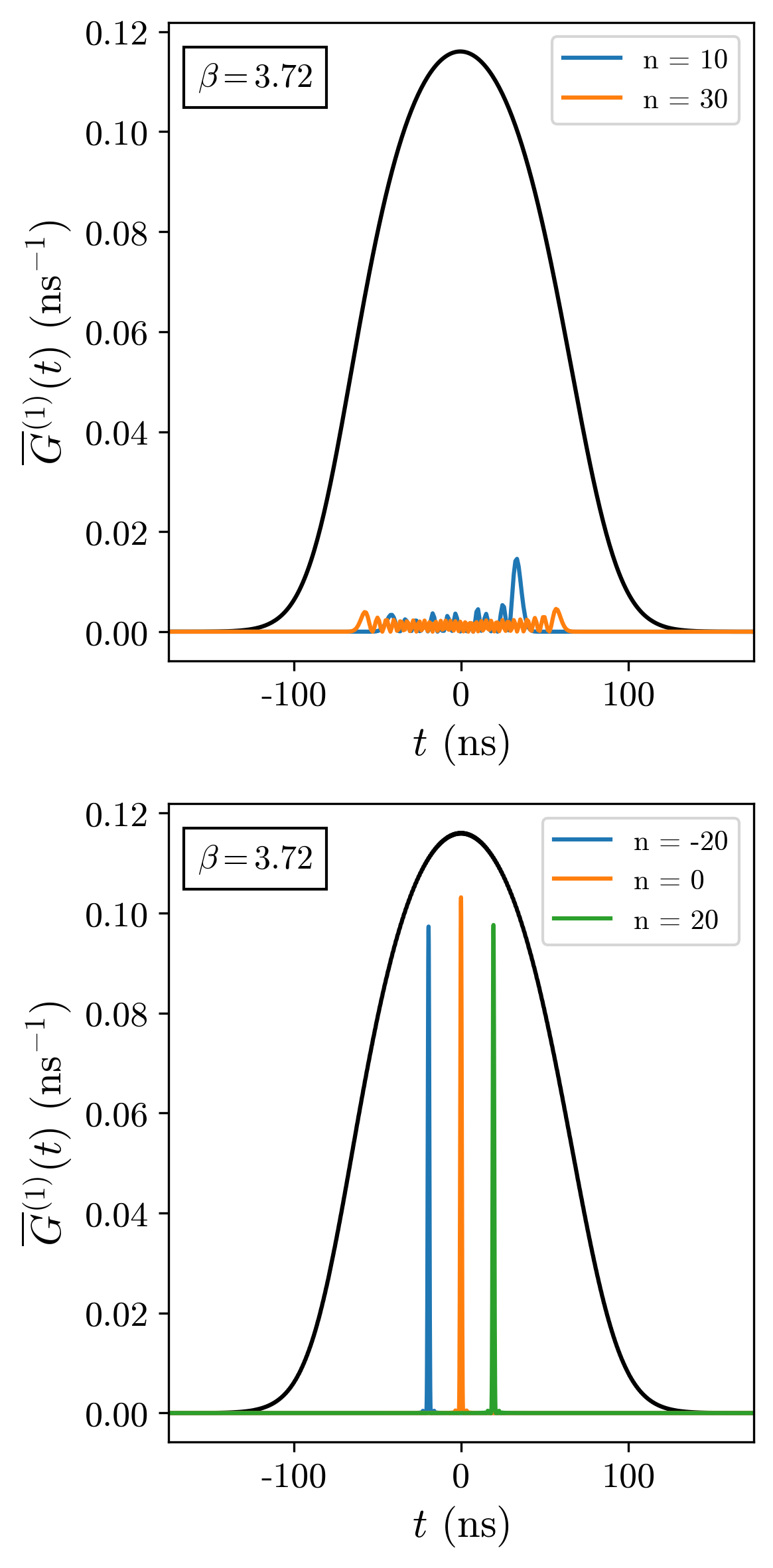}
    \caption{Plot of $\overline{G}^{(1)}(t)$ calculated using the Schmidt decomposition (top) and Whittaker-Shannon decomposition (bottom). }
    \label{fig:X3}
\end{figure}

In Fig. \ref{fig:X3} we plot $\overline{G}^{(1)}(t)$ for the Schmidt (top) and Whittaker-Shannon (bottom) decompositions, as well as a few contributions to each. We see a dramatic difference: At any particular time a huge number of Schmidt modes are required to capture the overall photon density, while in our ``packet'' decomposition only a small range of packets are required to describe the behaviour at any time. 

\begin{figure}
    \centering
    \includegraphics[width = 0.9\linewidth]{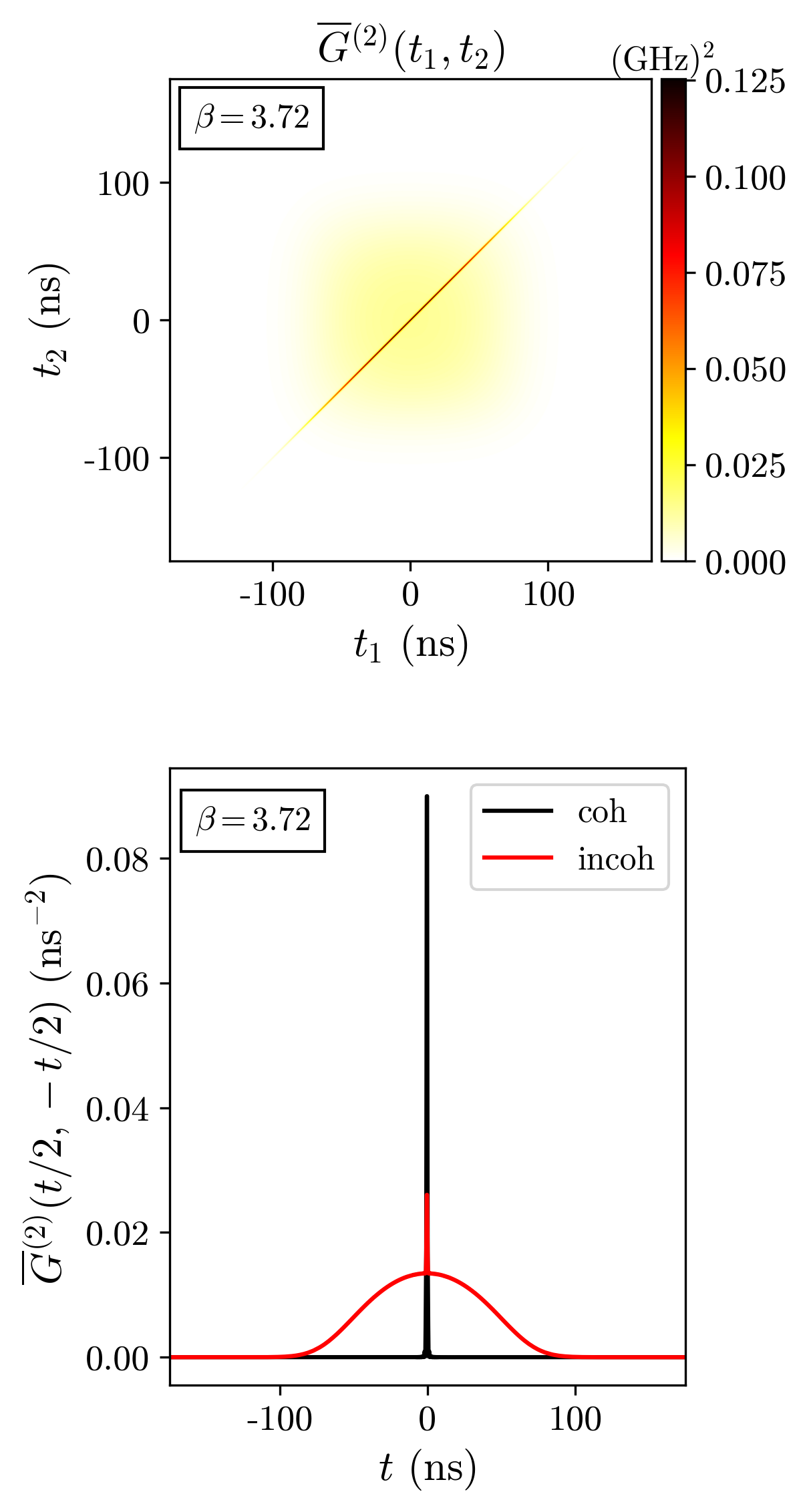}
    \caption{Plot of $G^{(2)}(t_1,t_2)$ (top) and the coherent and incoherent contribution to $G^{(2)}(t/2,-t/2)$ calculated using the Whittaker-Shannon decomposition. }
    \label{fig:X4}
\end{figure}

In Fig. \eqref{fig:X4} we plot $\overline{G}^{(2)}(t_1,t_2)$ (top) and the coherent and incoherent contributions to $\overline{G}^{(2)}(t/2,-t/2)$ (bottom) calculated using the Whittaker-Shannon decomposition; here we only plot the Whittaker-Shannon results because the Schmidt results look the same. The contributions to $\overline{G}^{(2)}(t_1,t_2)$ -- strong peak near $t_1 = t_2$ and smaller broad background -- match that of the double-Gaussian example for the three values of $\beta$. This is unsurprising because, although $\beta = 3.72$ in this example, $\mathring\beta = 0.28$, and so the state is weakly squeezed.

Since the state is weakly squeezed, the formalism provided in section \ref{sec:Local states and calculation of correlation functions} can be directly applied, providing a localized description of the pulse of light into a set of states with different time labels, each containing approximately two photons.


\section{Conclusion}
\label{eq:conclusion}
In this article we have developed a formalism to describe squeezed light with a large spectral-temporal correlation. As opposed to the usual strategy of employing the Schmidt decomposition, we feel it makes the physics more apparent. 
We began by characterizing general joint amplitudes by their timewidth $\mathcal{T}_p$ and bandwidth $\mathcal{B}_c$ (or equivalently the coherence time $\mathcal{T}_c = 1/\mathcal{B}_c$). Using the double-Gaussian joint amplitude as an example, we calculated the correlation functions using the Schmidt decomposition and found that for weak squeezing the form of $\overline{G}^{(2)}(t_1,t_2)$ matches that of the joint temporal amplitude and reaches its maximum value when $|t_1 -  t_2|<\mathcal{T}_c$; however, the Schmidt modes themselves extend on a much broader time scale given by $\mathcal{T}_p$. When calculating the correlation functions using the Schmidt decomposition, we found that a large amount of interference is present; we cannot associate a single Schmidt mode, or even a few, with a particular time. 

Next we considered another example, the sinc-hat joint amplitude, which demonstrated that this behaviour of the double-Gaussian is not unique. And although it is somewhat artificial, the sinc-hat joint amplitude is interesting in that its Schmidt amplitudes are nearly degenerate, allowing us to construct an approximate pseudo-Schmidt decomposition where the pseudo-Schmidt modes are displaced, localized ``sinc'' functions. Using this decomposition we could immediately identify contributions contained in $\overline{G}^{(1)}(t)$ or $\overline{G}^{(2)}(t_1,t_2)$ at a particular time as arising from a single pseudo-Schmidt mode, allowing us to ``take apart'' the photon statistics,  elucidating the physics. We also demonstrated that the weakly squeezed limit corresponds to $|\beta|/\sqrt{\mathcal{N}}\ll 1$, where $\mathcal{N}$ is the effective Schmidt number that identifies the number of pseudo-Schmidt modes required for the decomposition. This is useful because in the long pulse limit, where $|\beta|\gg1$ and $N_\T{pulse}\gg 1$, the quantity $|\beta|/\sqrt{\mathcal{N}}$ remains finite; despite there being a large number of photons in the pulse, the number of photons in each pseudo-Schmidt mode can be relatively small. 

To consider more general joint amplitudes, where there is a range of Schmidt amplitudes, we generalized the pseudo-Schmidt decomposition to any approximately bandlimited joint amplitude by using the Whittaker-Shannon interpolation formula. While the exponent in the squeezing operator then involves a double sum instead of the usual single sum, we can nevertheless define a packet expansion where each packet typically has a duration short compared to that of the pulse, with the packets thus analogous to the pseudo-Schmidt modes. In general, if the squeezing is weak to moderate, the correlation functions at a particular time are associated with only a few packets, allowing us to ``take apart'' the squeezed light and the resulting photon statistics. Finally we showed that if one is only interested in some finite time regions that form part of the pulse duration, an effective ket can be written as a product of kets associated with those time regions. Then in the weakly squeezed limit, which we can take to be set by the limit $|\mathring\beta|\ll 1$, there will be on average only a few photons within each time region, although in the long pulse limit the total number of photons will be very large.

In extensions of this work, we will consider two- and higher-mode squeezing, which is fairly straightforward, and we will apply this formalism to quantum optics-based experiments such as coincidence-accidental-detection ratios and SU(1,1) interferometry. And instead of describing squeezed light in the spectral-temporal domain, one could expand this formalism to describe squeezed light with a large correlation between the photon wavevector and the conjugate position variables in two, or three dimensions. Another interesting focus is the formulation of relationships between the Schmidt number, the effective Schmidt number discussed here, and the Shannon number (or time-bandwidth product) and its generalizations. Here there is the opportunity to apply the vast mathematical formalism that has already been developed to describe the information content of classical signals and images \cite{fedorov2006short,mikhailova2008biphoton,fedorov2008spontaneous,brecht2013characterizing,landau1962prolate,simons2009slepian,freeden2010handbook,miller2000communicating,pires2009direct,pors2008shannon,slepian1961prolate,wang2017review,whittaker1915xviii,shannon1949communication,butzer1992sampling,slepian1983some}.

With the continuous advancement of custom engineered nonlinear optical systems, the generation of nonclassical ``tri-photon'' states is slowly becoming a reality \cite{banic2022resonant}. Such states and their ``troint amplitudes'' are the generalization of squeezed states and the joint amplitudes discussed here. To characterize these states, generalizations of the Schmidt decomposition (singular-value-decomposition) need to be employed. One such example is the canonical polyadic decomposition (or CP decomposition) \cite{kolda2009tensor}; however, an orthogonal decomposition is not guaranteed to exist. Other generalizations of the Schmidt decomposition exist, such as the Tucker decomposition \cite{kolda2009tensor}, which gains orthogonality but loses the single-sum behaviour of the Schmidt decomposition. An alternative description can be provided using the Whittaker-Shannon interpolation formula. In the same way we generalized it to two-dimensional functions in this paper, it can be directly applied to any number of dimensions in a very straightforward way. Thus the formalism applied here can easily be generalized to situations where other decompositions are not possible.

\begin{acknowledgments}
We thank Xanadu Quantum Technologies for allowing access to their proprietary code repository used in section \ref{sec:A final example}, and Luke Helt for answering related questions. We also thank Marco Liscidini and Nicol\'as Quesada for valuable discussions. This work was supported by the Natural Sciences and
Engineering Research Council of Canada (NSERC). C. D. acknowledges an Ontario Graduate Scholarship.
\end{acknowledgments}

\onecolumngrid
\newpage
\appendix

\section{Field operator}
\label{sec:app:field operator}

For a quasi-1D structure, if $z$ is the direction in which light
is propagating, the electric field operator in the Heisenberg picture
takes the form \cite{quesada2022beyond}
\begin{equation}
     \begin{split}
         \boldsymbol{E}(x,y,z;t)&=\sum_{n}\int_{-\infty}^{\infty}\frac{dk}{\sqrt{2\pi}}\:\boldsymbol{e}_{n}(k;x,y)b_{n}(k)e^{ikz}e^{-i\omega t}+h.c.,
     \end{split}
\end{equation}
where the dependence of $\omega$ on $k$ is identifed by the dispersion
relation, the $\boldsymbol{e}_{n}(k;x,y)$ is a properly normalized
field profile for a transverse mode $n$ propagating with $k$ at frequency
$\omega$, and $b_{n}(k)$ is the associated lowering operator, $\left[b_{n}(k),b_{m}^{\dagger}(k')\right]=\delta_{nm}\delta(k-k').$
We assume that only one transverse mode is of interest and drop the
index $n$, and that only $k>0$ are of interest. Taking
$\omega=vk>0,$ where $v$ is the group velocity, we identify modes
by their frequency $\omega$, putting $c(\omega)=v^{-1/2}b(k)$ so
that $\left[c(\omega),c^{\dagger}(\omega')\right]=\delta(\omega-\omega')$
holds, and we can then write 
\begin{align}
 & \boldsymbol{E}(x,y,z;t)\rightarrow\int_{0}^{\infty}\frac{d\omega}{\sqrt{2\pi}}\frac{\boldsymbol{e}(\frac{\omega}{v};x,y)}{v^{1/2}}c(\omega)e^{i\omega z/v}e^{-i\omega t}+h.c.
\end{align}
Assuming that over the frequency range of interest $\boldsymbol{e}(\omega/v;x,y)$
varies little from its value $\boldsymbol{e}(\omega_{o}/v;x,y)$ at
a center frequency $\omega_{o}$, we can write 
\begin{equation}
    \begin{split}
        & \boldsymbol{E}(x,y,z;t)\rightarrow\frac{\boldsymbol{e}(\frac{\omega_{o}}{v};x,y)}{v^{1/2}}\int_{0}^{\infty}\hspace{-1mm}\frac{d\omega}{\sqrt{2\pi}}c(\omega)e^{i\omega z/v}e^{-i\omega t}+h.c.\\
        & =\frac{\boldsymbol{e}(\frac{\omega_{o}}{v};x,y)}{v^{1/2}}e^{i\omega_{o}z/v}e^{-i\omega_{o}t}\int_{0}^{\infty}\frac{d\omega}{\sqrt{2\pi}}c(\omega)e^{i(\omega-\omega_{o})z/v}e^{-i(\omega-\omega_{o})t}+h.c.
    \end{split}
\end{equation}
Then putting $a(\omega-\omega_{o})\equiv c(\omega),$ the commutation
relations (Eq. \eqref{eq:comm}) hold and for the range of frequencies much less than $\omega_{o}$ we have 
\begin{equation}
    \begin{split}
        \boldsymbol{E}(x,y,z;t)&=\frac{\boldsymbol{e}(\frac{\omega_{o}}{v};x,y)}{v^{1/2}}e^{i\omega_{o}z/v}e^{-i\omega_{o}t}\int_{-\infty}^{\infty}\frac{d\omega}{\sqrt{2\pi}}a(\omega)e^{i\omega z/v}e^{-i\omega t}+h.c.
    \end{split}
\end{equation}
This gives
\begin{align}
 & \boldsymbol{E}(x,y,0;t)=\frac{\boldsymbol{e}(\frac{\omega_{o}}{v};x,y)}{v^{1/2}}e^{-i\omega_{o}t}\overline{a}(t)+h.c.,
\end{align}
where 
\begin{align}
 & \overline{a}(t)=\int_{-\infty}^{\infty}\frac{d\omega}{\sqrt{2\pi}}a(\omega)e^{-i\omega t}.
\end{align}
Now since the Schrödinger operator for $\boldsymbol{E}(x,y,z)$ is
just the Heisenberg operator $\boldsymbol{E}(x,y,z;0),$ for that
Schrödinger operator we have 
\begin{align}
 & \boldsymbol{E}(x,y,z)=\frac{\boldsymbol{e}(\frac{\omega_{o}}{v};x,y)}{v^{1/2}}e^{i\omega_{o}z/v}\overline{a}(-\frac{z}{v})+h.c.
\end{align}
That is, the operator $\overline{a}(t)$ is associated with the electric
field at time $t$ and $z=0$, and as well with the electric field
at zero time and position $z=-vt$, as would be expected because of
the propagation with group velocity $v$. 

\section{Schematic of sinc-hat joint intensity}
\label{sec:app:Schematic of sinc-hat joint intensity}
\begin{figure}
    \centering
    \includegraphics[width = 0.8\linewidth]{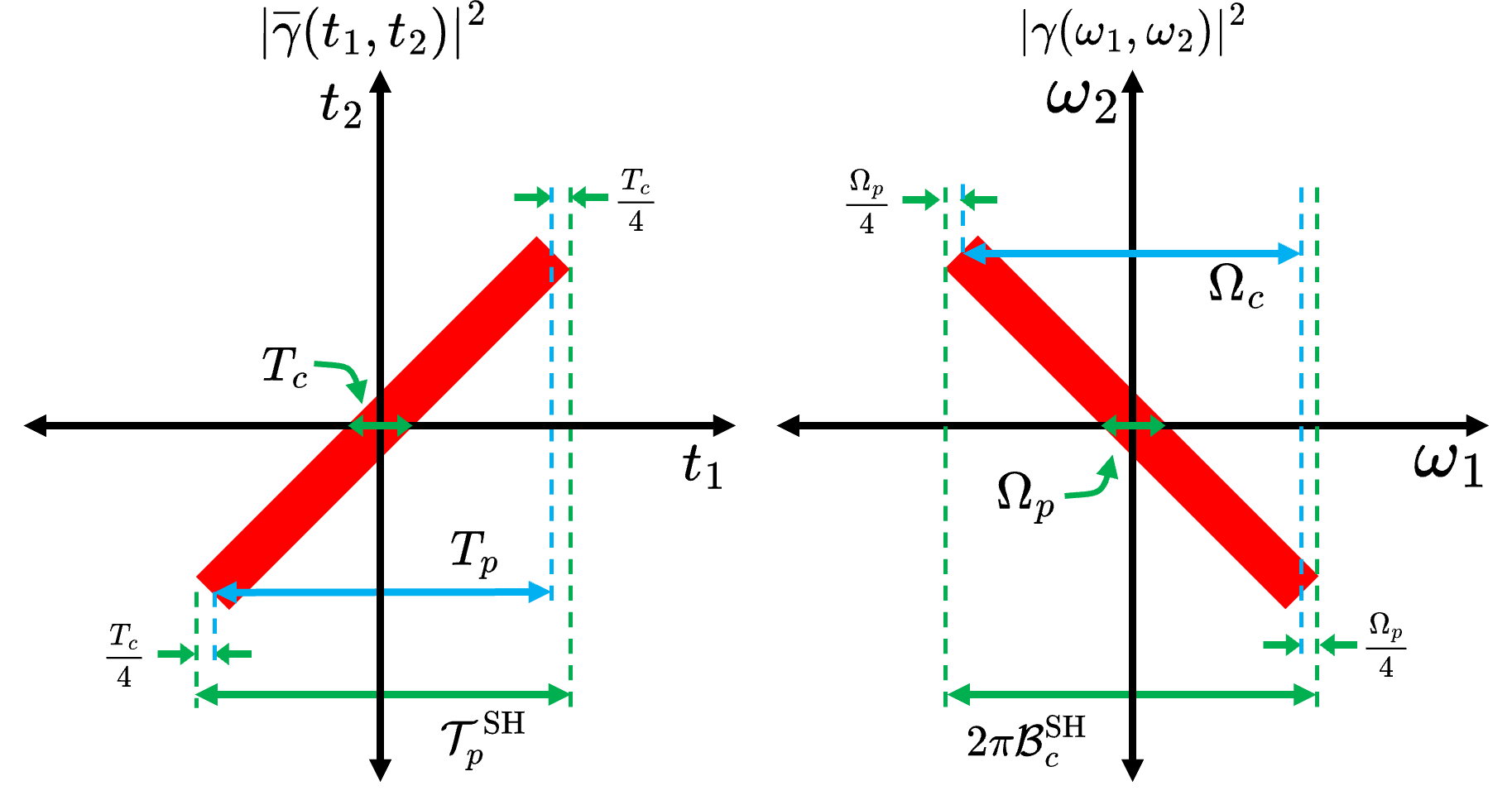}
    \caption{Schematic of the sinc-hat joint intensities showing the extra contributions to the horizontal widths in the respective corners.}
    \label{fig:Picture3}
\end{figure}

Consider the schematic sinc-hat joint intensity shown in Fig. \ref{fig:Picture3}, where we drop the sinc ``tails" along the anti-diagonal (diagonal) direction for the joint temporal (spectral) intensity. Of course, because of the sinc tails the joint intensities extend to infinity in either direction, but for $T_p/T_c$  sufficiently large these contributions are small enough that the behaviour in the anti-diagonal (diagonal) direction is effectively captured by the width $T_c$ ($\Omega_p$) set at $t_2 = 0$  ($\Omega_2 = 0$).

In the schematic one can see that along the lines $t_1=t_2$ ($\omega_1 = -\omega_2$) the joint temporal (spectral) intensity ranges over $-T_p/2 \to T_p/2$ ($-\Omega_c/2 \to \Omega_c/2$), however this is not the \emph{full} horizontal extent of the joint intensities. For the joint temporal intensity there are two extra contributions in the lower left-hand and upper right-hand corners. Using simple geometry one finds that each corner adds a duration of size $T_p/4$ to the horizontal width for a combined width of $\mathcal{T}_p^\mathrm{SH} = T_p + T_c/2$. A similar argument follows for the joint spectral intensity leading to a combined width of $2\pi\mathcal{B}_c^\mathrm{SH} = \Omega_c + \Omega_p/2$.

\section{Joint temporal amplitude scaling}
\label{sec:app:Joint temporal amplitude scaling}
\begin{figure}
    \centering
    \includegraphics{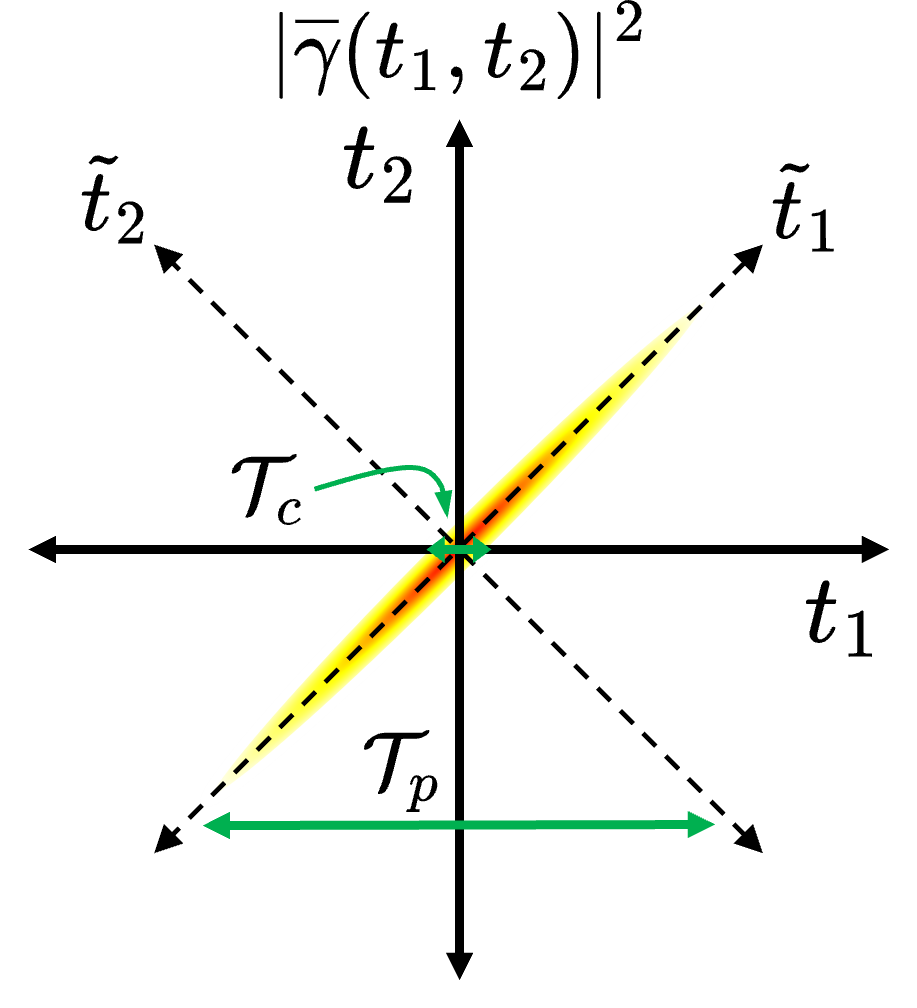}
    \caption{Schematic of a general joint temporal amplitude with a pulse duration and coherence time denoted by $\mathcal{T}_p$ and $\mathcal{T}_c$ respectively, in the original and rotated coordinate system.}
    \label{fig:Picture2}
\end{figure}

Consider the joint temporal amplitude schematically shown in Fig. \ref{fig:picture1}, characterized by $\mathcal{T}_p$ and $\mathcal{T}_c$. To show how the maximum value of the joint temporal amplitude scales with $\mathcal{T}_p$ and $\mathcal{T}_c$ we consider a change of variables 
\begin{equation}
    \tilde t_1 = \frac{1}{\sqrt{2}}\frac{t_1+t_2}{\mathcal{T}_p}, \hspace{5mm} \tilde t_2 = \frac{1}{\sqrt{2}}\frac{t_1-t_2}{\mathcal{T}_c}, \hspace{5mm} t_1 = \frac{1}{\sqrt{2}}(\tilde t_1 \mathcal{T}_p + \tilde t_2 \mathcal{T}_c), \hspace{5mm} t_2 = \frac{1}{\sqrt{2}}(\tilde t_1 \mathcal{T}_p - \tilde t_2 \mathcal{T}_c),
\end{equation}
which are aligned with the long and short axis of the joint temporal amplitude, see Fig. \ref{fig:Picture2}. The new variables $\tilde t_1, \tilde t_2$ are normalized and dimensionless; as they vary over the range $\tilde t_1, \tilde t_2\in [-1/\sqrt{2},1/\sqrt{2}]$, $t_1,t_2$ vary over the range of the joint amplitude specified by $\mathcal{T}_p$ and $\mathcal{T}_c$. The Jacobian of this coordinate transformation is $\text{det}(J) = \mathcal{T}_p\mathcal{T}_c$. Then setting 
\begin{equation}
\label{eq:tildejta}
    \tilde\gamma(\tilde t_1,\tilde t_2) = \sqrt{\mathcal{T}_p\mathcal{T}_c} \overline{\gamma}\left(\frac{1}{\sqrt{2}}(\tilde t_1 \mathcal{T}_p + \tilde t_2 \mathcal{T}_c), \frac{1}{\sqrt{2}}(\tilde t_1 \mathcal{T}_p - \tilde t_2 \mathcal{T}_c)\right) = \sqrt{\mathcal{T}_p\mathcal{T}_c}\overline{\gamma}\left(t_1,t_2\right),
\end{equation}
$\tilde\gamma(\tilde t_1,\tilde t_2)$ is a normalized and dimensionless joint amplitude that satisfies
\begin{equation}
    \int dt_1 dt_2 |\overline\gamma(t_1,t_2)|^2 = \int d\tilde t_1 d \tilde t_2 |\tilde\gamma(\tilde t_1,\tilde t_2)|^2 = 1.
\end{equation}
Since the rotated coordinates vary roughly over $\tilde t_1, \tilde t_2\in [-1/\sqrt{2},1/\sqrt{2}]$ and $|\tilde\gamma(\tilde t_1,\tilde t_2)|$ is always positive, we can infer that the approximate maximum of $|\tilde\gamma(\tilde t_1,\tilde t_2)|$ is on the order of one. Then rewritting Eq. \eqref{eq:tildejta}, we have
\begin{equation}
    \overline\gamma(t_1,t_2) = \frac{\tilde\gamma(\tilde t_1,\tilde t_2)}{\sqrt{\mathcal{T}_p\mathcal{T}_c}},
\end{equation}
and since the location of the maximum of $|\overline\gamma(t_1,t_2)|$ must be the same as that of $|\tilde\gamma(\tilde t_1,\tilde t_2)|$, and the latter maximum is of order unity, in general the maximum value of $\overline\gamma(t_1,t_2)$ scales with $1/\sqrt{\mathcal{T}_p\mathcal{T}_c}$.

\section{Relation between $K$ and $\mathcal{K}$}
\label{sec:app:relation between K and Delta}
We begin with an alternate expression \cite{fedorov2006short,mikhailova2008biphoton} for the Schmidt number,
\begin{equation}
\label{eq:alternateK}
    \frac{1}{K} = \int dt_1dt_2dt_1^\prime dt_2^\prime \overline\gamma(t_1,t_2)\overline\gamma^*(t_1^\prime,t_2)\overline\gamma(t_1^\prime,t_2^\prime)\overline\gamma(t_1,t_2^\prime).
\end{equation}
which can be confirmed by using the Schmidt decomposition (Eq. \eqref{eq:Schmidt}) in Eq. \eqref{eq:alternateK} and recalling the orthogonality of the Schmidt modes. 
If a Whittaker-Shannon decomposition (Eq. \eqref{eq:WSequations}) constructed with an appropriate bandwidth limit is put in Eq. \eqref{eq:alternateK}, with 
the use of the orthogonality relations of $\overline{\chi}_n(t)$ we find 
\begin{equation}
\label{eq:SchmidtWS}
    \frac{1}{K} = \tau^4 \sum_{a,b,c,d} \overline\gamma_{ab} \overline\gamma^*_{cb}\overline\gamma_{cd}\overline\gamma^*_{ad} = \tau^4\text{Tr}(\boldsymbol{\bar\gamma}\boldsymbol{\bar\gamma}^\dagger\boldsymbol{\bar\gamma}\boldsymbol{\bar\gamma}^\dagger), 
\end{equation}
where we have written $\overline\gamma_{ab}$ as short-hand for $\overline\gamma(a\tau,b\tau)$, and in the second equality we use $\boldsymbol{\bar\gamma}$ to indicate the matrix with components $\overline\gamma_{ab}$.

We now make use of the identity 
\begin{equation}
    |\text{Tr}(\mathbf{A}\mathbf{B}^\dagger)|^2\le \text{Tr}(\mathbf{A}\mathbf{A}^\dagger)\text{Tr}(\mathbf{B}\mathbf{B}^\dagger),
\end{equation}
which is the generalization of the Cauchy–Schwarz inequality to matrices under the trace inner product \cite{friedberg2014linear}. For $\mathbf{A}$ an $l\times l$ matrix, we put $\mathbf{B} = \boldsymbol{1}_{l\times l}$, the identity matrix of size $l\times l$, and find
\begin{equation}
    |\text{Tr}(\mathbf{A})|^2 \le \text{Tr}(\mathbf{A}\mathbf{A}^\dagger)l,
\end{equation}
or rather
\begin{equation}
\label{eq:traceinequality}
    \text{Tr}(\mathbf{A}\mathbf{A}^\dagger)\ge \frac{|\text{Tr}(\mathbf{A})|^2}{l}.
\end{equation}
Next we set $\mathbf{A} = \boldsymbol{\bar\gamma}\boldsymbol{\bar\gamma}^\dagger$, and since $\text{Tr}(\boldsymbol{\bar\gamma}\boldsymbol{\bar\gamma}^\dagger)\ge 0$ the inequality in Eq. \eqref{eq:traceinequality} leads to
\begin{equation}
    \text{Tr}(\boldsymbol{\bar\gamma}\boldsymbol{\bar\gamma}^\dagger\boldsymbol{\bar\gamma}\boldsymbol{\bar\gamma}^\dagger)\ge \frac{\text{Tr}(\boldsymbol{\bar\gamma}\boldsymbol{\bar\gamma}^\dagger)^2}{l}.
\end{equation}
After inputting this result into the equation \eqref{eq:SchmidtWS} for the Schmidt number, we arrive at
\begin{equation}
\label{eq:mainpoint}
    \frac{1}{K} \ge \frac{(\tau^2\text{Tr}(\boldsymbol{\bar\gamma}\boldsymbol{\bar\gamma}^\dagger))^2}{l}.
\end{equation}
This can be significantly simplified by noting that the joint amplitude is normalized to unity, and inputting the Whittaker-Shannon decomposition (Eq. \eqref{eq:WSequations}) into 
\begin{equation}
    \int dt_1dt_2 |\overline{\gamma}(t_1,t_2)|^2 = 1,
\end{equation}
we find $\tau^2\text{Tr}(\boldsymbol{\bar\gamma}\boldsymbol{\bar\gamma}^\dagger) = 1$,
and then from Eq. \eqref{eq:mainpoint} we have
\begin{equation}
    K\le l.
\end{equation}
Now $l$ is the dimension of the matrix $\boldsymbol{\gamma}$, but because $\overline{\chi}_n(t)$ has a width set by $\tau$ one needs on the order of $l = \mathcal{T}_p/\tau$ functions to interpolate the joint amplitude; see the discussion around Eq \eqref{eq:betabardef}. If we choose $\tau = \mathcal{T}_c$, then $l = \mathcal{T}_p/\mathcal{T}_c = \mathcal{K}$ and we find
\begin{equation}
    K \le \mathcal{K},
\end{equation}
which completes the argument.

\section{Correlation functions}
\label{sec:app:Correlation functions}
We begin by using the inverse relation given by Eq. \eqref{eq:inverseB_n} to write the correlation functions \eqref{eq:gs-1} as
\begin{subequations}
    \begin{gather}
        \overline{G}^{(1)}(t_1,t_2) = \overline{\chi}_x^*(t_1)\left\langle \Psi|B_x^\dagger B_y|\Psi\right\rangle\overline{\chi}_y(t_2),\\
  \overline{G}^{(2)}(t_{1},t_{2})=\overline{\chi}_w^*(t_1)\overline{\chi}_x^*(t_2)\left\langle \Psi|B_w\dg B_x\dg B_y B_z|\Psi\right\rangle\overline{\chi}_y(t_2)\overline{\chi}_z(t_1),
    \end{gather}
\end{subequations}
we use the same summation convention as in the text. Then using the transformation given by Eq. \eqref{eq:transformBn} we evaluate
\begin{equation}
    \begin{split}
        \left\langle \Psi|B_x^\dagger B_y|\Psi\right\rangle& = \bra{\T{vac}} \nu_{xa}^* B_a \nu_{yb} B_b\dg \vac\\
        &=\nu_{xa}^*\nu_{ya}\\
        &= (\m{\nu}^*\m{\nu}^T)_{xy}\\
        &=(\mathbf{U}^*(\T{sinh}\mathbf{P}^*) (\T{sinh}\mathbf{P}^T)\mathbf{U}^T)_{xy}\\
        &=(\T{sinh}^2\mathbf{P})_{xy},
    \end{split}
\end{equation}
and
\begin{equation}
    \begin{split}
        &\left\langle \Psi|B_w\dg B_x\dg B_y B_z|\Psi\right\rangle \\
        &= \bra{\T{vac}}\nu^*_{wa}B_a(\mu^*_{xb}B_b\dg + \nu^*_{xb}B_b)(\mu_{yc}B_c +\nu_{yc}B_c\dg)\nu_{zd}B_d\dg\vac\\
        &=S_{wa}^*\mu_{xb}^*\mu_{yc}\nu_{zd}\bra{\T{vac}}B_aB_b\dg B_cB_d\dg \vac + \nu_{wa}^*\nu_{xb}^*\nu_{yc}\nu_{zd}\bra{\T{vac}}B_aB_b B_c\dg B_d\dg \vac\\
        &= \nu_{wa}^*\mu_{xa}^*\mu_{yc}\nu_{zc} + \nu_{wa}^*\nu_{xb}^*\nu_{yb}\nu_{za} + \nu_{wa}^*\nu_{xb}^*\nu_{ya}\nu_{zb}\\
        &=(\m{\nu\mu}^T)^*_{wx} (\m{\nu\mu}^T)_{zy} + (\m{\nu}^*\m{\nu}^T)_{wz}(\m{\nu}^*\m{\nu}^T)_{xy} + (\m{\nu}^*\m{\nu}^T)_{wy}(\m{\nu}^*\m{\nu}^T)_{xz}\\
        &=(\mathbf{U}(\T{sinh}\mathbf{P})(\T{cosh}\mathbf{P}))^*_{wx}(\mathbf{U}(\T{sinh}\mathbf{P})(\T{cosh}\mathbf{P}))_{zy} + (\T{sinh}^2\mathbf{P})_{wz}(\T{sinh}^2\mathbf{P})_{xy}+(\T{sinh}^2\mathbf{P})_{wy}(\T{sinh}^2\mathbf{P})_{xz}.
    \end{split}
\end{equation}
So $\overline{G}^{(1)}(t_1,t_2)$ is given by
\begin{equation}
    \begin{split}
        \overline{G}^{(1)}(t_1,t_2)& = \overline{\chi}_x^*(t_1)(\T{sinh}^2\mathbf{P})_{xy}\overline{\chi}_y(t_2)\\
        & = \m{\overline{\chi}}\dg(t_1)(\T{sinh}^2\mathbf{P})\m{\overline{\chi}}(t_2),
    \end{split}
\end{equation}
where $\m{\overline{\chi}}(t) = (..., \overline{\chi}_{-1}(t), \overline{\chi}_0(t), \overline{\chi}_1(t),...)^T$ is the column vector formed from the set $\{\overline{\chi}_n(t)\}$ for a given $t$, and
\begin{equation}
    \begin{split}
        \overline{G}^{(2)}(t_1,t_2) &= \overline{\chi}_w^*(t_1)\overline{\chi}_x^*(t_2)(\mathbf{U}(\T{sinh}\mathbf{P})(\T{cosh}\mathbf{P}))^*_{wx}(\mathbf{U}(\T{sinh}\mathbf{P})(\T{cosh}\mathbf{P}))_{zy}\overline{\chi}_y(t_2)\overline{\chi}_z(t_1)\\
        &+ \overline{\chi}_w^*(t_1)\overline{\chi}_x^*(t_2)(\T{sinh}^2\mathbf{P})_{wz}(\T{sinh}^2\mathbf{P})_{xy}\overline{\chi}_y(t_2)\overline{\chi}_z(t_1)\\
        &+\overline{\chi}_w^*(t_1)\overline{\chi}_x^*(t_2)(\T{sinh}^2\mathbf{P})_{wy}(\T{sinh}^2\mathbf{P})_{xz}\overline{\chi}_y(t_2)\overline{\chi}_z(t_1)\\
        &=|\m{\overline{\chi}}^T(t_1)\mathbf{U}(\T{sinh}\mathbf{P})(\T{cosh}\mathbf{P})\m{\overline{\chi}}(t_2)|^2 + [\m{\overline{\chi}}\dg(t_1)(\T{sinh}^2\mathbf{P})\m{\overline{\chi}}(t_1)][\m{\overline{\chi}}\dg(t_2)(\T{sinh}^2\mathbf{P})\m{\overline{\chi}}(t_2)] + |\m{\overline{\chi}}\dg(t_1)(\T{sinh}^2\mathbf{P})\m{\overline{\chi}}(t_2)|^2\\
        &=|\m{\overline{\chi}}^T(t_1)\mathbf{U}(\T{sinh}\mathbf{P})(\T{cosh}\mathbf{P})\m{\overline{\chi}}(t_2)|^2 + \overline{G}^{(1)}(t_1)\overline{G}^{(1)}(t_2) + |\overline{G}^{(1)}(t_1,t_2)|^2.
    \end{split}
\end{equation}

\section{Mode-by-mode calculation}
\label{sec:app:Mode-by-mode calculation}
In this Appendix we show that the photon statistics of the sinc-hat joint amplitude can be described on a ``mode-by-mode" basis using the pseudo-Schmidt decomposition. We then generalize this to the Whittaker-Shannon decomposition, and show that at least approximately a ``mode-by-mode" description can be introduced.
\subsection{Pseudo-Schmidt decomposition}
We start by considering the analytical form of $\overline{G}^{(2)}(\Delta t)$ \eqref{eq:G2analyticlCWlimit} in the CW limit,
\begin{equation}
    \begin{split}
        \left.\overline{G}^{(2)}(\Delta t)\right|_{|\Delta t|\lesssim T_c} &\approx\overline{G}^{(2)}(\Delta t = 0) \\
        &=\frac{1}{T_c^2}(3N_\T{mode}^2 + N_\T{mode}),
    \end{split}
\end{equation}
where we have approximated the ``sinc'' function as unity and used Eq. \eqref{eq:Nmode}; as expected, we recover the usual standard result for $G^{(2)}(\Delta t = 0)$ \cite{loudon2000quantum,drago2022aspects}, and for $\Delta t>T_c$ we have $\overline{G}^{(2)}(\Delta t)\le \overline{G}^{(2)}(\Delta t = 0)$, the condition for ``bunched light'' \cite{walls1983squeezed}. Since $\overline{G}^{(2)}(\Delta t)$ varies little over a length of time $|\Delta t|\lesssim T_c$, the quantity $T_c \overline{G}^{(2)}(\Delta t)$ is a ``coincidence rate,'' and 
\begin{equation}
\label{eq:G2timesN}
    T_p \left(T_e\left.\overline{G}^{(2)}(\Delta t)\right|_{\Delta t \lesssim T_c} \right) \approx \mathcal{N}(3N_\T{mode}^2 + N_\T{mode})
\end{equation}
is the total coincidence count over the duration of the pulse. 

To gain insight into the scaling of $\overline{G}^{(2)}(\Delta t)$ with $N_\mathrm{mode}$, we note that for a single pseudo-Schmidt mode labeled by $n$, the probability of detecting $2x$ photons is \cite{loudon2000quantum}
\begin{equation}
    P_n(2x) = \frac{1}{c}\left( \frac{\sqrt{(2x)!}}{2^x x!}\right)^2 \left(\frac{s}{c}\right)^{2x},
\end{equation}
where we put $s \equiv\mathrm{sinh}(r)$ and $c \equiv \mathrm{cosh}(r)$, and $r = |\beta|/\sqrt{\mathcal{N}}$ is the squeezing parameter which is independent of $n$ for the values of $n$ for which it is nonzero (see Eq. \eqref{eq:app_p_n}). Since each pseudo-Schmidt mode is normalized we have
\begin{equation}
    1 = \sum_{x=0}^\infty P_n(2x) \Rightarrow c = \sum_{x=0}^\infty \left( \frac{\sqrt{(2x)!}}{2^x x!}\right)^2 \left(\frac{s}{c}\right)^{2x}.
\end{equation}
Taking the derivative of both sides with respect to $r$, and simplifying using hyperbolic trigonometry identities followed by a re-indexing of the sum, we find
\begin{equation}
    \sum_{x=0}^\infty (2x)P_n(2x) = s^2 = N_\T{mode},
\end{equation}
as expected; the average number of photons in each pseudo-Schmidt mode is given by $N_\T{mode}$. Notice that we can re-write the expectation value as 
\begin{equation}
    N_\T{mode} = \sum_{x=0}^\infty (2x)P_n(2x) =  \sum_{x=0}^\infty \frac{(2x)!}{(2x-1)!}P_n(2x) = \sum_{x=0}^\infty \mathcal{P}(2x, 1)P_n(2x),
\end{equation}
where $\mathcal{P}(2x, 1)$ is the number of permutations of $2x$ objects when we select one object. Again taking the derivative of both sides with respect to $r$,
after further hyperbolic trigonometry manipulations and a re-indexing of the sum, we find
\begin{equation}
    \sum_{x=0}^\infty (2x)(2x-1)P_n(2x) = 3N_\T{mode}^2 + N_\T{mode},
\end{equation}
where the right hand side is the familiar scaling of $\overline{G}^{(2)}(\Delta t)$ with $N_\mathrm{mode}$. However, playing the same game, we can re-write this as
\begin{equation}
    \sum_{x=0}^\infty (2x)(2x-1)P_n(2x) = \sum_{x=0}^\infty \frac{(2x)!}{(2x-2)!}P_n(2x) = \sum_{x=0}^\infty \mathcal{P}(2x,2)P_n(2x) = 3N_\T{mode}^2 + N_\T{mode},
\end{equation}
where $\mathcal{P}(2x,2)$ is the number of permutations of $2x$ objects when we select two of them. 

Thus for short time differences, on the order of $T_c$, the coincidence rate is the expectation value of the total number of ways of ``picking'' two photons from any pseudo-Schmidt mode, and to calculate the total coincidence count we multiply by $\mathcal{N}$ (\eqref{eq:G2timesN}), the total number of pseudo-Schmidt modes (cf. \eqref{eq:Nmode} and \eqref{eq:Npulse}). Indeed, for time differences less than $T_c$ we can set $n=m$  for both terms in Eq. \eqref{eq:Gwork_approx}, so that to good approximation we have
\begin{equation}
\label{eq:G2PSsignlesum}
    G^{(2)}(t_1,t_2) \approx (3N_\T{mode}^2 + N_\T{mode})\sum_n |\overline{\eta}_n(t_1)|^2 |\overline{\eta}_n(t_2)|^2. 
\end{equation}
For such time differences the coincidence count at a particular $t_1, t_2$ is associated with only a few pseudo-Schmidt modes, unlike in the Schmidt decomposition, where the inclusion of the contributions of a large number of Schmidt modes is needed.

\subsection{Whittaker-Shannon decomposition}
We now consider the Whittaker-Shannon decomposition, starting with the exact results for the correlation functions (\ref{eq:WSG1},\ref{eq:WSG2}). The photon density is given by
\begin{equation}
    \begin{split}
    \label{eq:WSG1_diag}
        \overline{G}^{(1)}(t)& = \m{\overline{\chi}}\dg(t)(\T{sinh}^2\mathbf{P})\m{\overline{\chi}}(t)\\
        & = \sum_{n,m} \overline{\chi}_n^*(t)(\T{sinh}^2\mathbf{P})_{nm} \overline{\chi}_m(t)\\
        &\approx\sum_n (\T{sinh}^2\mathbf{P})_{nn} |\overline{\chi}_n(t)|^2,
    \end{split}
\end{equation}
where in the second line we expanded the matrix multiplication, and in the third line approximated the double sum as just the diagonal contributions. In reducing the double sum to a single sum we are losing significant contributions to the photon density, because typically $(\T{sinh}^2\mathbf{P})_{nm}$ will be nonzero over a small range of $|n-m|$; further, the product of two neighbouring Whittaker-Shannon modes is not negligible. While the final result of Eq. \eqref{eq:WSG1_diag} provides a crude approximate at best, it is similar to the pseudo-Schmidt result given by Eq. \eqref{eq:Gresults_approx}. This suggests that we should identify 
\begin{equation}
    \Gamma_n^2 = (\T{sinh}^2\mathbf{P})_{nn}
\end{equation}
as the number of photons in each mode, and indeed we find this to be the case in Section  \ref{sec:Packet expansion}. 

We now turn our attention to $\overline{G}^{(2)}(t_1,t_2)$, starting with the coherent contribution
\begin{equation}
    \begin{split}
        \overline{G}_\text{coh}^{(2)}(t_1,t_2)& = |\m{\overline{\chi}}^T(t_1)\mathbf{U}(\T{sinh}\mathbf{P})(\T{cosh}\mathbf{P})\m{\overline{\chi}}(t_2)|^2\\\
        & = \sum_{n,m,p,q} \overline{\chi}^*_n(t_1) (\mathbf{U}\T{sinh}\mathbf{P}\T{cosh}\mathbf{P})_{nm}^*\overline{\chi}^*_m(t_2) \overline{\chi}_p(t_1) (\mathbf{U}\T{sinh}\mathbf{P}\T{cosh}\mathbf{P})_{pq}\overline{\chi}_q(t_2)\\
        &\approx \sum_{n,m}  \left|(\mathbf{U}\T{sinh}\mathbf{P}\T{cosh}\mathbf{P})_{nm}\right|^2|\overline{\chi}_n(t_1)|^2|\overline{\chi}_m(t_2)|^2,
    \end{split}
\end{equation}
where in the last line we used the same approximation that we did for the photon density. Then if we consider times $t_1,t_2$ such that $|t_1 - t_2| \lesssim \tau$, the significant contribution to the sum is when $n=m$, and we have
\begin{equation}
    \begin{split}
        \overline{G}_\text{coh}^{(2)}(t_1,t_2)& \approx \sum_{n}  \left|(\mathbf{U}\T{sinh}\mathbf{P}\T{cosh}\mathbf{P})_{nn}\right|^2|\overline{\chi}_n(t_1)|^2|\overline{\chi}_n(t_2)|^2.
    \end{split}
\end{equation}
Following the same steps for the incoherent contribution, for times $t_1,t_2$ such that $|t_1 - t_2| \lesssim \tau$, $\overline{G}^{(2)}(t_1,t_2)$ is given by
\begin{equation}
    \overline{G}^{(2)}(t_1,t_2) = \sum_n (\left|(\mathbf{U}\T{sinh}\mathbf{P}\T{cosh}\mathbf{P})_{nn}\right|^2 + 2\Gamma_n^4) |\overline{\chi}_n(t_1)|^2|\overline{\chi}_n(t_2)|^2.
\end{equation}
While again a crude approximation at best, we do note the similarity between the final result and the pseudo-Schmidt result in Eq. \eqref{eq:G2PSsignlesum}. 

\section{}
\label{sec:app:D}
We start with equation \eqref{eq:rhodef} for $\overline{\rho}_n(t)$, and invert it by multiplying by $(\T{sinh}\mathbf{P})^{-1} = \T{csch}\mathbf{P}$
\begin{equation}
\label{eq:chi=rho}
    \overline{\chi}_n(t) = \sum_{m} (\T{csch}\mathbf{P})_{nm}\Gamma_m\overline{\rho}_m(t).
\end{equation}
Taking the term inside the squared norm for the coherent contribution (Eq. \eqref{eq:WScohcont}) and writing it in its index representation, we insert the form of $\overline{\chi}_n(t)$ in terms of $\overline{\rho}_m(t)$ using \eqref{eq:chi=rho}, so that
\begin{equation}
    \begin{split}
    \m{\overline{\chi}}^T(t_1)\mathbf{U}(\T{sinh}\mathbf{P})(\T{cosh}\mathbf{P})\m{\overline{\chi}}(t_2)& = \overline{\chi}_x(t_1) U_{xa}(\T{sinh}\mathbf{P})_{ab}(\T{cosh}\mathbf{P})_{by}\overline{\chi}_y(t_2) \\
    &=(\T{csch}\mathbf{P})_{xm}\Gamma_m\overline{\rho}_m(t_1)U_{xa}(\T{sinh}\mathbf{P})_{ab}(\T{cosh}\mathbf{P})_{by}(\T{csch}\mathbf{P})_{yn}\Gamma_n\overline{\rho}_n(t_2)\\
    &=\Gamma_n\Gamma_m\overline{\rho}_m(t_1)\overline{\rho}_n(t_2)((\T{csch}\mathbf{Q})\mathbf{U}(\T{sinh}\mathbf{P})(\T{cosh}\mathbf{P})(\T{csch}\mathbf{P}))_{mn}\\
    &=\Gamma_n\Gamma_m\overline{\rho}_m(t_1)\overline{\rho}_n(t_2)(\mathbf{U}\T{coth}\mathbf{P})_{mn},
    \end{split}
\end{equation}
where in deriving this result we used the properties of $\mathbf{P}, \mathbf{Q}$ and $\mathbf{U}$ discussed in section \ref{sec:Employing the Whittaker-Shannon decomposition}, and we set $\T{coth}\mathbf{P} = (\T{tanh}\mathbf{P})^{-1} = (\T{cosh}\mathbf{P})(\T{csch}\mathbf{P})$. The resulting coherent contribution to the correlation function is then given by
\begin{equation}
    \overline{G}^{(2)}_\T{coh}(t_1,t_2) = \sum_{n,m}|\Gamma_n\Gamma_m(\mathbf{U}\T{coth}\mathbf{P})_{mn}\overline{\rho}_m(t_1)\overline{\rho}_n(t_2)|^2.
\end{equation}

\bibliography{apssamp}

\providecommand{\noopsort}[1]{}\providecommand{\singleletter}[1]{#1}%
\begin{thebibliography}{38}%
\makeatletter
\providecommand \@ifxundefined [1]{%
 \@ifx{#1\undefined}
}%
\providecommand \@ifnum [1]{%
 \ifnum #1\expandafter \@firstoftwo
 \else \expandafter \@secondoftwo
 \fi
}%
\providecommand \@ifx [1]{%
 \ifx #1\expandafter \@firstoftwo
 \else \expandafter \@secondoftwo
 \fi
}%
\providecommand \natexlab [1]{#1}%
\providecommand \enquote  [1]{``#1''}%
\providecommand \bibnamefont  [1]{#1}%
\providecommand \bibfnamefont [1]{#1}%
\providecommand \citenamefont [1]{#1}%
\providecommand \href@noop [0]{\@secondoftwo}%
\providecommand \href [0]{\begingroup \@sanitize@url \@href}%
\providecommand \@href[1]{\@@startlink{#1}\@@href}%
\providecommand \@@href[1]{\endgroup#1\@@endlink}%
\providecommand \@sanitize@url [0]{\catcode `\\12\catcode `\$12\catcode
  `\&12\catcode `\#12\catcode `\^12\catcode `\_12\catcode `\%12\relax}%
\providecommand \@@startlink[1]{}%
\providecommand \@@endlink[0]{}%
\providecommand \url  [0]{\begingroup\@sanitize@url \@url }%
\providecommand \@url [1]{\endgroup\@href {#1}{\urlprefix }}%
\providecommand \urlprefix  [0]{URL }%
\providecommand \Eprint [0]{\href }%
\providecommand \doibase [0]{https://doi.org/}%
\providecommand \selectlanguage [0]{\@gobble}%
\providecommand \bibinfo  [0]{\@secondoftwo}%
\providecommand \bibfield  [0]{\@secondoftwo}%
\providecommand \translation [1]{[#1]}%
\providecommand \BibitemOpen [0]{}%
\providecommand \bibitemStop [0]{}%
\providecommand \bibitemNoStop [0]{.\EOS\space}%
\providecommand \EOS [0]{\spacefactor3000\relax}%
\providecommand \BibitemShut  [1]{\csname bibitem#1\endcsname}%
\let\auto@bib@innerbib\@empty
\bibitem [{\citenamefont {Pirandola}\ \emph {et~al.}(2018)\citenamefont
  {Pirandola}, \citenamefont {Bardhan}, \citenamefont {Gehring}, \citenamefont
  {Weedbrook},\ and\ \citenamefont {Lloyd}}]{pirandola2018advances}%
  \BibitemOpen
  \bibfield  {author} {\bibinfo {author} {\bibfnamefont {S.}~\bibnamefont
  {Pirandola}}, \bibinfo {author} {\bibfnamefont {B.~R.}\ \bibnamefont
  {Bardhan}}, \bibinfo {author} {\bibfnamefont {T.}~\bibnamefont {Gehring}},
  \bibinfo {author} {\bibfnamefont {C.}~\bibnamefont {Weedbrook}},\ and\
  \bibinfo {author} {\bibfnamefont {S.}~\bibnamefont {Lloyd}},\ }\bibfield
  {title} {\bibinfo {title} {Advances in photonic quantum sensing},\
  }\href@noop {} {\bibfield  {journal} {\bibinfo  {journal} {Nature Photonics}\
  }\textbf {\bibinfo {volume} {12}},\ \bibinfo {pages} {724} (\bibinfo {year}
  {2018})}\BibitemShut {NoStop}%
\bibitem [{\citenamefont {Kolobov}(2007)}]{kolobov2007quantum}%
  \BibitemOpen
  \bibfield  {author} {\bibinfo {author} {\bibfnamefont {M.~I.}\ \bibnamefont
  {Kolobov}},\ }\href@noop {} {\emph {\bibinfo {title} {Quantum imaging}}}\
  (\bibinfo  {publisher} {Springer Science \& Business Media},\ \bibinfo {year}
  {2007})\BibitemShut {NoStop}%
\bibitem [{\citenamefont {Bourassa}\ \emph {et~al.}(2021)\citenamefont
  {Bourassa}, \citenamefont {Alexander}, \citenamefont {Vasmer}, \citenamefont
  {Patil}, \citenamefont {Tzitrin}, \citenamefont {Matsuura}, \citenamefont
  {Su}, \citenamefont {Baragiola}, \citenamefont {Guha}, \citenamefont
  {Dauphinais} \emph {et~al.}}]{bourassa2021blueprint}%
  \BibitemOpen
  \bibfield  {author} {\bibinfo {author} {\bibfnamefont {J.~E.}\ \bibnamefont
  {Bourassa}}, \bibinfo {author} {\bibfnamefont {R.~N.}\ \bibnamefont
  {Alexander}}, \bibinfo {author} {\bibfnamefont {M.}~\bibnamefont {Vasmer}},
  \bibinfo {author} {\bibfnamefont {A.}~\bibnamefont {Patil}}, \bibinfo
  {author} {\bibfnamefont {I.}~\bibnamefont {Tzitrin}}, \bibinfo {author}
  {\bibfnamefont {T.}~\bibnamefont {Matsuura}}, \bibinfo {author}
  {\bibfnamefont {D.}~\bibnamefont {Su}}, \bibinfo {author} {\bibfnamefont
  {B.~Q.}\ \bibnamefont {Baragiola}}, \bibinfo {author} {\bibfnamefont
  {S.}~\bibnamefont {Guha}}, \bibinfo {author} {\bibfnamefont {G.}~\bibnamefont
  {Dauphinais}}, \emph {et~al.},\ }\bibfield  {title} {\bibinfo {title}
  {Blueprint for a scalable photonic fault-tolerant quantum computer},\
  }\href@noop {} {\bibfield  {journal} {\bibinfo  {journal} {Quantum}\ }\textbf
  {\bibinfo {volume} {5}},\ \bibinfo {pages} {392} (\bibinfo {year}
  {2021})}\BibitemShut {NoStop}%
\bibitem [{\citenamefont {Quesada}\ \emph {et~al.}(2022)\citenamefont
  {Quesada}, \citenamefont {Helt}, \citenamefont {Menotti}, \citenamefont
  {Liscidini},\ and\ \citenamefont {Sipe}}]{quesada2022beyond}%
  \BibitemOpen
  \bibfield  {author} {\bibinfo {author} {\bibfnamefont {N.}~\bibnamefont
  {Quesada}}, \bibinfo {author} {\bibfnamefont {L.}~\bibnamefont {Helt}},
  \bibinfo {author} {\bibfnamefont {M.}~\bibnamefont {Menotti}}, \bibinfo
  {author} {\bibfnamefont {M.}~\bibnamefont {Liscidini}},\ and\ \bibinfo
  {author} {\bibfnamefont {J.}~\bibnamefont {Sipe}},\ }\bibfield  {title}
  {\bibinfo {title} {Beyond photon pairs—nonlinear quantum photonics in the
  high-gain regime: a tutorial},\ }\href@noop {} {\bibfield  {journal}
  {\bibinfo  {journal} {Advances in Optics and Photonics}\ }\textbf {\bibinfo
  {volume} {14}},\ \bibinfo {pages} {291} (\bibinfo {year} {2022})}\BibitemShut
  {NoStop}%
\bibitem [{\citenamefont {Quesada}\ and\ \citenamefont
  {Sipe}(2014)}]{quesada2014effects}%
  \BibitemOpen
  \bibfield  {author} {\bibinfo {author} {\bibfnamefont {N.}~\bibnamefont
  {Quesada}}\ and\ \bibinfo {author} {\bibfnamefont {J.}~\bibnamefont {Sipe}},\
  }\bibfield  {title} {\bibinfo {title} {Effects of time ordering in quantum
  nonlinear optics},\ }\href@noop {} {\bibfield  {journal} {\bibinfo  {journal}
  {Physical Review A}\ }\textbf {\bibinfo {volume} {90}},\ \bibinfo {pages}
  {063840} (\bibinfo {year} {2014})}\BibitemShut {NoStop}%
\bibitem [{\citenamefont {Glauber}(1963)}]{glauber1963quantum}%
  \BibitemOpen
  \bibfield  {author} {\bibinfo {author} {\bibfnamefont {R.~J.}\ \bibnamefont
  {Glauber}},\ }\bibfield  {title} {\bibinfo {title} {The quantum theory of
  optical coherence},\ }\href@noop {} {\bibfield  {journal} {\bibinfo
  {journal} {Physical Review}\ }\textbf {\bibinfo {volume} {130}},\ \bibinfo
  {pages} {2529} (\bibinfo {year} {1963})}\BibitemShut {NoStop}%
\bibitem [{\citenamefont {Chebotarev}\ and\ \citenamefont
  {Teretenkov}(2014)}]{chebotarev2014singular}%
  \BibitemOpen
  \bibfield  {author} {\bibinfo {author} {\bibfnamefont {A.~M.}\ \bibnamefont
  {Chebotarev}}\ and\ \bibinfo {author} {\bibfnamefont {A.~E.}\ \bibnamefont
  {Teretenkov}},\ }\bibfield  {title} {\bibinfo {title} {Singular value
  decomposition for the takagi factorization of symmetric matrices},\
  }\href@noop {} {\bibfield  {journal} {\bibinfo  {journal} {Applied
  Mathematics and Computation}\ }\textbf {\bibinfo {volume} {234}},\ \bibinfo
  {pages} {380} (\bibinfo {year} {2014})}\BibitemShut {NoStop}%
\bibitem [{\citenamefont {Fedorov}\ \emph {et~al.}(2006)\citenamefont
  {Fedorov}, \citenamefont {Efremov}, \citenamefont {Volkov},\ and\
  \citenamefont {Eberly}}]{fedorov2006short}%
  \BibitemOpen
  \bibfield  {author} {\bibinfo {author} {\bibfnamefont {M.}~\bibnamefont
  {Fedorov}}, \bibinfo {author} {\bibfnamefont {M.}~\bibnamefont {Efremov}},
  \bibinfo {author} {\bibfnamefont {P.}~\bibnamefont {Volkov}},\ and\ \bibinfo
  {author} {\bibfnamefont {J.}~\bibnamefont {Eberly}},\ }\bibfield  {title}
  {\bibinfo {title} {Short-pulse or strong-field breakup processes: a route to
  study entangled wave packets},\ }\href@noop {} {\bibfield  {journal}
  {\bibinfo  {journal} {Journal of Physics B: Atomic, Molecular and Optical
  Physics}\ }\textbf {\bibinfo {volume} {39}},\ \bibinfo {pages} {S467}
  (\bibinfo {year} {2006})}\BibitemShut {NoStop}%
\bibitem [{\citenamefont {Mikhailova}\ \emph {et~al.}(2008)\citenamefont
  {Mikhailova}, \citenamefont {Volkov},\ and\ \citenamefont
  {Fedorov}}]{mikhailova2008biphoton}%
  \BibitemOpen
  \bibfield  {author} {\bibinfo {author} {\bibfnamefont {Y.~M.}\ \bibnamefont
  {Mikhailova}}, \bibinfo {author} {\bibfnamefont {P.}~\bibnamefont {Volkov}},\
  and\ \bibinfo {author} {\bibfnamefont {M.}~\bibnamefont {Fedorov}},\
  }\bibfield  {title} {\bibinfo {title} {Biphoton wave packets in parametric
  down-conversion: Spectral and temporal structure and degree of
  entanglement},\ }\href@noop {} {\bibfield  {journal} {\bibinfo  {journal}
  {Physical Review A}\ }\textbf {\bibinfo {volume} {78}},\ \bibinfo {pages}
  {062327} (\bibinfo {year} {2008})}\BibitemShut {NoStop}%
\bibitem [{\citenamefont {Fedorov}\ \emph {et~al.}(2008)\citenamefont
  {Fedorov}, \citenamefont {Efremov}, \citenamefont {Volkov}, \citenamefont
  {Moreva}, \citenamefont {Straupe},\ and\ \citenamefont
  {Kulik}}]{fedorov2008spontaneous}%
  \BibitemOpen
  \bibfield  {author} {\bibinfo {author} {\bibfnamefont {M.}~\bibnamefont
  {Fedorov}}, \bibinfo {author} {\bibfnamefont {M.}~\bibnamefont {Efremov}},
  \bibinfo {author} {\bibfnamefont {P.}~\bibnamefont {Volkov}}, \bibinfo
  {author} {\bibfnamefont {E.}~\bibnamefont {Moreva}}, \bibinfo {author}
  {\bibfnamefont {S.}~\bibnamefont {Straupe}},\ and\ \bibinfo {author}
  {\bibfnamefont {S.}~\bibnamefont {Kulik}},\ }\bibfield  {title} {\bibinfo
  {title} {Spontaneous parametric down-conversion: Anisotropical and
  anomalously strong narrowing of biphoton momentum correlation
  distributions},\ }\href@noop {} {\bibfield  {journal} {\bibinfo  {journal}
  {Physical Review A}\ }\textbf {\bibinfo {volume} {77}},\ \bibinfo {pages}
  {032336} (\bibinfo {year} {2008})}\BibitemShut {NoStop}%
\bibitem [{\citenamefont {Brecht}\ and\ \citenamefont
  {Silberhorn}(2013)}]{brecht2013characterizing}%
  \BibitemOpen
  \bibfield  {author} {\bibinfo {author} {\bibfnamefont {B.}~\bibnamefont
  {Brecht}}\ and\ \bibinfo {author} {\bibfnamefont {C.}~\bibnamefont
  {Silberhorn}},\ }\bibfield  {title} {\bibinfo {title} {Characterizing
  entanglement in pulsed parametric down-conversion using chronocyclic wigner
  functions},\ }\href@noop {} {\bibfield  {journal} {\bibinfo  {journal}
  {Physical Review A}\ }\textbf {\bibinfo {volume} {87}},\ \bibinfo {pages}
  {053810} (\bibinfo {year} {2013})}\BibitemShut {NoStop}%
\bibitem [{\citenamefont {Landau}\ and\ \citenamefont
  {Pollak}(1962)}]{landau1962prolate}%
  \BibitemOpen
  \bibfield  {author} {\bibinfo {author} {\bibfnamefont {H.~J.}\ \bibnamefont
  {Landau}}\ and\ \bibinfo {author} {\bibfnamefont {H.~O.}\ \bibnamefont
  {Pollak}},\ }\bibfield  {title} {\bibinfo {title} {Prolate spheroidal wave
  functions, fourier analysis and uncertainty—iii: the dimension of the space
  of essentially time-and band-limited signals},\ }\href@noop {} {\bibfield
  {journal} {\bibinfo  {journal} {Bell System Technical Journal}\ }\textbf
  {\bibinfo {volume} {41}},\ \bibinfo {pages} {1295} (\bibinfo {year}
  {1962})}\BibitemShut {NoStop}%
\bibitem [{\citenamefont {Simons}(2009)}]{simons2009slepian}%
  \BibitemOpen
  \bibfield  {author} {\bibinfo {author} {\bibfnamefont {F.~J.}\ \bibnamefont
  {Simons}},\ }\bibfield  {title} {\bibinfo {title} {Slepian functions and
  their use in signal estimation and spectral analysis},\ }\href@noop {}
  {\bibfield  {journal} {\bibinfo  {journal} {arXiv preprint arXiv:0909.5368}\
  } (\bibinfo {year} {2009})}\BibitemShut {NoStop}%
\bibitem [{\citenamefont {Freeden}\ \emph {et~al.}(2010)\citenamefont
  {Freeden}, \citenamefont {Nashed},\ and\ \citenamefont
  {Sonar}}]{freeden2010handbook}%
  \BibitemOpen
  \bibfield  {author} {\bibinfo {author} {\bibfnamefont {W.}~\bibnamefont
  {Freeden}}, \bibinfo {author} {\bibfnamefont {M.~Z.}\ \bibnamefont
  {Nashed}},\ and\ \bibinfo {author} {\bibfnamefont {T.}~\bibnamefont
  {Sonar}},\ }\href@noop {} {\emph {\bibinfo {title} {Handbook of
  geomathematics}}}\ (\bibinfo  {publisher} {Springer Science \& Business
  Media},\ \bibinfo {year} {2010})\BibitemShut {NoStop}%
\bibitem [{\citenamefont {Miller}(2000)}]{miller2000communicating}%
  \BibitemOpen
  \bibfield  {author} {\bibinfo {author} {\bibfnamefont {D.~A.}\ \bibnamefont
  {Miller}},\ }\bibfield  {title} {\bibinfo {title} {Communicating with waves
  between volumes: evaluating orthogonal spatial channels and limits on
  coupling strengths},\ }\href@noop {} {\bibfield  {journal} {\bibinfo
  {journal} {Applied Optics}\ }\textbf {\bibinfo {volume} {39}},\ \bibinfo
  {pages} {1681} (\bibinfo {year} {2000})}\BibitemShut {NoStop}%
\bibitem [{\citenamefont {Pires}\ \emph {et~al.}(2009)\citenamefont {Pires},
  \citenamefont {Monken},\ and\ \citenamefont {Van~Exter}}]{pires2009direct}%
  \BibitemOpen
  \bibfield  {author} {\bibinfo {author} {\bibfnamefont {H.~D.~L.}\
  \bibnamefont {Pires}}, \bibinfo {author} {\bibfnamefont {C.}~\bibnamefont
  {Monken}},\ and\ \bibinfo {author} {\bibfnamefont {M.}~\bibnamefont
  {Van~Exter}},\ }\bibfield  {title} {\bibinfo {title} {Direct measurement of
  transverse-mode entanglement in two-photon states},\ }\href@noop {}
  {\bibfield  {journal} {\bibinfo  {journal} {Physical Review A}\ }\textbf
  {\bibinfo {volume} {80}},\ \bibinfo {pages} {022307} (\bibinfo {year}
  {2009})}\BibitemShut {NoStop}%
\bibitem [{\citenamefont {Pors}\ \emph {et~al.}(2008)\citenamefont {Pors},
  \citenamefont {Oemrawsingh}, \citenamefont {Aiello}, \citenamefont
  {Van~Exter}, \citenamefont {Eliel}, \citenamefont {Woerdman} \emph
  {et~al.}}]{pors2008shannon}%
  \BibitemOpen
  \bibfield  {author} {\bibinfo {author} {\bibfnamefont {J.}~\bibnamefont
  {Pors}}, \bibinfo {author} {\bibfnamefont {S.}~\bibnamefont {Oemrawsingh}},
  \bibinfo {author} {\bibfnamefont {A.}~\bibnamefont {Aiello}}, \bibinfo
  {author} {\bibfnamefont {M.}~\bibnamefont {Van~Exter}}, \bibinfo {author}
  {\bibfnamefont {E.}~\bibnamefont {Eliel}}, \bibinfo {author} {\bibfnamefont
  {J.}~\bibnamefont {Woerdman}}, \emph {et~al.},\ }\bibfield  {title} {\bibinfo
  {title} {Shannon dimensionality of quantum channels and its application to
  photon entanglement},\ }\href@noop {} {\bibfield  {journal} {\bibinfo
  {journal} {Physical review letters}\ }\textbf {\bibinfo {volume} {101}},\
  \bibinfo {pages} {120502} (\bibinfo {year} {2008})}\BibitemShut {NoStop}%
\bibitem [{\citenamefont {Loudon}(2000)}]{loudon2000quantum}%
  \BibitemOpen
  \bibfield  {author} {\bibinfo {author} {\bibfnamefont {R.}~\bibnamefont
  {Loudon}},\ }\href@noop {} {\emph {\bibinfo {title} {The quantum theory of
  light}}}\ (\bibinfo  {publisher} {OUP Oxford},\ \bibinfo {year}
  {2000})\BibitemShut {NoStop}%
\bibitem [{\citenamefont {Dayan}(2007)}]{dayan2007theory}%
  \BibitemOpen
  \bibfield  {author} {\bibinfo {author} {\bibfnamefont {B.}~\bibnamefont
  {Dayan}},\ }\bibfield  {title} {\bibinfo {title} {Theory of two-photon
  interactions with broadband down-converted light and entangled photons},\
  }\href@noop {} {\bibfield  {journal} {\bibinfo  {journal} {Physical Review
  A}\ }\textbf {\bibinfo {volume} {76}},\ \bibinfo {pages} {043813} (\bibinfo
  {year} {2007})}\BibitemShut {NoStop}%
\bibitem [{\citenamefont {Edrelyi}\ \emph {et~al.}(1953)\citenamefont {Edrelyi}
  \emph {et~al.}}]{edrelyi1953higher}%
  \BibitemOpen
  \bibfield  {author} {\bibinfo {author} {\bibfnamefont {A.}~\bibnamefont
  {Edrelyi}} \emph {et~al.},\ }\bibfield  {title} {\bibinfo {title} {Higher
  transcendental functions, vol. ii mcgraw-hill book co},\ }\href@noop {}
  {\bibfield  {journal} {\bibinfo  {journal} {Inc. New York}\ } (\bibinfo
  {year} {1953})},\ \bibinfo {note} {see Page 194.}\BibitemShut {Stop}%
\bibitem [{{\relax DLMF}()}]{NIST:DLMF}%
  \BibitemOpen
  {\relax DLMF},\ \href {https://dlmf.nist.gov/} {\bibinfo {title} {{\it NIST
  Digital Library of Mathematical Functions}}},\ \bibinfo {howpublished}
  {\url{https://dlmf.nist.gov/}, Release 1.1.11 of 2023-09-15},\ \bibinfo
  {note} {f.~W.~J. Olver, A.~B. {Olde Daalhuis}, D.~W. Lozier, B.~I. Schneider,
  R.~F. Boisvert, C.~W. Clark, B.~R. Miller, B.~V. Saunders, H.~S. Cohl, and
  M.~A. McClain, eds. Eq. 18.18.28}\BibitemShut {NoStop}%
\bibitem [{\citenamefont {Sakurai}\ and\ \citenamefont
  {Commins}(1995)}]{sakurai1995modern}%
  \BibitemOpen
  \bibfield  {author} {\bibinfo {author} {\bibfnamefont {J.~J.}\ \bibnamefont
  {Sakurai}}\ and\ \bibinfo {author} {\bibfnamefont {E.~D.}\ \bibnamefont
  {Commins}},\ }\href@noop {} {\bibinfo {title} {Modern quantum mechanics,
  revised edition}} (\bibinfo {year} {1995})\BibitemShut {NoStop}%
\bibitem [{\citenamefont {Raymer}\ and\ \citenamefont
  {Landes}(2022)}]{raymer2022theory}%
  \BibitemOpen
  \bibfield  {author} {\bibinfo {author} {\bibfnamefont {M.~G.}\ \bibnamefont
  {Raymer}}\ and\ \bibinfo {author} {\bibfnamefont {T.}~\bibnamefont
  {Landes}},\ }\bibfield  {title} {\bibinfo {title} {Theory of two-photon
  absorption with broadband squeezed vacuum},\ }\href@noop {} {\bibfield
  {journal} {\bibinfo  {journal} {Physical Review A}\ }\textbf {\bibinfo
  {volume} {106}},\ \bibinfo {pages} {013717} (\bibinfo {year}
  {2022})}\BibitemShut {NoStop}%
\bibitem [{\citenamefont {Cutipa}\ and\ \citenamefont
  {Chekhova}(2022)}]{cutipa2022bright}%
  \BibitemOpen
  \bibfield  {author} {\bibinfo {author} {\bibfnamefont {P.}~\bibnamefont
  {Cutipa}}\ and\ \bibinfo {author} {\bibfnamefont {M.~V.}\ \bibnamefont
  {Chekhova}},\ }\bibfield  {title} {\bibinfo {title} {Bright squeezed vacuum
  for two-photon spectroscopy: simultaneously high resolution in time and
  frequency, space and wavevector},\ }\href@noop {} {\bibfield  {journal}
  {\bibinfo  {journal} {Optics Letters}\ }\textbf {\bibinfo {volume} {47}},\
  \bibinfo {pages} {465} (\bibinfo {year} {2022})}\BibitemShut {NoStop}%
\bibitem [{\citenamefont {Slepian}\ and\ \citenamefont
  {Pollak}(1961)}]{slepian1961prolate}%
  \BibitemOpen
  \bibfield  {author} {\bibinfo {author} {\bibfnamefont {D.}~\bibnamefont
  {Slepian}}\ and\ \bibinfo {author} {\bibfnamefont {H.~O.}\ \bibnamefont
  {Pollak}},\ }\bibfield  {title} {\bibinfo {title} {Prolate spheroidal wave
  functions, fourier analysis and uncertainty—i},\ }\href@noop {} {\bibfield
  {journal} {\bibinfo  {journal} {Bell System Technical Journal}\ }\textbf
  {\bibinfo {volume} {40}},\ \bibinfo {pages} {43} (\bibinfo {year}
  {1961})}\BibitemShut {NoStop}%
\bibitem [{\citenamefont {Wang}(2017)}]{wang2017review}%
  \BibitemOpen
  \bibfield  {author} {\bibinfo {author} {\bibfnamefont {L.-L.}\ \bibnamefont
  {Wang}},\ }\bibfield  {title} {\bibinfo {title} {A review of prolate
  spheroidal wave functions from the perspective of spectral methods},\
  }\href@noop {} {\bibfield  {journal} {\bibinfo  {journal} {J. Math. Study}\
  }\textbf {\bibinfo {volume} {50}},\ \bibinfo {pages} {101} (\bibinfo {year}
  {2017})}\BibitemShut {NoStop}%
\bibitem [{\citenamefont {Cohen}\ and\ \citenamefont
  {Louie}(2016)}]{cohen2016fundamentals}%
  \BibitemOpen
  \bibfield  {author} {\bibinfo {author} {\bibfnamefont {M.~L.}\ \bibnamefont
  {Cohen}}\ and\ \bibinfo {author} {\bibfnamefont {S.~G.}\ \bibnamefont
  {Louie}},\ }\href@noop {} {\emph {\bibinfo {title} {Fundamentals of condensed
  matter physics}}}\ (\bibinfo  {publisher} {Cambridge University Press},\
  \bibinfo {year} {2016})\BibitemShut {NoStop}%
\bibitem [{\citenamefont {Whittaker}(1915)}]{whittaker1915xviii}%
  \BibitemOpen
  \bibfield  {author} {\bibinfo {author} {\bibfnamefont {E.~T.}\ \bibnamefont
  {Whittaker}},\ }\bibfield  {title} {\bibinfo {title} {Xviii.—on the
  functions which are represented by the expansions of the
  interpolation-theory},\ }\href@noop {} {\bibfield  {journal} {\bibinfo
  {journal} {Proceedings of the Royal Society of Edinburgh}\ }\textbf {\bibinfo
  {volume} {35}},\ \bibinfo {pages} {181} (\bibinfo {year} {1915})}\BibitemShut
  {NoStop}%
\bibitem [{\citenamefont {Shannon}(1949)}]{shannon1949communication}%
  \BibitemOpen
  \bibfield  {author} {\bibinfo {author} {\bibfnamefont {C.~E.}\ \bibnamefont
  {Shannon}},\ }\bibfield  {title} {\bibinfo {title} {Communication in the
  presence of noise},\ }\href@noop {} {\bibfield  {journal} {\bibinfo
  {journal} {Proceedings of the IRE}\ }\textbf {\bibinfo {volume} {37}},\
  \bibinfo {pages} {10} (\bibinfo {year} {1949})}\BibitemShut {NoStop}%
\bibitem [{\citenamefont {Butzer}\ and\ \citenamefont
  {Stens}(1992)}]{butzer1992sampling}%
  \BibitemOpen
  \bibfield  {author} {\bibinfo {author} {\bibfnamefont {P.~L.}\ \bibnamefont
  {Butzer}}\ and\ \bibinfo {author} {\bibfnamefont {R.~L.}\ \bibnamefont
  {Stens}},\ }\bibfield  {title} {\bibinfo {title} {Sampling theory for not
  necessarily band-limited functions: a historical overview},\ }\href@noop {}
  {\bibfield  {journal} {\bibinfo  {journal} {SIAM review}\ }\textbf {\bibinfo
  {volume} {34}},\ \bibinfo {pages} {40} (\bibinfo {year} {1992})}\BibitemShut
  {NoStop}%
\bibitem [{\citenamefont {Slepian}(1983)}]{slepian1983some}%
  \BibitemOpen
  \bibfield  {author} {\bibinfo {author} {\bibfnamefont {D.}~\bibnamefont
  {Slepian}},\ }\bibfield  {title} {\bibinfo {title} {Some comments on fourier
  analysis, uncertainty and modeling},\ }\href@noop {} {\bibfield  {journal}
  {\bibinfo  {journal} {SIAM review}\ }\textbf {\bibinfo {volume} {25}},\
  \bibinfo {pages} {379} (\bibinfo {year} {1983})}\BibitemShut {NoStop}%
\bibitem [{\citenamefont {Lo}\ and\ \citenamefont
  {Sollie}(1993)}]{lo1993generalized}%
  \BibitemOpen
  \bibfield  {author} {\bibinfo {author} {\bibfnamefont {C.}~\bibnamefont
  {Lo}}\ and\ \bibinfo {author} {\bibfnamefont {R.}~\bibnamefont {Sollie}},\
  }\bibfield  {title} {\bibinfo {title} {Generalized multimode squeezed
  states},\ }\href@noop {} {\bibfield  {journal} {\bibinfo  {journal} {Physical
  Review A}\ }\textbf {\bibinfo {volume} {47}},\ \bibinfo {pages} {733}
  (\bibinfo {year} {1993})}\BibitemShut {NoStop}%
\bibitem [{\citenamefont {Ma}\ and\ \citenamefont
  {Rhodes}(1990)}]{ma1990multimode}%
  \BibitemOpen
  \bibfield  {author} {\bibinfo {author} {\bibfnamefont {X.}~\bibnamefont
  {Ma}}\ and\ \bibinfo {author} {\bibfnamefont {W.}~\bibnamefont {Rhodes}},\
  }\bibfield  {title} {\bibinfo {title} {Multimode squeeze operators and
  squeezed states},\ }\href@noop {} {\bibfield  {journal} {\bibinfo  {journal}
  {Physical Review A}\ }\textbf {\bibinfo {volume} {41}},\ \bibinfo {pages}
  {4625} (\bibinfo {year} {1990})}\BibitemShut {NoStop}%
\bibitem [{\citenamefont {Banic}\ \emph {et~al.}(2022)\citenamefont {Banic},
  \citenamefont {Liscidini},\ and\ \citenamefont {Sipe}}]{banic2022resonant}%
  \BibitemOpen
  \bibfield  {author} {\bibinfo {author} {\bibfnamefont {M.}~\bibnamefont
  {Banic}}, \bibinfo {author} {\bibfnamefont {M.}~\bibnamefont {Liscidini}},\
  and\ \bibinfo {author} {\bibfnamefont {J.}~\bibnamefont {Sipe}},\ }\bibfield
  {title} {\bibinfo {title} {Resonant and nonresonant integrated third-order
  parametric down-conversion},\ }\href@noop {} {\bibfield  {journal} {\bibinfo
  {journal} {Physical Review A}\ }\textbf {\bibinfo {volume} {106}},\ \bibinfo
  {pages} {013710} (\bibinfo {year} {2022})}\BibitemShut {NoStop}%
\bibitem [{\citenamefont {Kolda}\ and\ \citenamefont
  {Bader}(2009)}]{kolda2009tensor}%
  \BibitemOpen
  \bibfield  {author} {\bibinfo {author} {\bibfnamefont {T.~G.}\ \bibnamefont
  {Kolda}}\ and\ \bibinfo {author} {\bibfnamefont {B.~W.}\ \bibnamefont
  {Bader}},\ }\bibfield  {title} {\bibinfo {title} {Tensor decompositions and
  applications},\ }\href@noop {} {\bibfield  {journal} {\bibinfo  {journal}
  {SIAM review}\ }\textbf {\bibinfo {volume} {51}},\ \bibinfo {pages} {455}
  (\bibinfo {year} {2009})}\BibitemShut {NoStop}%
\bibitem [{\citenamefont {Friedberg}\ \emph {et~al.}(2014)\citenamefont
  {Friedberg}, \citenamefont {Insel},\ and\ \citenamefont
  {Spence}}]{friedberg2014linear}%
  \BibitemOpen
  \bibfield  {author} {\bibinfo {author} {\bibfnamefont {S.}~\bibnamefont
  {Friedberg}}, \bibinfo {author} {\bibfnamefont {A.}~\bibnamefont {Insel}},\
  and\ \bibinfo {author} {\bibfnamefont {L.}~\bibnamefont {Spence}},\ }\href
  {https://books.google.ca/books?id=KyB0DAAAQBAJ} {\emph {\bibinfo {title}
  {Linear Algebra}}}\ (\bibinfo  {publisher} {Pearson Education},\ \bibinfo
  {year} {2014})\BibitemShut {NoStop}%
\bibitem [{\citenamefont {Drago}\ and\ \citenamefont
  {Sipe}(2022)}]{drago2022aspects}%
  \BibitemOpen
  \bibfield  {author} {\bibinfo {author} {\bibfnamefont {C.}~\bibnamefont
  {Drago}}\ and\ \bibinfo {author} {\bibfnamefont {J.}~\bibnamefont {Sipe}},\
  }\bibfield  {title} {\bibinfo {title} {Aspects of two-photon absorption of
  squeezed light: The continuous-wave limit},\ }\href@noop {} {\bibfield
  {journal} {\bibinfo  {journal} {Physical Review A}\ }\textbf {\bibinfo
  {volume} {106}},\ \bibinfo {pages} {023115} (\bibinfo {year}
  {2022})}\BibitemShut {NoStop}%
\bibitem [{\citenamefont {Walls}(1983)}]{walls1983squeezed}%
  \BibitemOpen
  \bibfield  {author} {\bibinfo {author} {\bibfnamefont {D.~F.}\ \bibnamefont
  {Walls}},\ }\bibfield  {title} {\bibinfo {title} {Squeezed states of light},\
  }\href@noop {} {\bibfield  {journal} {\bibinfo  {journal} {nature}\ }\textbf
  {\bibinfo {volume} {306}},\ \bibinfo {pages} {141} (\bibinfo {year}
  {1983})}\BibitemShut {NoStop}%
\end{thebibliography}%
\end{document}